\documentclass[iop]{emulateapj}
\begin{document} 

\def\nms{\mathsurround=0pt}
\def\oversim#1#2{\lower 2pt\vbox{\baselineskip 0pt \lineskip 1pt \ialign{$\nms#1\hfil##\hfil$\crcr#2\crcr\sim\crcr}}}
\def\ltsim{\mathrel{\mathpalette\oversim<}}
\def\gtsim{\mathrel{\mathpalette\oversim>}}

\title{The Metal Aversion of LGRBs}
\author{J. F. Graham}
\affil{Space Telescope Science Institute, 3700 San Martin Drive, Baltimore MD 21218 \\ Department of Physics and Astronomy, Johns Hopkins University, Baltimore, MD 21218}
\author{A. S. Fruchter}
\affil{Space Telescope Science Institute, 3700 San Martin Drive, Baltimore MD 21218}

\begin{abstract}

Recently, it has been suggested that the metallicity aversion of long-duration gamma-ray bursts (LGRBs) is not intrinsic to their formation, but rather a consequence of the anti-correlation between star-formation and metallicity seen in the general galaxy population.  To investigate this proposal, we compare the metallicity of the hosts of LGRBs, broad-lined Type Ic (Ic-bl) supernovae (SNe), and Type II SNe to each other and to the metallicity distribution of star-forming galaxies using the Sloan DIgital Sky Survey (SDSS) to represent galaxies in the local universe and the Team Keck Redshift Survey (TKRS) for galaxies at intermediate redshifts.

The differing metallicity distributions of LGRB hosts and the star formation in local galaxies forces us to conclude that the low-metallicity preference of LGRBs is not primarily driven by the anti-correlation between star-formation and metallicity, but rather must be overwhelmingly due to the astrophysics of the LGRBs themselves.  Three quarters of our LGRB sample are found at metallicities below 12+log(O/H) < 8.6, while less than a tenth of local star-formation is at similarly low metallicities.  However, our supernova samples are statistically consistent with the metallicity distribution of the general galaxy population.  Additionally, we show that the star-formation rate distribution of the LGRB and SNe host populations are consistent with the star-formation rate distribution of the SDSS galaxy sample.  This provides further evidence that the low-metallicity distribution of LGRBs is not caused by the general properties of star-forming galaxies.  Using the TKRS population of galaxies, we can exclude the possibility that the LGRB host metallicity aversion is caused by the decrease in galaxy metallicity with redshift, as this effect is clearly much smaller than the observed LGRB host metallicity bias over the redshift span of our sample.  The presence of the strong metallicity difference between LGRBs and Type Ic-bl SNe largely eliminates the possibility that the observed LGRB metallicity bias is a byproduct of a difference in the initial mass functions of the galaxy populations.  Rather, metallicity below half-solar must be a fundamental component of the evolutionary process that separates LGRBs from the vast majority of Type Ic-bl SNe and from the bulk of local star-formation.

\end{abstract}

\section{Introduction}

Shortly after long soft gamma-ray bursts (LGRBs) were identified as extragalactic events spanning cosmological distances, it became apparent that they are frequently found in a particular type of host galaxy: blue irregulars \citep{Fruchter1999, Fruchter, LeFlochblue, LeFlochblue2002}.  LGRBs have shown a strong preference for occurring in starforming galaxies \citep{Fruchter1999, Christensen, LeFloch2006}, which often exhibit bright emission lines \citep{Bloom, Vreeswijk, Levesque051022} indicative of substantial populations of young, massive stars.  LGRBs have also frequently been associated with Type Ic supernova \citep{Stanek2003, Hjorth2003, Woosley} whose spectral lines show broadening from their high-velocity ($\sim 15,000$ km s$^{-1}$) ejecta.  While a broad-lined Type Ic (Ic-bl) supernova (SN) has been found underlying the light of nearly every LGRB in which a deep spectroscopic search could be performed.  The converse however, is most certainly not the case.  Type Ic-bl SNe events without LGRBs are vastly more common than those with an accompanying LGRB.  Thus LGRBs are now seen as a rare subtype of core-collapse supernovae (CCSNe) set along a progressive narrowing of stellar types which raise a number of astrophysical questions: What makes some CCSNe Type Ic's?  What makes some Type Ic's relativistic broad-lined events, and what makes some of these relativistic broad-lined events also have a gamma-ray burst?  The search for the physical constraints that produce stars along this narrowing pyramid of phenomena is presently a primary focus of the study of massive stellar evolution.  

Several years ago, \cite{Fruchter} performed a detailed study of the LGRB host galaxy population, using the hosts of the Great Observatories Origins Deep Survey (GOODS) CCSNe sample as a comparative group.  The results of this work showed a surprising difference between the two populations.  While half of the GOODs CCSNe occurred in grand design spirals, with the other half in irregulars, however only one out of 42 LGRB host galaxies was in a grand design spiral.  If one constrains the LGRB host population to a redshift of 1.2 or less so as to match the redshift distribution of the GOODS SNe sample this drops to one out of 18, still a rather surprising result.  The remainder of the LGRB host population are composed of generally faint, blue, irregular galaxies.  This strong bias galaxy type has been supported by later work using an even larger comparison sample of CCSNe \citep{Svensson}.

In addition to the galaxy type preference, \cite{Fruchter} showed a strong preference for LGRB's occurring in the brightest, and hence likely the most star forming regions of their hosts.  This suggests that LGRBs are formed from very massive progenitors (O and B stars) which often do not have time to travel far from their birth sites before exploding.  However, CCSNe follow the blue light distribution of their host galaxies.  This suggests that they do not generally require as massive progenitors as LGRBs.  However, given that there is no evidence for a difference in the stellar IMF between blue irregulars and spirals \citep{Bastian}, massive stellar progenitors should be just as available per unit star-formation in spirals as they are in irregulars.  \cite{Fruchter} therefore concluded that LGRB formation likely requires a low metallicity progenitor.  The bias towards irregular galaxies is then a result of their generally low metallicity as expected by the mass-metallicity relation.  

A similar conclusion was reached by \cite{Stanek2007}, who showed that the very nearest LGRB hosts all have low metallicity when compared to similar magnitude galaxies in the Sloan Digital Sky Survey (SDSS) sample.  Furthermore, \cite{Kewley2007} found the LGRB host sample to be comparable to extremely metal-poor galaxies in luminosity-metallicity relation, star-formation rate (SFR), and internal extinction.  Additionally, \cite{Wolf} formally demonstrated that a metallicity dependence could indeed explain the difference in morphological type between the hosts of LGRBs and the more general CCSNe population.

The suggestion that LGRBs form preferentially in low-metallicity hosts, however, is not uniformly accepted.  Using a wide redshift sample of LGRBs, \cite{Savaglio} argued that LGRB hosts lie on the same mass-metallicity relation as regular galaxies.  Indeed, \cite{Berger2006} have used this claim to argue that because the host of LGRB 020127 is unusually bright it must also be metal rich.  \cite{Peeples} however, show the existence of low metallicity outliers on the luminosity-metallicity relation, and argue against the assignment of metallicities (to individual galaxies) based only on their luminosities.  In particular they highlight the morphologically similarities of their bright outliers to the brighter hosts in the \cite{Fruchter} sample.

In a comparison of CCSNe hosts, \cite{Modjaz2008} showed that LGRBs are observed to occur in host galaxies with much lower metallicities than either the hosts of Type Ic-bl SNe or the bulk of the star-forming galaxies in SDSS.  Given the dramatic difference in metallicity between the Type Ic-bl and LGRB samples, which persisted even when hosts of similar luminosity were compared, \cite{Modjaz2008} concluded that this is due to a metallicity avoidance among the LGRBs.  As Type Ic-bl SNe are frequently observed to be coincident with many LGRB events this seemed to suggest a metallicity dependent step in either the formation of the requisite gamma-ray jet or in its ability to escape their progenitor during a highly relativistic core collapse SNe of an already outer layer stripped massive star \citep{Langer}.

Recently however, \cite{Mannucci} has suggested that the well know mass metallicity relationship (c.f. \citealt{T04}) should be extended to a mass, metallicity, star-formation relationship.  In this relationship the metallicity of a galaxy of a given stellar mass is anti-correlated with its SFR.  Indeed, \cite{Mannucci} suggests that this single relationship may be able to extend to quite high redshift, with the generally lower-metallicities of galaxies at higher redshift corresponding well to their higher SFR.

\cite{Kocevski} and \cite{MannucciLGRBs} independently argue that the apparent LGRB preference for low-metallicity hosts is thus due to a more fundamental mass, metallicity, \& SFR relation.  LGRB hosts are low-metallicity because they are effectively selected based on the basis of star-formation.  However this does not explain why, as already noted by \cite{Modjaz2008}, the Type Ic-bl SNe without associated LGRBs do not show a preference for low-metallicity hosts.  Nor does it explain why CCSNe in the GOODs sample should live in a different galaxy host population than LGRBs \citep{Fruchter, Svensson}.  One would expect both of these SN populations to be biased towards star-forming galaxies, and thus low-metallicity.

In order to more clearly address this issue we compare the metallicity distribution of the hosts of LGRBs with that of the hosts of several similar indicators of star-formation: LGRBs, Type Ic-bl, and Type II SNe.  Finally, we compare all of these hosts galaxies with more general catalogs of star-forming galaxies, compiled from the SDSS and Team Keck Redshift Survey (TKRS).  This broad-range of star-formation indicators gives a good and varied base from which to determine whether LGRBs are indeed biased toward low metallicity even in excess of that expected due to the high SFRs of their hosts.

In the next sections we describe in detail the samples used for these comparisons.  We then present the methods employed to place all of the samples on a consistent metallicity scale.  Finally we present the direct comparison of the metallicities of these populations.

\section{The Samples} \label{sample_pops}

The fundamental question of this paper is whether LGRBs track star-formation, or whether their choice of host is biased by metallicity.  In order to determine this, we will want to compare the locations of LGRBs with other tracers of star-formation.  

The most obvious tracer of star-formation to compare with LGRBs are star-forming galaxies themselves.  We use the Sloan Digital Sky Survey (SDSS - \citealt{SDSSdr7}) to obtain a sample of star-forming galaxies at low-redshift and the Team Keck Redshift Survey (TKRS - \citealt{TKRS}) to allow us to extend our galaxy star-forming sample out to $z \sim 1$.  These samples are described in Sections \ref{SDSS} and \ref{TKRS}, respectively.

Type II SNe also provide a very direct estimate of star-formation.  While one might worry that some Type II SNe are lost to extinction, the same is true of the H$\alpha$ emission which is used to estimate the star-formation of galaxies and of (to a lesser extent) LGRBs themselves (a radio or X-ray position is often sufficient to identify the host of a LGRB).  

Many LGRBs are associated with Type Ic-bl SNe.  Indeed in almost every case where one estimates that one should have been able to discern the light (or in quite a few cases the spectrum) of a Type Ic underlying the afterglow, were it there, evidence for a Type Ic has been found (see \citealt{Cano_thesis} for a detailed discussion of this).  However, where spectral observations have been done, they reveal Type Ic with broad-lines due to a very energetic explosion imparting large velocities to the ejecta.  Therefore, we also compare the locations of LGRBs with Type II and Type Ic-bl SNe, which are not known to be associated with LGRBs.

Finally, we present our sample of LGRBs we have assembled using both our own spectra and spectra available in the literature.

Below we describe all our samples in detail.

\subsection{Star-forming Galaxies from the SDSS} \label{SDSS} \label{ggp} \label{ggp2}

The SDSS is a combined photometric and spectroscopic survey of approximately 14,555 square degrees with spectroscopy over 930,000 galaxies.  The SDSS gives accurate astrometry, photometry and line fluxes.  The survey has already been used for numerous studies, and in particular the work of \cite{T04} examining the distribution of metallicities of the galaxies.  Here we will use this catalog as the basis of our comparison, but we rederive metallicities using the R$_{23}$ method (as for the LGRB sample).  Our metallicity derivation will be discussed in more detail in Section \ref{metal_measurement}.  

The SDSS provides a pre-selected sample of star-forming galaxies using the {\it SUBCLASS} starforming or starburst \citep{SDSS_spec_class}.  These galaxies have lines indicative of star-formation, but do not appear to be dominated by an Active Galactic Nucleus (AGN).  However, we further restrict these galaxies based on line strengths in order to assure we can obtain reasonably accurate metallicities.  

As described in Section \ref{R23} the $R_{23}$ diagnostic requires the 3727 {\AA} [O II] doublet, the 4861 {\AA} H$\beta$ line and the 4959 {\AA} \& 5007 {\AA} [O III] lines.  The 3727 {\AA} [O II] doublet, commonly used by the $R_{23}$ method, is actually a combination of the 3726 {\AA} and 3729 {\AA} lines, which are frequently unresolved in long slit spectroscopy.  Here these lines are summed and their errors combined statistically to form the 3727 {\AA} line before applying the signal to noise cut in order to better replicate the long slit methodology.  The [O III] lines are in a 3:1 flux ratio; therefore we do an optimal combination to generate improved line fluxes and respective error before applying the signal to noise cut.  In both cases independent errors between lines are assumed.  The 4861 {\AA} H$\beta$ and 6563 {\AA} H$\alpha$ lines are processed with a simple S/N $>$ 8 cut.  No similar constraints are placed on the 6583 {\AA} [N II] line to avoid introducing a metallicity bias (i.e. under detection of low metallically systems due to discarding galaxies with low [N II] S/N).  Since the 6583 {\AA} [N II] line is in such close proximity to H$\alpha$, the H$\alpha$ reductions are sufficient to remove bad cases on [N II].  Also with the exception of the [N II] all specified lines must be nonzero.  It should be noted that the requirement for detection of the 3727 {\AA} line and the limits of the SDSS spectral wavelength coverage limit the comparison population to a redshift of 0.0209 or higher.  

The SDSS catalog we have actually employed is the Max-Planck-Institut f\"{u}r Astrophysik - John Hopkins University (MPA-JHU) emission line analysis for SDSS Data Release 7 (\citealt{SDSS-mpg, T04}.\footnote{See http://www.mpa-garching.mpg.de/SDSS/ for the data products, their descriptions, and a more detailed citations list}  In order to fully reproduce our sample again the user should specify SPECTROTYPE galaxy, SUBCLASS starforming or starburst, and also require non-zero ugriz CMODEL values, which we required for subsequent photometry.  The resulting sample consists of approximately 137 thousand galaxies.

From our full SDSS sample, we also create two sub-selected samples.  The first employs a redshift cut to create a volume limited sample.  This is done to minimize the incompleteness of faint galaxies in the sample.  The second sample is a weighted sample of the first, where the weighting chooses galaxies as SNe and LGRBs may do: by their SFR.

\subsubsection{redshift of z $<$ 0.04 SDSS general star-forming galaxies} \label{sdss_z_range} \label{zcut}

The SDSS is a magnitude limited sample.  This means that large galaxies (generally at significant redshifts) are overrepresented compared to what one finds in a fixed volume.  This has the effect of skewing the metallicity distribution, due to the previously mentioned mass-metallicity or nearly equivalently, luminosity-metallicty relationship of galaxies.  To obtain an unbiased sample of metallicities a volume limited sample is preferable.

\begin{figure*}[ht]
\begin{center}

\includegraphics[width=.49\textwidth]{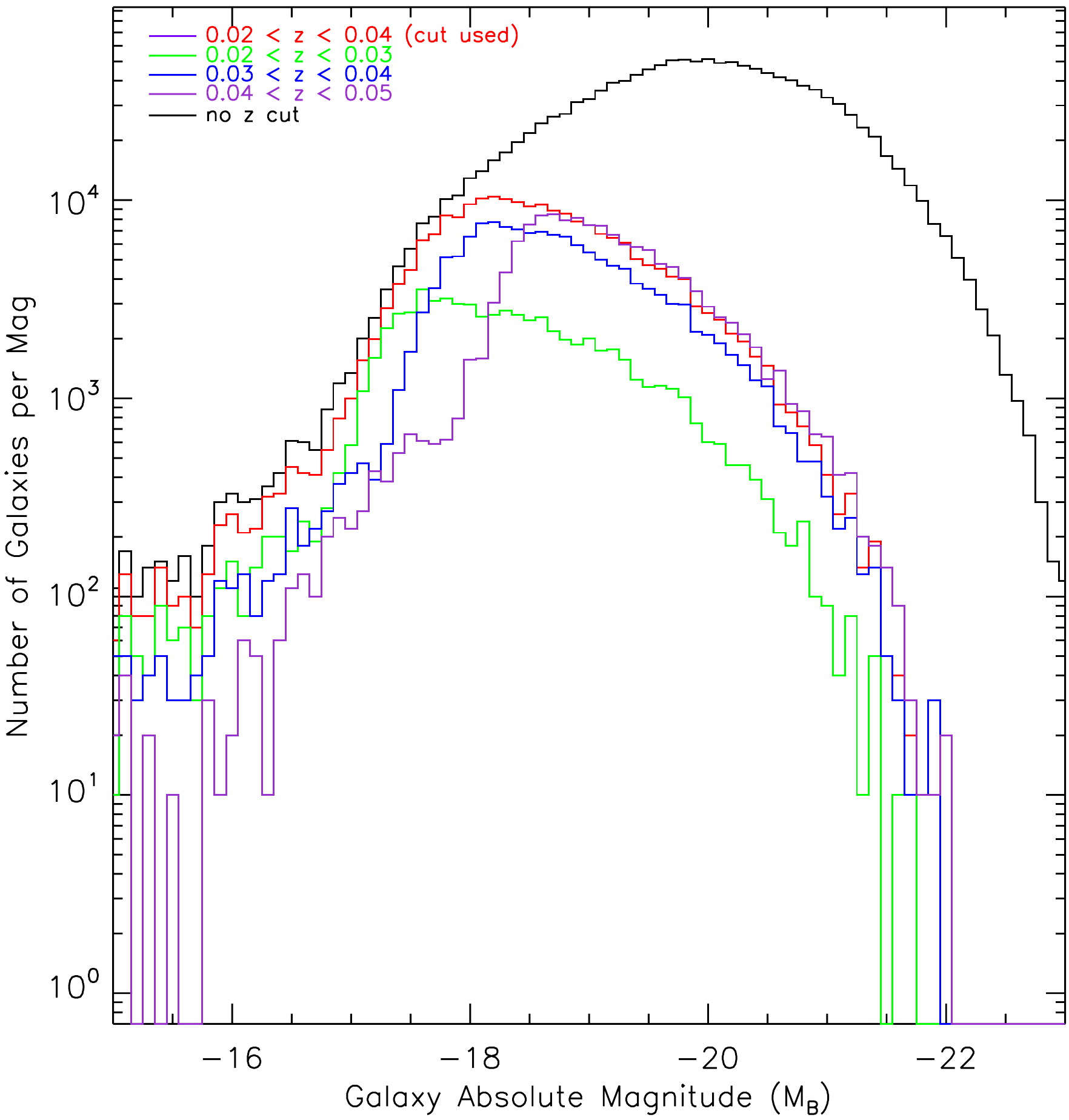}
\includegraphics[width=.49\textwidth]{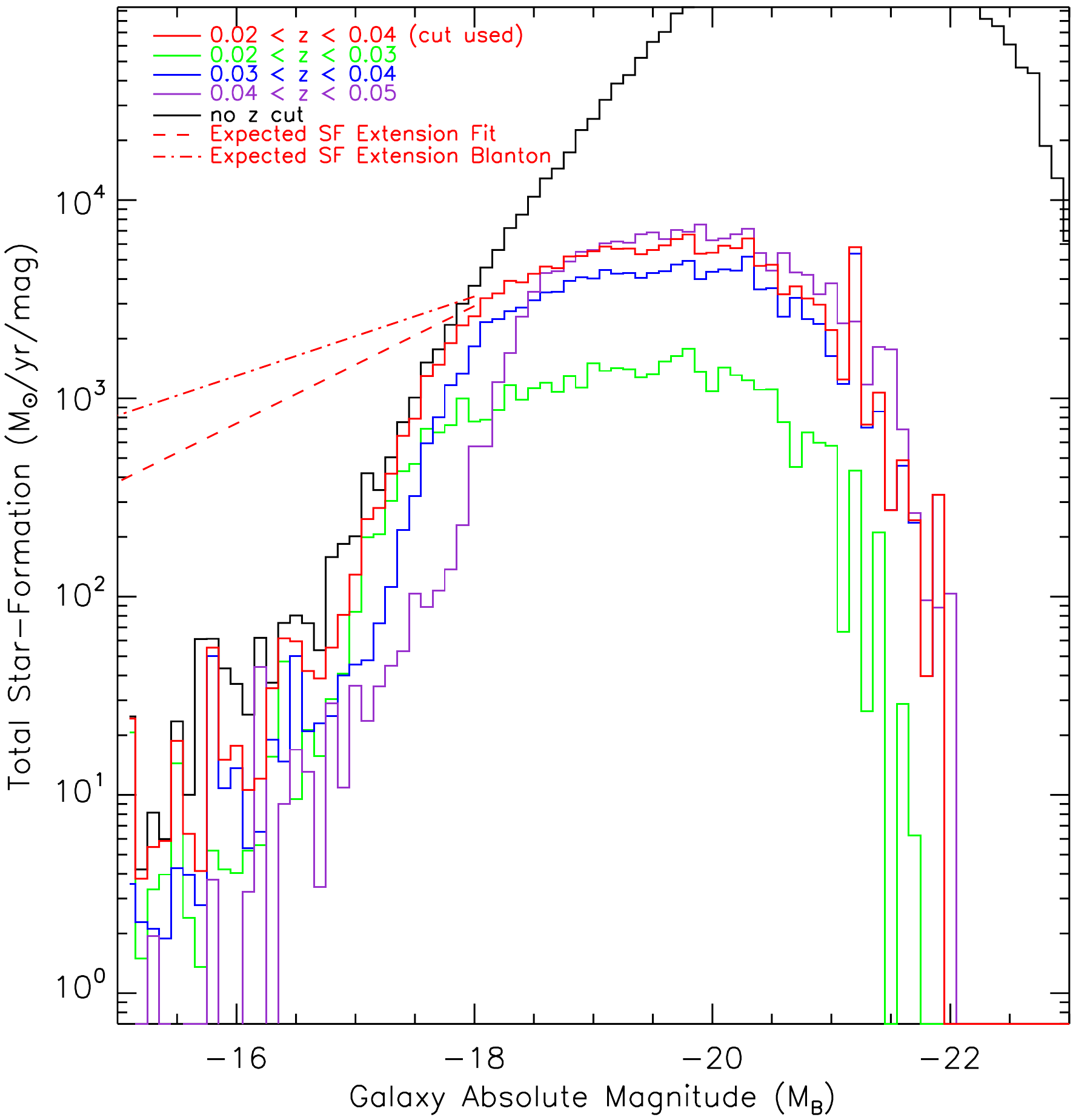}

\caption{\label{zcut_distro} Histogram plots of the number (left) and star-formation (right) distribution of the SDSS population at various redshift ranges.  Note that there is a lack of any significant star-formation contribution in galaxies dimmer then -17 M$_B$ and this contribution drops off around -18 M$_B$ for the redshift cut populations shown.  For the 0.02 $<$ z $<$ 0.04 redshift cut adopted we fit star-formation in the -18 $>$ M$_B$ $>$ -19 range and extrapolate to estimate the M$_B$ $>$ -18 star-formation using both the slope as fitted and the slope as estimated from the $\alpha_2 \sim -1.5$ value estimated in \cite{Blanton}.}

\end{center}
\end{figure*}

\begin{figure*}[ht]
\begin{center}

\includegraphics[width=.32\textwidth]{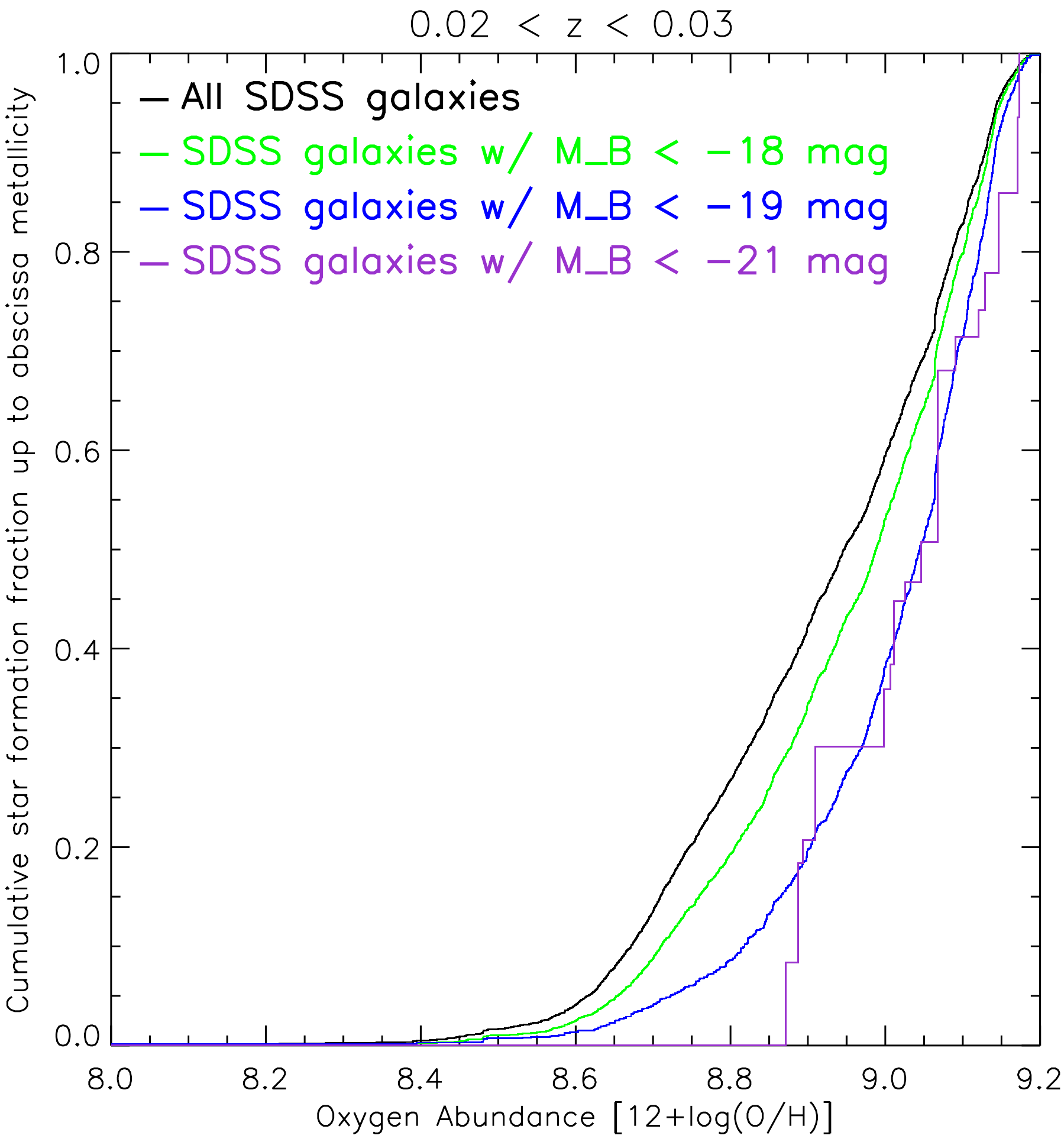}
\includegraphics[width=.32\textwidth]{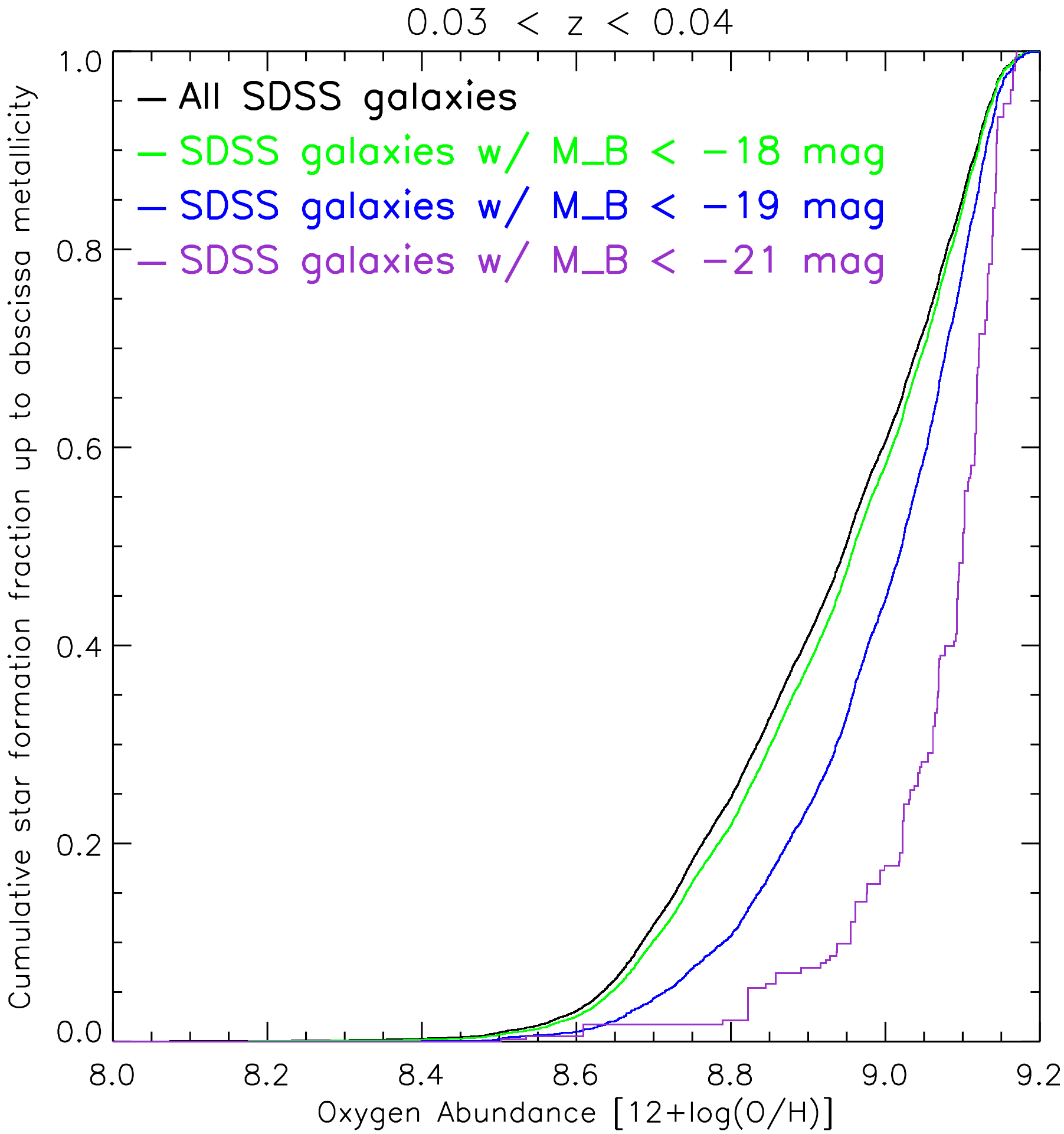}
\includegraphics[width=.32\textwidth]{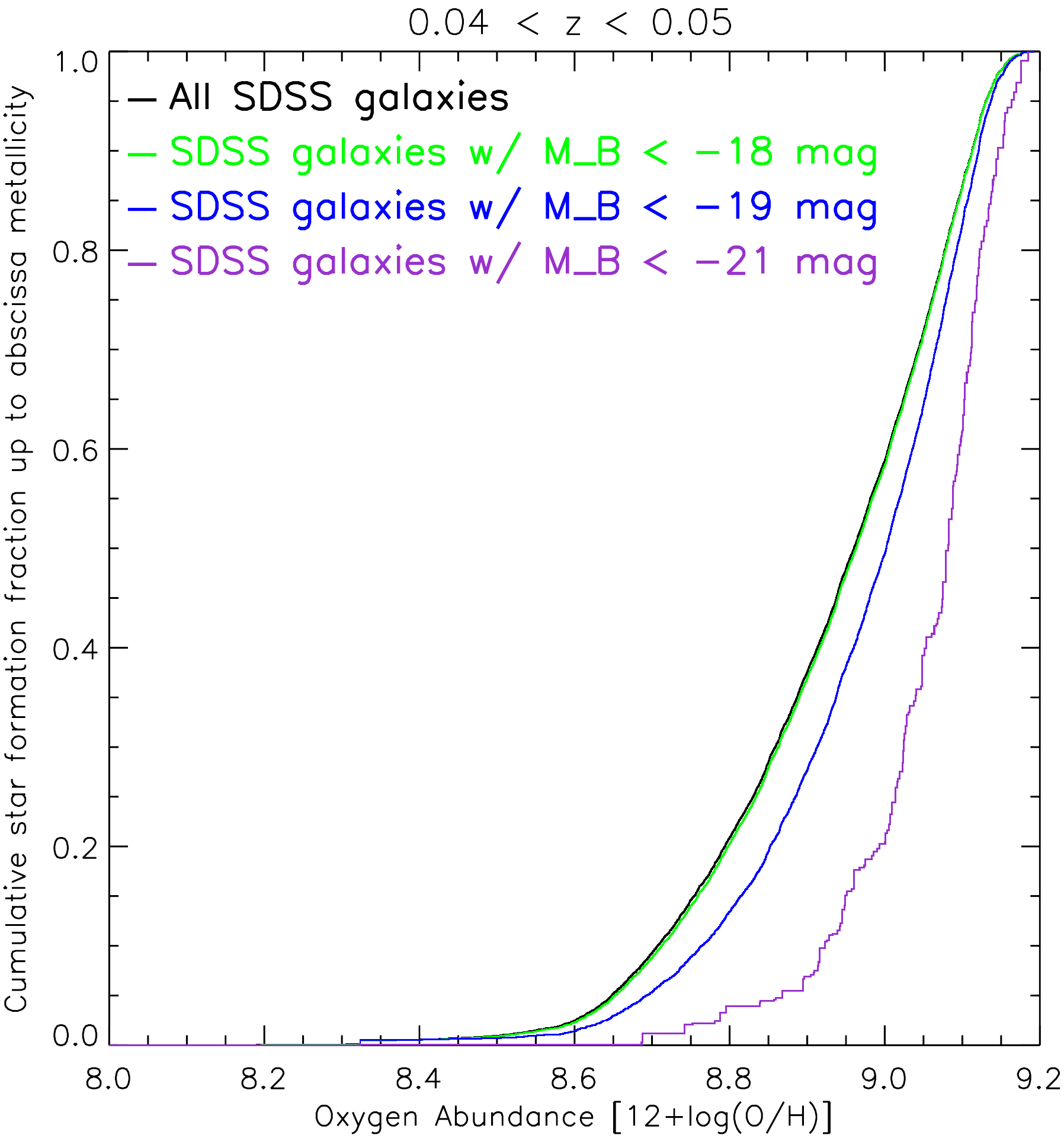}

\caption{\label{zcut_plots} Fractional cumulative distribution plots of star-formation vs.\thinspace \thinspace metallicity for various redshift cuts in the SDSS general star forming galaxy population.  From this it is apparent that the volume limited (0.02 $<$ z $<$ 0.04) SDSS population is complete only for objects brighter then -18 M$_B$.}

\end{center}
\end{figure*}

 We are forced to consider only galaxies with a redshift greater then $z \sim 0.02$ by our requirement of the 3727 {\AA} [O II] doublet for the $R_{23}$ method.  At lower redshifts, the doublet falls to the blue of the SDSS spectral range.  Our ability to reach high redshifts is limited by the magnitude limit of the SDSS survey.  The SDSS spectroscopic survey is only complete to m$_g$ = 18, and becomes very sparse at fainter magnitudes (some fibers, but not many, were devoted to objects fainter than m$_g$ = 18).  The effect of this magnitude cut on the available comparison SDSS sample as a function of absolute magnitude can be seen in Figure \ref{zcut_distro}.  Similarly, in Figure \ref{zcut_plots} we show the distribution of star-formation in our SDSS sample as a function of metallicity for several redshift cuts.  A minimal redshift range extending $0.02 < z < 0.03$ would give completeness down to a luminosity of M$_B \sim$ -17.5, however our ability to constrain the metallicity distribution of galaxies brighter than M$_B \sim$ -21 is limited by a small number statics.  Extending the redshift range to z $<$ 0.04 gives a sufficient sample size on the bright end and completeness down to a luminosity of M$_B \sim$ -18.  Further extending the redshift range to z $<$ 0.05 gives no significant improvement on the bright end and would reduce the completeness down to a luminosity of M$_B \sim$ -18.5 and is thus not advantageous.  Our resulting redshift range is identical to that chosen by \cite{Prieto} in their supernova host survey for similar reasons.  This leaves approximately 21 thousand galaxies within the 0.02 $<$ z $<$ 0.04 redshift cut.

If we fit a power-law to the star-formation function shown in Figure \ref{zcut_distro} from M$_B \sim$ -19 to the completeness luminosity limit or redshift range of M$_B \sim$ -18, we find a slope of log(SFR) $\propto$ -0.3 $\times$ M$_B$.  Our star-formation numbers are derived from H$\alpha$ luminosities, but if one assumes that star-formation is proportional to galaxy luminosity, this corresponds to a Schechter function slope of $\alpha = -1.3$, or equal to that derived by \cite{Blanton}.  This agrees well with that determined by \cite{Blanton}.  It implies that approximately 21\% of star-formation lies in galaxies with an M$_B$ dimmer then -18.  However, \cite{Blanton} argues that the actual slope at these fainter magnitudes might actually be as steep as $\alpha = -1.5$, in which case approximately 31\% of star-formation lies in M$_B$ $>$ -18 galaxies.  The numbers are in contrast with our incomplete sample with only about 6\% of star-formation occurring in galaxies dimmer than an absolute magnitude of -18.  While in many cases we limit our discussion to galaxies brighter than our -18 effective limit, but where we do not, these fractions should be kept in mind.  Imposing an M$_B$ brighter then -18 magnitude cut on the volume limited SDSS sample yields approximately 15 thousand galaxies.

\subsubsection{Synthetic star-formation weighted galaxy sample} \label{synthetic}

The SDSS selects galaxies by number.  All galaxies large or small, if detected, appear as a point on the SDSS plot.  However, SNe and LGRBs (if, for example, \citealt{MannucciLGRBs} are correct), choose their galaxies with a probability proportional to the star-formation of the host.  It is therefore useful in this paper for illustrative purposes to create a sample of galaxies weighted according to star-formation.  To do this, we take our SDSS volume-limited sample, and select galaxies from it with a probability of selection proportion to the SFR of the galaxy.  In practice this means that in our random weighted sample, a given galaxy from the original sample may not appear, or it may appear more than once.

\subsection{$ z < 1$ Star-Forming Galaxy Population Via TKRS} \label{TKRS}

Our desire to obtain a volume limited galaxy sample with a good representation of faint galaxies has limited our comparison SDSS sample to very low redshift ($z < 0.04$).  However, in order to get a reasonable sample of LGRBs, we include all LGRBs with measured emission line metallicities.  While this sample has a median redshift of 0.3 (the effective extent of the {\it full} SDSS sample), it extends up to a redshift of 0.8.  Therefore if we are to also take into account the evolution of galaxy properties with redshift in our comparison, the SDSS sample is an inadequate sole comparison.

While no available sample perfectly meets our requirements, the TKRS metallicity sample provided in \cite{KobulnickyKewley} is excellent in many respects.  The sample has an effective magnitude (or more appropriately line-strength limit) similar to that of the LGRB sample that we have obtained.  It extends out to $z \sim 1$, and thus agrees well with the redshift range of our LGRB sample, and is already on the \cite{KobulnickyKewley} metallicity scale, which we use.

The primary drawback of the TKRS sample is that the [N II]/H$\alpha$ observations necessary to resolve the $R_{23}$ metallicity degeneracy (see Section \ref{R23}) are not available in all cases -- primarily on galaxies at z $\gtsim$ 0.5 where the [N II] and H$\alpha$ lines move into the infrared.  In these cases, where the branch of the $R_{23}$ metallicity diagnostics is uncertain, the TKRS sample assumes the upper branch.  This is a clear bias.  However, at redshifts z $\ltsim$ 0.5 only about 10\% are in the lower branch.  While higher redshift galaxies of similar luminosity tend to move to lower metallicities, this magnitude limited sample also moves to higher luminosities at higher redshifts.  Thus it is reasonable for us to assume that at redshifts above 0.5 perhaps 10\% of the sample, all in the lower branch, are missed.

\subsection{Supernovae Host Galaxy Populations}

In addition to the LGRB hosts and the low redshift general star-forming galaxies, we also include supernovae (SNe) host galaxies in our analyzes and resulting comparisons.  SNe are direct tracers of the deaths massive stars -- and due to the short lifetimes of massive stars, their deaths trace star-formation.  Broad-lined Type Ic's (Ic-bl's) are often found under the decaying afterglow of an LGRB.  Thus studying the metallicity distribution of these explosions provides a direct test of whether the metallicity bias that may be found in LGRBs is only seen in LGRBs, or whether it is also shared by a larger class of particularly energetic exploding stars.  However, in order to provide an unbiased comparison sample we supplement the Type Ic-bl SNe with the more readily available Type II events.  The majority of SNe are of Type II and these are perhaps the archetypical SNe.  Recently it has been shown that not only do these SNe track blue light between galaxies; their formation locations also track the blue light within galaxies \citep{Fruchter, Kelly}.  They may therefore be the most-unbiased tracers of star-formation besides the massive stars themselves.  For maximum consistently in both data and subsequent analysis, we use only Type II whose hosts are in our volume limited SDSS star-forming galaxy population.  This is a luxury that the relative frequency of Type II SNe allows.  Comparing between the SNe types allows us to be sensitive to potential biases introduced by the broader redshift range for the broad-lined Ic's or the different data sourcing for the Type Ic-bl population as well as any intrinsic differences between the SNe types.

\subsubsection{Broad-lined Type Ic SNe hosts} \label{Ic_SNe}

For our broad-lined Type Ic (Ic-bl) sample we build upon the sample of \cite{Modjaz2008} and expand it with a few additional objects that are now available \citep{Sanders}.  As addressed in \cite{Modjaz2008} biases are introduced when SNe are found via targeted surveys.  In these surveys, such as the Lick Observatory Supernovae Search (LOSS) with the Katzman Automatic Imaging Telescope (KAIT), a galaxy population pool is selected in advance then repeatedly searched for the occurrence of SNe.  This obviously has potential to bias the SNe population with the selection criteria used in compiling the galaxy search pool.  Since this has the potential to affect the metallicity luminosity relation of those SNe we distinguish between SNe found in targeted and untargeted surveys whenever displaying the objects in plots.  

Also addressed in \cite{Modjaz2008} are differences between the metallicity at the SNe location with in the host and the metallicity at the host center.  Since in most cases we lack measurements of both, we follow the general methodology of \cite{Modjaz2008} employing a metallicity gradient to estimate the metallicity at the SNe location when it was not specifically measured.  (See Appendix \ref{metal_grad_fit} for details on how this fit was determined and applied).  In the case of SN 2002ap, occurring in M74 / NGC 628, we do not need to estimate the metallicity gradient of the host.  Instead we adopt spectra from \cite{M74} using the closest spectra from the core to calculate the central galaxy metallicity and estimate the SNe site metallicity from spectra taken at a similar radius from the galaxy core.

\subsubsection{Type II SNe hosts} \label{type_II_pop} \label{II_non_targ} \label{II_targ}

A large sample of Type II SNe hosts are available from the work of \cite{Prieto}.  Limiting the sample to Type II SNe within the redshift range outlined in Section \ref{zcut} and hosts present in the SDSS general star-forming galaxies sample yields 16 objects.  However 7 of these SNe were discovered via targeted galaxy surveys and thus may potentially be biased to the brighter host galaxies (as seen in the targeted Type Ic-bl population).  In an attempt to track this bias we also employ a population of non-targeted Type II SNe for comparison.

To expand our untargeted Type II sample, we use the SDSS \citep{SDSS-SNe1,SDSS-SNe2} and Supernova Factory (SNFactory - \citealt{SNfactory}) SNe searches selecting only Type II events with hosts in our volume limited SDSS star-forming galaxy population.  Unfortunately, this yields only 3 and 7 objects from the SDSS and SNFactory surveys respectively (of which 4 were already in the \citealt{Prieto} SNe sample).  Furthermore, there is evidence that these surveys are biased towards preferentially detecting events on low surfaces brightness backgrounds as described in Appendix \ref{problem}.

\subsection{LGRB host galaxies} \label{LGRBs}

To provide as comprehensive an LGRB host galaxy population as possible we combine our own spectroscopic observations of LGRB Hosts (\citealt{pregraham} in prep) with those from the literature.  Our combined sample contains all hosts known to us which have emission line spectroscopy capable of giving a metallicity value using the R$_{23}$ method.  When using observation from the literature, we compute our own metallicity values via the published line strengths rather then adopt published metallicities so as to obtain better consistency between samples.  The line strengths used as well as the derived quantities such as calculated metallicity and SFR are given found in Tables \ref{GRB_table}, \ref{Ic_table} \& \ref{II_table} presented in Section \ref{datatables}.

The metallicity value of the location of a LGRB or SN will generally not be exactly that of the center of the host.  However, LGRBs frequently occur at or near the brightest regions of their generally irregular host galaxies (\citealt{Fruchter, Svensson}).  Therefore, when a spatially resolved spectrum of a LGRB host is not available, it is reasonable to use the average host value as the value for both the galaxy center and the site of the explosion.

There are however three LGRBs in our sample where the LGRB site region and host galaxy center are well separated at ground-based resolution: LGRBs 980425, 020819B, 060505.  These deserve special treatment.  

\begin{enumerate}
\item LGRB 980425/SN 1998bw is the nearest known LGRB.  It lies on a dwarf irregular, but because of its low redshift (z $\sim$ 0.008 - \citealt{IAUC6895}) the host is very well resolved from the ground.  There are several published spectra of the LGRB site (\citealt{Sollerman, Hammer, Christensen980425}) from which we determine metallicity values of log(O/H)+12 = 8.48 \citep{Sollerman}, 8.64 \citep{Hammer}, and 8.23 \citep{Christensen980425}.  We adopt the median value, 8.48 from \cite{Sollerman}, for the site metallicity.  We note, however, that a disagreement between the R$_{23}$ and [N II]/[O II] diagnostics for the \cite{Sollerman} spectrum suggests that the errors on that spectrum may be large.

The IFU data of \cite{Christensen980425} allows us to make a rough estimate of the central metallicity of the host.  The fibers in the SDSS, subtend on a scale of $\sim$ 1.2 to 2.3 kpc in our volume limited sample.  Sampling the IFU elements within radii 7.4" and 14.2" of the core, we compute metallicities of 8.59 and 8.55 respectively.  If we take our LGRB site metallicity value and apply the metallicity gradient method described in Appendix \ref{mshifting} to estimate the central metallicity we also find a value of 8.55 which we adopt as our central metallicity value for LGRB 980425.

\item The host of LGRB 020819B is a large spiral \citep{Jakobsson,pregraham} with a diameter of about 7" on the sky.  The LGRB was dark, but a radio position with an error circle of 1" diameter \citep{Jakobsson} places it on a small ``blob" at the edge of the spiral (this could either be an outlying star-forming region of the spiral or possibly a satellite).  The metallicity of the center of the host has been well determined \citep{Levesque020819B}.  However, a problem with the blue chip of the spectrograph on the same observing run, meant that a full R$_{23}$ metallicity could not be obtained for the "blob" under the location of the burst.  However, the [N II]/H$\alpha$ diagnostic can be used to give a rough estimate of the local metallicity \citep{Levesque020819B}.  We will discuss this object further in \cite{pregraham} in prep.

\item The host LGRB 060505 (z $\sim$ 0.089) is a well resolved spiral \citep{060505}.  The LGRB occurred right on top of an HII region.  No supernova was seen and strong limits have been placed on the magnitude of the underlying supernova \citep{Ofek}.  There has been some discussion as to whether this burst was actually a long or a short burst \citep{Ofek, 060505}.  However, due to the location of the burst on an HII region, and the bursts $\sim$ 4s duration, we have included it in our long burst sample.  \cite{060505} gives us metallicities for both the center of the host (Bm in their table) and burst location.

\end{enumerate}

\section{Methods \& Determination of analyzed properties}

The majority of analyses used within this paper are based on values of four physical galaxy properties: metallicity (both event local and galaxy central), total SFR, absolute B band magnitude (M$_B$), and redshift.  Here we describe how these values (except for redshift) are determined.

\subsection{Metallicity} \label{Metallicity} \label{metal_measurement}

Due to differences between metallicity diagnostics and their various calibrations, a true comparison of metallicity requires using a common scale and if possible a consistent methodology and diagnostic.  Here we adopt the \cite{KobulnickyKewley} scale exclusively and the R$_{23}$ diagnostic to the extent possible.  To ensure consistency of method and scale we calculate metallicity values ourselves from published fluxes and our own work (\citealt{pregraham} in prep).  Metallicity calculations are preformed with an improved version of the IDL code outlined in \cite{kd2002} (updated to the \citealt{KobulnickyKewley} scale).  For conversion to other scales and discussion of associated issues we refer the reader to \cite{KewleyEllison} and the metallicity diagnostic transformations given therein.

Based on the \cite{solar} 6300 {\AA} [O I] line measurements of the Sun, solar metallicity is estimated to be log(O/H)+12 = 8.69 $\pm$ 0.05 \citep{solar} in the \cite{KobulnickyKewley} scale.  It should be noted that converting solar metallicities to emission line H II region abundances is exceedingly difficult.  Thus the \citealt{KobulnickyKewley} scale is more accurate internally than to any absolute reference to solar value and such comparisons should be avoided or limited to broad generalizations.

\subsubsection{$R_{23}$ Diagnostic} \label{R23}

The R$_{23}$ method is a commonly used metallicity diagnostic based on the electron temperature sensitivity of the oxygen spectral lines, achieved using the ratio of oxygen line strengths to a hydrogen spectral feature, which is independent of metallicity .  Doubly ionized oxygen, [O III], has strong lines at 4959 and 5007 {\AA} and [O II] has a particularly strong, typically unresolved, doublet at 3727 {\AA}.  The metallicity independent 4861 {\AA} H$\beta$ line is conveniently placed between these sets of oxygen lines.  This close placement minimizes errors due to misestimates of internal extinction and possibly even systematic effects of equipment.  The ratio of the sum of the fluxes of these oxygen lines divided by the H$\beta$ flux gives the equation for R$_{23}$ used in the classical application of this diagnostic.  

\begin{equation}
\label{eqn}
 R_{23} = {I_{3727} + I_{4959} + I_{5007} \over I_{H\beta}}
\end{equation}

First proposed by Bernard Pagel in 1979 \citep{Pagel1979, Pagel1980}, R$_{23}$ has become the primary metallicity diagnostic for galaxies where the faint [O III] 4363 {\AA} line is not measurable, and thus has been used extensively for work at moderately high redshifts, where the majority of our LGRB sample is found.

In order to correlate the flux of a line belonging to an individual atomic ionization level with the total abundance of that element it is necessary to know what fraction of the element is ionized to the level in question.  Here, in the case of oxygen, this is achieved by measuring the flux ratio between the [O II] and [O III] lines.  This gives the relative population in the O II and O III ionization states, and allows one to fit the metallicity for that specific ionization state ratio.

Thus in the classical application of this diagnostic, the R$_{23}$ value would be calculated from the measured line ratios and then compared along an [O III] to [O II] line ratio contour.  This classical application, however, treats ionization as a parameter independent of metallicity and ignores the effect metallicity has on the ionization state of the lines.  \cite{kd2002} solve this issue by iterative fitting of the metallicity and ionization parameter.  

While we have so far mentioned only oxygen line fluxes, in some cases observational difficulties may mean one has only equivalent line widths for a given spectrum.  However, \cite{KobulnickyPhillips} has established that the R$_{23}$ method can be directly used on equivalent width values (instead of reddening corrected flux values) and is found to be more accurate than flux ratios when reddening information, and in particular the Balmer decrement, is not available.

The R$_{23}$ method diagnostics suffer a degeneracy due to different effects being dominant at different regimes.  In the low metallicity regime the effects of the metals on the electron temperature of the system can be ignored due to their low relative abundance.  Thus the more metals in the electron gas, the more collisional excitations and more resultant flux in the metal lines.  As the metallicity rises however, emission from infrared fine-structure lines becomes significant and serves as a cooling mechanism, lowering the electron temperature, the electron velocity, the number of collisional excitations, and thus the metal line flux.

This temperature dependence causes two metallicity values (one high, one low) to generate the same R$_{23}$ line ratio.  Unless one of the degenerate values can be obviously excluded (or the two values are within the error range of the R$_{23}$ calibration) new empirical data is the only accurate way to break the degeneracy.  This is usually accomplished, as we do here, by application of the [N II]/H$\alpha$ diagnostic.  

\subsubsection{[N II]/H$\alpha$ Diagnostic} \label{NII_Halpha}

The [N II]/H$\alpha$ line ratio provides a crude metallicity indicator \citep{kd2002}.  However, due to its strong dependence on the ionization parameter it gives only a gross estimate of abundance unless the ionization is known.  Since determining the ionization parameter requires measurement of lines which themselves constitute better metallicity diagnostics, the application of an ionization parameter correction to this diagnostic is of limited utility.  Also the diagnostic is easily distorted by contamination from shock excitation or hard ionizing radiation from AGN.  Thus its primary application is selecting between the degenerate upper and lower branch values provided by the R$_{23}$ diagnostic.

However use of the H$\alpha$ and 6583 {\AA} [N II] lines can be advantageous especially when a low redshift coupled with instrumentation limitations preclude measurement of the 3727 {\AA} [O II] line.  Additionally the small separation between the 6584 {\AA} [N II] line and H$\alpha$ lines renders this method immune to the effects of reddening and enables it to be applied without any flux calibration required.  

While the degeneracy of the R$_{23}$ diagnostic can be broken, as we do in this work, by application of the [N II]/H$\alpha$ diagnostic, this has the obvious disadvantage of requiring additional measurements considerably outside the R$_{23}$ wavelength range This in turn often necessitates a separate observation and for higher redshift objects use of a separate near-infrared, instrument.  Nonetheless, the [N II]/H$\alpha$ diagnostic is one of the most reliable methods for breaking the R$_{23}$ degeneracy.

\subsection{Star-Formation Rates}

We estimate SFRs from galaxy H$\alpha$ fluxes using the \cite{K98SFR} H$\alpha$ SFR diagnostic.  When an H$\alpha$ flux is unavailable (for instance a LGRB host above z = 0.6 without calibrated infrared observations), we estimate the H$\alpha$ flux using the H$\beta$ flux via the Balmier decrement).  Unlike metallicity, estimation of SFRs requires absolute (as apposed to relative) line flux measurements over the entire galaxy.  This requirement significantly complicates our analysis, as we are typically dealing with slit or fiber spectroscopy which does not cover the entire galaxy.  Therefore, in addition to a standard absolute calibration of the spectroscopy, we must correct for slit or fiber losses.  We discuss these corrections below.

\subsubsection{Slit loss correction for LGRB host galaxies} \label{LGRB_slc}

To obtain slit loss estimates for the LGRB population in cases where the host is not well-resolved from the ground, we convolve HST images of the hosts to ground based seeing and then determine the fraction of blue light that would land in a slit sized box.  When possible these slit loss fractions were computed for the size of slit used (typically 1" and assumed to be 1" when not stated explicitly) after convolving the image to match the described seeing (or typical seeing for the telescope used when not stated explicitly) in two orientations (one maximally the other minimally favorable).

Typical slit loss estimates were between 25 to 40 \% with a typical difference between the orientations of under 5\%.  Given the small size of the orientation dependance the average between the orientations is used for the correction.  Due to the stability of the flux loss, in cases where this correction can not be computed (i.e.  sufficient images are not available) the average value is used.  This approach does have the obvious limitation of assuming that the measured line strength is representative of the object as a whole (central galaxy spectroscopy is used when possible).  Cases without suitable spectroscopic estimate of star-formation are dropped from analyses that require the actual host SFR; otherwise the entire sample is used

\subsubsection{fiber loss correction for SDSS galaxy derived populations}

Just as with slit spectroscopy, it is necessary to correct for the difference in total flux of the SDSS objects with the flux observed propagating down the fibers.  Fiber loss estimates are computed using the flux ratio of the fiber plug magnitudes to the estimate of the total galaxy magnitude, which is reported as CMODEL magnitudes in the MPA-JHU emission line analysis (described in Section \ref{ggp}).  Galaxies without CMODEL magnitudes are discarded.  This methodology does have potential issues with biasing the galaxy SFR to the center of the galaxy intensity.  However while this source of error can not be effectively eliminated, is also a slight issue with the LGRB population as well, and is countered as best as possible by using the galaxy central properties for comparison.  Since the Type II SNe hosts samples as well as the synthetic star-formation waited random selection are both subsets of the SDSS spectroscopic sample, they are corrected as well as part of this procedure.

\subsubsection{SDSS spectroscopy substitution for Type Ic-bl SNe hosts} \label{noIcSFRs}

The Type Ic-bl SNe hosts population is highly inhomogeneous.  The orientations and sizes of the slits are often not well documented, and due to the the proximity of these hosts, the slits subtend very small fractions of the galaxies.  Due to the absence of accurate slit loss estimates on the Type Ic-bl host spectroscopy and the low redshift / large angular size / low slit coverage / high slit loss of this population it is not directly usable in analyses that require the actual host SFR (Sections \ref{frac} \& Figure \ref{all_SFR}).  However rather then omit the Type Ic-bl SNe hosts completely we search the SDSS sample for spectroscopy on their host galaxies and use that to calculate SFRs as in the previous section.  As this spectroscopy is used only for SFRs (and not metallicities) coverage of the 3727 {\AA} [O II] line is not required thus allowing application of this methodology without the z $>$ 0.0209 redshift limitation.  This yields 7 Type Ic-bl SNe hosts with usable SFRs.

\subsubsection{M$_B$}

For the SDSS galaxy derived populations absolute B band magnitudes are determined from the redshift and CMODEL ugriz magnitudes provided in the MPA-JHU-SDSS data products via the SDSS2BESSELL routine of the KCORRECT IDL software package \citep{kcorrect}.  The provided CMODEL magnitudes are integrated galaxy apparent magnitude estimates for the spectroscopic targets based on the SDSS photometric data.  The SDSS2BESSELL software package converts the ugriz magnitudes into the UBVRI magnitude system (and requires input values in all 5 SDSS bands).  Conversion from apparent to absolute magnitude using the SDSS redshift value is preformed using the LF\_DISTMOD IDL software package.

M$_B$ values for the LGRB host galaxies and Type Ic-bl SNe are taken from the literature (sometimes as apparent magnitudes and converted to absolute via redshift, in the same manner via LF\_DISTMOD, as needed).  M$_B$ values for the TKRS sample are taken directly from \cite{KobulnickyKewley} (along with all other TKRS data used).

We choose to employ the luminosity metallicity relation in place of the mass metallicity relation primarily due the availability of B band luminosity data across our populations.  Many of our objects, both LGRBs and Type Ic-bl SNe hosts, do not have the necessary photometric coverage to allow for a robust determination of mass values.  Attempting to extrapolate photometric coverage from spectroscopic continuums is difficult given the faint continuums present in most of our emission line dominated spectra, especially of the LGRB host galaxies.  The Type Ic-bl SNe also have the difficulty of being close enough that long slit spectroscopy gives only limited spacial coverage of the host (which in many cases is not at the galaxy center) thus the colors would likely not be indicative of an host galaxy as a whole.  Luminosity also has the added benefit of avoiding coverage issues and biases at different redshifts (i.e. ugriz coverage with SDSS gives very different rest frame coverage at z $\sim$ 0 vs. z $\sim$ 0.8 while the B band luminosity can be easily extrapolated throughout our entire redshift range).  Since \citealt{Modjaz2008} also used M$_B$ values this facilitated easer adoption of and comparison with her data.

\subsection{Data Tables} \label{datatables}

In Tables \ref{GRB_table}, \ref{Ic_table} \& \ref{II_table} we present our line fluxes and estimated physical galaxy properties in tabular form.  Line fluxes are presented with Galactic extinction removed but without correction for internal extinction.  When the literature fluxes were not already corrected for Galactic extinction the value from the \cite{Schlegel} dust maps was applied.  We do this, since some metallicity determination methodologies also fit the internal extinction simultaneously with the metallicity (i.e.\thinspace\thinspace\citealt{T04}), values are given without removing internal extinction.

Internal extinction is calculated by determining an extinction value which matches the measured H$\alpha$ to H$\beta$ line flux ratio to the balmier decrement.  If the H$\alpha$ flux is not available but the H$\gamma$ flux is, the extinction is similarly calculated from the H$\beta$ to H$\gamma$ ratio.  The H$\alpha$ flux can then be estimated from the H$\beta$ flux via the balmier decrement.  This estimated H$\alpha$ flux can then be used to obtain the SFR.

In some cases, we have calibrated optical spectra that do not contain the H$\alpha$ line, along with an uncalibrated near-infrared spectra of the H$\alpha$ and 6583 {\AA} [N II] lines for the same object.  In these cases, we can use the estimated H$\alpha$ flux and the observed [N II] / H$\alpha$ count ratio to estimate the 6583 {\AA} [N II] line flux.  

When the 4959 {\AA} [O III] line was not available and the 5007 {\AA} [O III] is, the 4959 {\AA} line is assumed to have $\frac{1}{3}$ the 5007 {\AA} lines flux as quantum mechanically required (see \citealt{070714Bpaper}).  Values given are measured without slit \& fiber loss correction.  The ``spectrum" column denotes whether we are presenting directly measured fluxes, fluxes normalized to H$\beta$ or equivalent width values.  If the H$\beta$ column of an H$\beta$ normalized flux is not 1 or 100 this is due to the galactic extinction correction being applied to the normalized values in the literature.  The numeric superscript on the spectrum column indicated the reference(s) for the spectroscopy.  Similarly, the superscripts on M$_B$, and redshift columns provide references for those values.  The reference key is given in Section \ref{tabref} which is separate from the general references of this paper.

\begin{table*}[t]
\begin{center}
\begin{tabular}{|c|c|c|c|c|c|c|c|c|c|c|c|c|}
\hline
 & & & & 3727 {\AA} & 4861 {\AA} & 4959 {\AA} & 5007 {\AA} & 6563 {\AA} & 6584 {\AA} & \\
Object & Metallicity & Redshift & M$_B$ & [O II] line & H$\beta$ line & [O III] line & [O II] line & H$\alpha$ line & [N II] line & Spectrum\\
\hline
GRB 991208 & 8.05 & 0.706$^{1}$ & -18.68$^{2}$ & 4.581 & 3.493 & 2.463 & 5.848 & 17.04 & 0.852 & flux$^{3}$ \\
GRB 980425 & 8.55 & 0.0085$^{4}$ & -18.09$^{2}$ & 576.6 & 219.69 & 278.31 & 851.4 & 713.7 & 68.7 & Hbeta$^{5}$ \\
GRB 010921 & 8.34 & 0.451$^{1}$ & -19.87$^{2}$ & 38.43 & 9.518 & 5.602 & 29.65 & 40.11 & 1.858 & flux$^{6,3}$ \\
GRB 011121 & 8.20 & 0.362$^{7}$ & -19.75$^{2}$ & 57.97 & 17.13 & 16.79 & 16.64 & 65.61 & 2.0 & flux$^{7}$ \\
GRB 020903 & 8.38 & 0.251$^{1}$ & -19.34$^{2}$ & 68.9 & 44.0 & 74.0 & 335.0 & 168.0 & 7.2 & flux$^{8}$ \\
GRB 020819B & 8.97 & 0.411$^{9}$ & -21.53$^{2}$ & 30.22 & 19.45 & ... & 9.974 & 106.1 & 45.21 & flux$^{9}$ \\
GRB 030329 & 8.12 & 0.1685$^{4}$ & -16.52$^{2}$ & 1.61 & 1 & 1.12 & 3.40 & 2.74 & 0.1 & Hbeta$^{4,10}$\\
GRB 031203 & 8.27 & 0.1055$^{4}$ & -18.52$^{2}$ & 1.06 & 1 & 2.11 & 6.36 & 2.82 & 0.15 & Hbeta$^{4}$ \\
GRB 050824 & 8.39 & 0.828$^{3}$ & -19.02$^{2}$ & 3.791 & 2.529 & 6.313 & 15.45 & 7.6 & 0.2797 & flux$^{11}$ \\
GRB 050826 & 8.83 & 0.296$^{3}$ & -20.28$^{2}$ & 91.66 & 28.65 & 12.46 & 36.22 & 85.22 & 14.41 & flux$^{3}$ \\
GRB 051022 & 8.77 & 0.80625$^{12}$ & -21.23$^{2}$ & 67 & 25.29 & 22.24 & 59.57 & 104.99 & 15.97 & eqw$^{12}$ \\
GRB 060218 & 8.24 & 0.034$^{1}$ & -15.92$^{2}$ & 161.5 & 68.83 & 93.43 & 229.9 & 170.5 & 5.122 & flux$^{1}$ \\
GRB 060505 & 8.64 & 0.0889$^{13}$ & -19.38$^{13,14}$ & 18.30 & 4.553 & 2.331 & 5.500 & 22.85 & 5.185 & flux$^{13}$ \\
GRB 070612A & 8.17 & 0.671$^{3}$ & -20.86$^{15}$ & 83.02 & 36.60 & 13.67 & 41.01 & 152.0 & 1.520 & flux$^{3}$ \\
\hline
\end{tabular}
\end{center}
\caption{\label{GRB_table}Table of LGRB hosts.  Metallicity values are computed from the tabular line fluxes via the \cite{KobulnickyKewley} R$_{23}$ diagnostic.  When the 4959 {\AA} [O III] line is not listed and the 5007 {\AA} [O III] is available the 4959 {\AA} line is assumed to have $\frac{1}{3}$ the 5007 {\AA} lines flux as quantum mechanically required (see \citealt{070714Bpaper} for a more detailed description).  References for data in this table are indicated by numeric superscript with the key given in Section \ref{tabref} and are separate from the general references of this paper.  All given flux values have been corrected for galactic but not internal extinction (when the literature fluxes were not already corrected for galactic extinction the value from the \citealt{Schlegel} dust maps was applied).  The Spectrum column indicates type of spectral normalization with reference in superscript.  Flux spectra are in units of $10^{17}$ ergs s$^{-1}$ cm$^{-2}$}
\end{table*}

\begin{table*}[t]
\begin{center}
\begin{tabular}{|c|c|c|c|c|c|c|c|c|c|c|c|c|}
\hline
 & Central & Site & & & 3727 {\AA} & 4861 {\AA} & 4959 {\AA} & 5007 {\AA} & 6563 {\AA} & 6584 {\AA} & \\
Object & Metallicity & Metallicity & Redshift & M$_B$ & [O II] line & H$\beta$ line & [O III] line & [O II] line & H$\alpha$ line & [N II] line & Spectrum\\
\hline
SN 1997dq$^{T}$ & {\bf 9.10} & 9.02 & 0.0033$^{16}$ & -20.1$^{16}$ & ... & 68 & ... & ... & 496 & 280 & flux$^{16}$ \\
SN 1997ef$^{T}$ & 9.09 & {\bf 9.00} & 0.0117$^{16}$ & -20.2$^{16}$ & 1164 & 553.6 & 63.49 & 257.0 & 1911 & 641.6 & flux$^{16}$ \\
SN 1998ey$^{T}$ & 9.08 & {\bf 9.08} & 0.0161$^{16}$ & -21.8$^{16}$ & ... & 73 & ... & ... & 277 & 133 & flux$^{16}$ \\
SN 2002ap$^{T}$ & {\bf 9.05} & {\bf 8.72} & 0.0022$^{16}$ & -20.6$^{16}$ & 1.813 & 1.285 & 0.118 & 0.212 & 6.007 & 1.812 & Hbeta$^{17}$ \\
SN 2002bl$^{T}$ & {\bf 8.96} & 8.81 & 0.01591$^{18}$ & -20.3$^{18}$ & ... & ... & ... & ... & 111.24 & 49.98 & flux$^{19}$ \\
SN 2003bg$^{T}$ & {\bf 9.03} & 8.76 & 0.0044$^{16}$ & -17.5$^{16}$ & ... & ... & ... & ... & 2.35 & 1.02 & flux$^{16}$ \\
SN 2003jd$^{T}$ & 8.92 & {\bf 8.78} & 0.0188$^{16}$ & -20.3$^{16}$ & 876.6 & 266.7 & 134.6 & 396.3 & 833.9 & 108.6 & flux$^{16}$ \\
SN 2005kr$^{N}$ & {\bf 8.78} & 8.75 & 0.1345$^{16}$ & -17.4$^{16}$ & 20.51 & 11.05 & 11.24 & 32.40 & 34.55 & 2.929 & flux$^{16}$ \\
SN 2005ks$^{N}$ & {\bf 8.90} & 8.81 & 0.0987$^{16}$ & -19.2$^{16}$ & 216.6 & 83.40 & 16.34 & 60.95 & 306.6 & 103.9 & flux$^{16}$ \\
SN 2005nb$^{N}$ & 8.73 & {\bf 8.72} & 0.0238$^{16}$ & -21.3$^{16}$ & 1384 & 427.2 & 173.3 & 514.7 & 1646 & 325.8 & flux$^{16}$ \\
SN 2006nx$^{N}$ & {\bf 8.59} & 8.54 & 0.1370$^{16}$ & -18.9$^{16}$ & 25.30 & 10.06 & 9.923 & 30.05 & 42.39 & 13.49 & flux$^{16}$ \\
SN 2006qk$^{N}$ & {\bf 8.82} & 8.80 & 0.0584$^{16}$ & -17.9$^{16}$ & 149.0 & 58.14 & ... & 20.15 & 278.4 & 104.2 & flux$^{16}$ \\
SN 2007I$^{N}$ & {\bf 8.70} & 8.65 & 0.0216$^{16}$ & -16.9$^{16}$ & ... & 28.7 & 11.9 & 57.2 & 119 & 20 & flux$^{16}$ \\
SN 2007ce$^{N}$ & 8.37 & {\bf 8.27} & 0.046$^{20}$ & -17.69$^{15}$ & 132.5 & 108.4 & 216.5 & 594.8 & 285.1 & 7.813 & Hbeta$^{20}$ \\
SN 2008iu$^{N}$ & 8.64 & {\bf 8.46} & 0.13$^{20}$ & -16.59$^{15}$ & 136.1 & 117.1 & 245.0 & 745.2 & 255.1 & 43.25 & Hbeta$^{20}$ \\
SN 2009bb$^{N}$ & 9.04 & {\bf 9.02} & 0.009877$^{21}$ & -19.98$^{22}$ & 58.57 & 37.36 & 5.456 & 17.56 & 155.3 & 51.59 & flux$^{22}$ \\
SN 2010ah$^{N}$ & 8.92 & {\bf 8.61} & 0.05$^{20}$ & -17.22$^{15}$ & 400.3 & 104.0 & 62.35 & 186.9 & 359.3 & 33.88 & Hbeta$^{20}$ \\
SN 2010ay$^{N}$ & {\bf 8.68} & 8.67 & 0.067$^{20}$ & -18.30$^{15}$ & 620.5 & 271.6 & 305.1 & 905.0 & 837.2 & 67.20 & flux$^{19}$ \\
\hline
\end{tabular}
\end{center}
\caption{\label{Ic_table}Table of broad-lined Type Ic SNe hosts.  The spectral values displayed are from spectra either of the the center of the galaxy or the site of the SN explosion.  The metallicities are derived from these spectral values via the \cite{KobulnickyKewley} R$_{23}$ diagnostic when the required lines are available.  Otherwise, the [N II]/H$\alpha$ diagnostic is used.  Metallicities directly calculated from the tabular values are shown in bold.  The metallicity values not shown in bold are shifted from the calculated (bold) values via the metallicity gradient method described in Appendix \ref{metal_grad_fit}.  As in Table \ref{GRB_table} references are given in Section \ref{tabref} and are separate from the general references of this paper.  Spectra column indicates type of spectra with reference in superscript.  Flux spectra are in units of $10^{17}$ ergs s$^{-1}$ cm$^{-2}$.  The superscript by the SN name denotes whether the SN was detected in a targeted or non targeted manner as listed below:}
\vspace{-4pt}
N SNe found in a non targeted manner\newline
T SNe found in a targeted manner
\end{table*}

\begin{table*}[t]
\begin{center}
\begin{tabular}{|c|c|c|c|c|c|c|c|c|c|c|c|c|}
\hline
 & Central & & & 3727 {\AA} & 4861 {\AA} & 4959 {\AA} & 5007 {\AA} & 6563 {\AA} & 6584 {\AA}\\
Object & Metallicity & Redshift & M$_B$ & [O II] line & H$\beta$ line & [O III] line & [O II] line & H$\alpha$ line & [N II] line\\
\hline
SN 1988Q$^{a,N}$ & 8.84 & 0.0351 & -19.16 & 443.2 & 155.3 & 54.87 & 164.4 & 543.5 & 118.4 \\
SN 1997cs$^{a,N}$ & 8.84 & 0.0369 & -19.99 & 740.2 & 239.5 & 46.92 & 140.5 & 882.7 & 240.4 \\
SN 1999ab$^{a,T}$ & 8.82 & 0.0321 & -19.07 & 238.3 & 76.19 & 24.21 & 72.61 & 265.6 & 58.60 \\
SN 2001dy$^{a,T}$ & 9.09 & 0.0300 & -20.11 & 186.6 & 275.8 & 53.83 & 161.1 & 1113 & 426.3 \\
SN 2001fb$^{a,N}$ & 8.85 & 0.0321 & -20.09 & 947.7 & 410.8 & 133.7 & 400.3 & 1774 & 428.4 \\
SN 2002ew$^{a,T}$ & 8.87 & 0.0299 & -19.27 & 763.0 & 226.4 & 63.12 & 188.6 & 671.9 & 166.0 \\
SN 2003ci$^{a,T}$ & 9.12 & 0.0303 & -21.28 & 122.8 & 193.3 & 15.42 & 46.15 & 1054 & 353.8 \\
SN 2003la$^{a,T}$ & 9.07 & 0.0307 & -20.34 & 462.9 & 345.7 & 27.50 & 82.46 & 1454 & 526.4 \\
SN 2004eb$^{a,T}$ & 8.82 & 0.0285 & -20.49 & 5105 & 1815 & 665.3 & 1993 & 6733 & 1564 \\
SN 2004hx$^{b,N}$ & 8.89 & 0.0381 & -18.63 & 81.95 & 27.40 & 6.465 & 19.37 & 88.28 & 23.31 \\
SN 2005em$^{a,T}$ & 8.28 & 0.0251 & -14.70 & 750.9 & 206.4 & 144.4 & 431.8 & 553.6 & 47.13 \\
SN 2005mb$^{a,T}$ & 9.11 & 0.0238 & -21.15 & 196.1 & 281.5 & 30.61 & 91.76 & 1284 & 548.9 \\
SN 2006iw$^{a,b,N}$ & 8.86 & 0.0307 & -18.64 & 108.1 & 38.01 & 8.819 & 26.40 & 139.4 & 37.59 \\
SN 2007dp SNF20070427-003$^{c,N}$ & 8.76 & 0.0328 & -18.27 & 272.8 & 65.21 & 20.05 & 60.07 & 194.7 & 41.26 \\
SN 2007es SNF20070626-009$^{c,N}$ & 9.07 & 0.0274 & -20.11 & 92.75 & 87.61 & 14.16 & 42.43 & 357.5 & 190.3 \\
SN 2007fg SNF20070703-008$^{c,N}$ & 8.81 & 0.0258 & -18.08 & 328.8 & 110.2 & 56.19 & 168.5 & 346.7 & 47.30 \\
SN 2007fe SNF20070630-010$^{a,c,N}$ & 8.79 & 0.0331 & -18.87 & 366.1 & 102.4 & 30.07 & 90.18 & 338.2 & 73.01 \\
SN 2007hm SNF20070831-018$^{a,c,N}$ & 8.92 & 0.0250 & -18.87 & 176.4 & 59.93 & 15.28 & 45.75 & 179.3 & 46.93 \\
SN 2007ib$^{a,b,N}$ & 9.08 & 0.0344 & -19.90 & 196.3 & 134.0 & 12.46 & 37.31 & 466.8 & 155.8 \\
none SNF20080323-006$^{c,N}$ & 9.09 & 0.0306 & -19.28 & 105.3 & 89.81 & 8.291 & 24.83 & 354.6 & 129.6 \\
none SNF20080614-002$^{c,N}$ & 8.86 & 0.0262 & -17.79 & 132.7 & 45.97 & 17.98 & 53.92 & 144.4 & 23.87 \\
SN 2009dm$^{a,T}$ & 8.66 & 0.0245 & -17.85 & 423.9 & 139.9 & 111.5 & 334.1 & 485.7 & 57.35 \\
\hline
\end{tabular}
\end{center}
\caption{\label{II_table}Table of Type II SNe hosts.  Values are displayed as in Table \ref{GRB_table}.  All of our Type II SNe have hosts in the SDSS Catalog.  All of the values for these hosts, with the exception of the metallicity which is recomputed, come exclusively from the SDSS-mpg data products.  Flux spectra are in units of $10^{17}$ ergs s$^{-1}$ cm$^{-2}$.  The superscript a, b, \& c by the SN name denotes the survey population in which the SN was originally detected as listed below (multiable listings indicate that the SN was detected in more then one population): As in Figure \ref{Ic_table} the superscript ``T" or ``N" denotes whether the SN was detected in a targeted or non targeted manner respectively.}
\vspace{-4pt}
a \cite{Prieto} compilation (contains SNe found in both targeted and non targeted surveys)\newline
b SDSS Supernova Survey (non targeted survey)\newline
c Nearby Supernova Factory (non targeted survey)
\end{table*}
\subsubsection{Tabular References} \label{tabref}
\noindent 
1 Levesque, E.~M., Berger, E., Kewley, L.~J., \& Bagley, M.~M.\ 2010, \aj, 139, 694\\
2 Svensson, K.~M., Levan, A.~J., Tanvir, N.~R., Fruchter, A.~S., \& Strolger, L.-G.\ 2010, \mnras, 405, 57\\
3 Levesque, E.~M., Kewley, L.~J., Berger, E., \& Zahid, H.~J.\ 2010, \aj, 140, 1557\\
4 Sollerman, J., {\"O}stlin, G., Fynbo, J.~P.~U., et al. 2005, New A, 11, 103\\
5 Christensen, L., Vreeswijk, P.~M., Sollerman, J., et al. 2008, \aap, 490, 45\\
6 Price, P.~A., Kulkarni, S.~R., Berger, E., et al. 2002, \apjl, 571, L121\\
7 Garnavich, P.~M., Stanek, K.~Z., Wyrzykowski, L., et al. 2003, \apj, 582, 924\\
8 Hammer, F., Flores, H., Schaerer, D., et al. 2006, \aap, 454, 103\\
9 Levesque, E.~M., Kewley, L.~J., Graham, J.~F., \& Fruchter, A.~S.\ 2010, \apjl, 712, L26\\
10 Levesque, E.~M., Kewley, L.~J., Berger, E., \& Zahid, H.~J.\ 2010, \aj, 140, 1557\\
11 McGlynn, S., Clark, D.~J., Dean, A.~J., et al. 2007, \aap, 466, 895\\
12 Graham et al. 2012 in prep\\
13 Th{\"o}ne, C.~C., Fynbo, J.~P.~U., {\"O}stlin, G., et al. 2008, \apj, 676, 1151\\
14 computed from data given therein\\
15 calculated from SDSS mags and redshift\\
16 Modjaz, M., Kewley, L., Kirshner, R.~P., et al. 2008, \aj, 135, 1136\\
17 Ferguson, A.~M.~N., Gallagher, J.~S., \& Wyse, R.~F.~G.\ 1998, \aj, 116, 673\\
18 Modjaz, M., Kewley, L., Bloom, J.~S., et al. 2011, \apjl, 731, L4\\
19 SDSS-mpg\\
20 Sanders, N.~E., Soderberg, A.~M., Levesque, E.~M., et al. 2012, arXiv:1206.2643\\
21 NED\\
22 Levesque, E.~M., Soderberg, A.~M., Foley, R.~J., et al. 2010, \apjl, 709, L26\\

\section{Analysis \& Results}

We begin our analysis by comparing the metallicity luminosity relation of LGRB and broad-lined Type Ic (Ic-bl) SNe hosts (Section \ref{siteLZ}) as was originally done in the seminal work of \cite{Modjaz2008}.  Using a much smaller sample of LGRBs, \cite{Modjaz2008} provided some of the first strong evidence that LGRBs show a strong metallicity aversion.  \cite{Modjaz2008} compared the luminosities metallicities of LGRB and Type Ic-bl hosts to the SDSS population.  However, the SDSS suffers two major problems when used in this way as a comparison sample.  First, it is not a volume limited sample, but a magnitude limited sample, and thus it is heavily weighted to bright (and therefore typically metal rich) galaxies.  Secondly, it is primarily a very local sample with median and maximum redshifts several times smaller than our LGRB sample, which extends nearly out to $z \sim 1$.  We attack these problems with two basic steps.  Out of the main SDSS we create a volume limited survey, to allow comparison with the (often very faint) LGRB hosts (see Section \ref{zcut}), and we use the TKRS to provide a higher redshift galaxy comparison sample (see Section \ref{TKRS}).  Further we expand upon the Modjaz comparison sample of broad-line Type Ic SNe, by adding in Type II SNe, as well as a synthetic star-formation weighted population as further indicators of the distribution of star-formation.  

We do not rely upon the two-dimensional luminosity metallicity plot as the basis of our main analysis.  Instead, using the volume limited sample, we analyze how how each class of objects is distributed in metallicity with respect to its underlying star-formation and whether, once metallicity is accounted for, the probability of an object going off in a galaxy is proportional to the SFR of that galaxy.  

Our analysis shows that the offset LGRBs from the standard mass-metallicity relation for galaxies is robust.  Inclusion of the bias towards lower metallicity caused by star-formation is not sufficient to explain the stark deviation of LGRBs.  While the SNe track the star-formation distribution of the local universe, eighty percent of LGRBs are found in the ten percent of star-formation with the lowest metallicity.  Nonetheless, the probability of an LGRB going off in a galaxy is directly proportional to its SFR, once metallicity has been taken into account.  Indeed, due to an apparent similarity in star-formation distributions between low and higher metallicity galaxies, the distribution of LGRBs, just like the SNe, tracks the SFR of the entire sample of star-forming galaxies.  Finally we show that the expected metallicity evolution with redshift of the LGRB host sample is much smaller then the LGRB metallicity aversion.  Thus we conclude that the only remaining viable explanation for the metallicity distribution of LGRBs is an intrinsic preference for low (i.e.  one-third solar) metallicity environments, which does not completely exclude LGRB formation at solar metallicity and above.

\subsection{Metallicity Luminosity Relation of Progenitor Sites} \label{siteLZ} \label{hostLZ}

\begin{figure*}[p!]
\begin{center}
\includegraphics[width=1.0\textwidth]{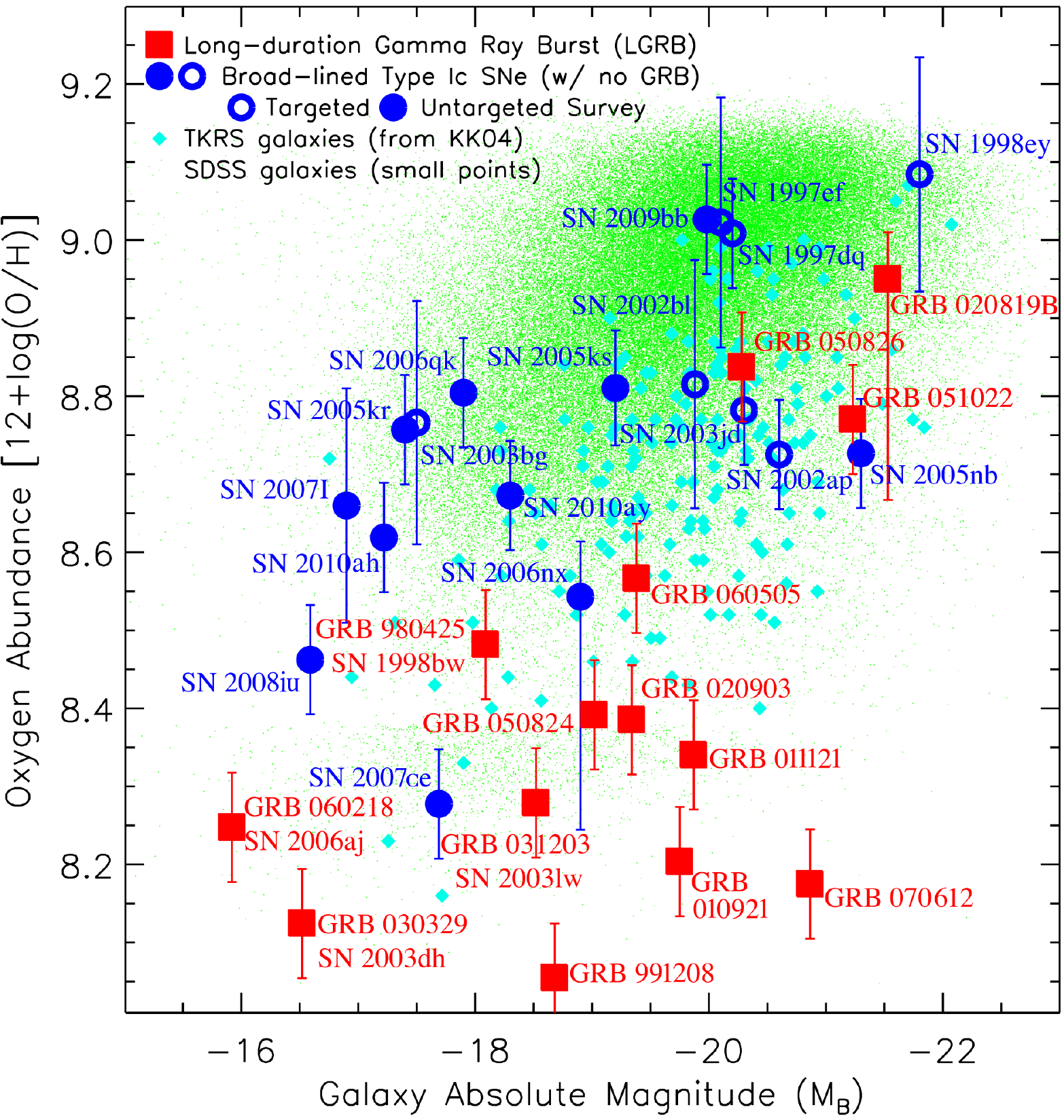}
\caption{\label{LZ} Metallicity vs.~absolute B band galaxy luminosity of LGRB (squares) and broad-lined Type Ic SNe hosts (circles).  Filled circles represent SNe selected in an untaragted manner whereas open circlers are from galaxy targeted SNe surveys (and thus may be biased in galaxy properties by target selection).  Star-forming galaxies from the SDSS (small dots) and TKRS (diamonds) are plotted in the background to provide a low and high redshift comparison sample respectively.  Where measured site metallicities are not available the central host metallicity had been downshifted to estimate the site value using the metallicity gradient methodology outlined in Appendix \ref{mshifting} (also see the Appendix for Figure \ref{measured_plots} which shows values as measured).  Note the profound difference between the LGRB metallicity values and those of the Type Ic-bl SNe}.  

\end{center}
\end{figure*}

As already noted, we begin our analysis by comparing the site metallicites and host luminosities of LGRBs and Type Ic-bl SNe with the central metallicities
of the SDSS star-forming galaxy population.  However, the SDSS is a very local sample.  After the cuts based on galaxy emission line signal to noise, described in Section \ref{zcut}, the median redshift of of our SDSS star-forming galaxy sample is z = 0.0728, with no objects having a redshift above 0.4 (with only few objects with a redshift above 0.3).  Our LGRB population however, has a median redshift of z = 0.36 and a maximum redshift of z = 0.83.  Thus to extend the redshift range of the star-forming galaxy population, in this work we will also include the TKRS population described in Section \ref{TKRS} for much of our analysis.

R$_{23}$ metallicites are used for all LGRB hosts and star-forming galaxies, and SN hosts where available.  For the fraction of SN hosts where R$_{23}$ lines are not available, the less accurate [N II]/H$\alpha$ metallicity is used.  All metallicities were calculated by us using the available emission lines, and in all cases are placed on the \cite{KobulnickyKewley} metallicity scale as described in Section \ref{metal_measurement}.  Due to the absence of line flux error data for many of the objects, a standard error of $\pm$0.07 dex is assumed for R$_{23}$ metallicities and $\pm$0.15 for [N II]/H$\alpha$ metallicities was assumed.  Compared to the \cite{Modjaz2008} sample, our sample has an approximate three fold increase in the number of LGRBs, one additional SNe, and a second general star forming galaxy population (the TKRS) to better reflect the redshift distribution of the LGRBs.  In order to avoid unknowingly biasing our SNe host properties, we differentiate between SNe that were discovered in an untargeted search of the sky as opposed to SNe discovered in targeted surveys of specific galaxies.  The properties of the hosts of the targeted SNe may be biased by the selection criteria used in selecting the galaxies within which to search for SNe events.

Nearly all LGRBs that have been studied sufficiently well to assure detection of an underlying Type Ic SN have shown evidence of a Type Ic-bl SN.  Thus we might expect the progenitors of the general population of Type Ic-bl SNe to be closely related to those of LGRBs.  A comparison of the hosts of LGRBs and those of Type Ic-bl SNe could give us some indication if this assumption is indeed true.

In Figure \ref{LZ} we plot metallicity vs.~B band absolute galaxy luminosity for LGRBs, Type Ic-bl SNe hosts without associated LGRB events, and general star-forming galaxies (at low redshift from the SDSS and at high redshift from the TKRS).  As \cite{Modjaz2008} noted, the LGRB and SNe metallicity site values are more indicative of the progenitor local environment's metallicity than the galaxy core values and therefore we plot site values here, even though we only have core values for the background populations.  We will address the issue of comparing between site metallicities of our LGRB and SNe populations and the central metallicites of our general star-forming galaxies in Section \ref{perils}.

Immediately obvious is the profound difference between the LGRB metallicities and those of both the (non LGRB) broad-line Type Ic SNe and general star forming galaxies as originally shown by \cite{Modjaz2008}.  This effect is clear even though our larger LGRB sample contains three LGRB hosts at typical galaxy and SNe host metallicities.  These objects, LGRBs 051022, 020819B, \& 050826, are the focus of a companion paper, \cite{pregraham} in prep.  Nonetheless, the LGRB population prefers much lower metallicity host environments than the environments typically found in the general star-forming galaxy population or indeed the general Type Ic-bl SNe population.  Thus, the primary conclusion that there is an intrinsic preference for low metallicity in LGRB formation remains true.  We will address this preference for low-metallicity statistically in Section \ref{popstat}.

\subsection{The Current Host Metallicity Luminosity Relation} \label{snapshot}

In this section we make two primary changes to the analysis already presented.  We restrict the redshift range of the SDSS, and we add two new comparison populations, Type II SNe and a star-formation weighted galaxy sample.  Limiting the SDSS to the redshift cut of 0.02 $\leq$ z $\leq$ 0.04 described in Section \ref{zcut} avoids two major problems.  First, the SDSS is not a volume limited sample, but essentially a magnitude limited sample.  It is therefore heavily weighted to bright (and therefore typically metal rich) galaxies and thus neglects fainter galaxies more akin to LGRB hosts.  Secondly the SDSS sample extends out to roughly $z \sim 0.3$, a range large enough that cosmic abundance evolution will start to become noticeable.  However, with the narrow 0.02 $\leq$ z $\leq$ 0.04 redshift constraint, the galaxies evolution issue is rendered negligible and the high completeness of the SDSS within this redshift range for galaxies brighter then -18 M$_B$ converts the population to a volume limited survey allowing suitable comparison with the (often very faint) LGRB hosts.

Unfortunately, sample size limitations preclude the universal application of this redshift cut across all populations.  Obviously the LGRB population cannot be usefully constrained to the stated redshift range (this would reduce the LGRB hosts to a sample of one).  Since evolution of galaxies increases their metallicity over time this will introduce a low metallicity bias in the higher redshift LGRB host population which we will subsequently demonstrate (using the TKRS sample) to be a much smaller effect than the low metallicity preference of the LGRBs in Section \ref{redshift}.  (Since the TKRS population is all at z $>$ 0.3 they are not considered here).  The broad-line Type Ic SNe sample also largely lies outside of the 0.02 $\leq$ z $\leq$ 0.04 redshift range (with a significant fraction below).  However since this population is naturally constrained in redshift (median redshift of z = 0.0216 and a maximum redshift of z = 0.137), any biases in metallicity and host luminosity are limited, particularly as these objects have received deep follow-up observations.

\begin{figure*}[p!]
\begin{center}
\includegraphics[width=1.0\textwidth]{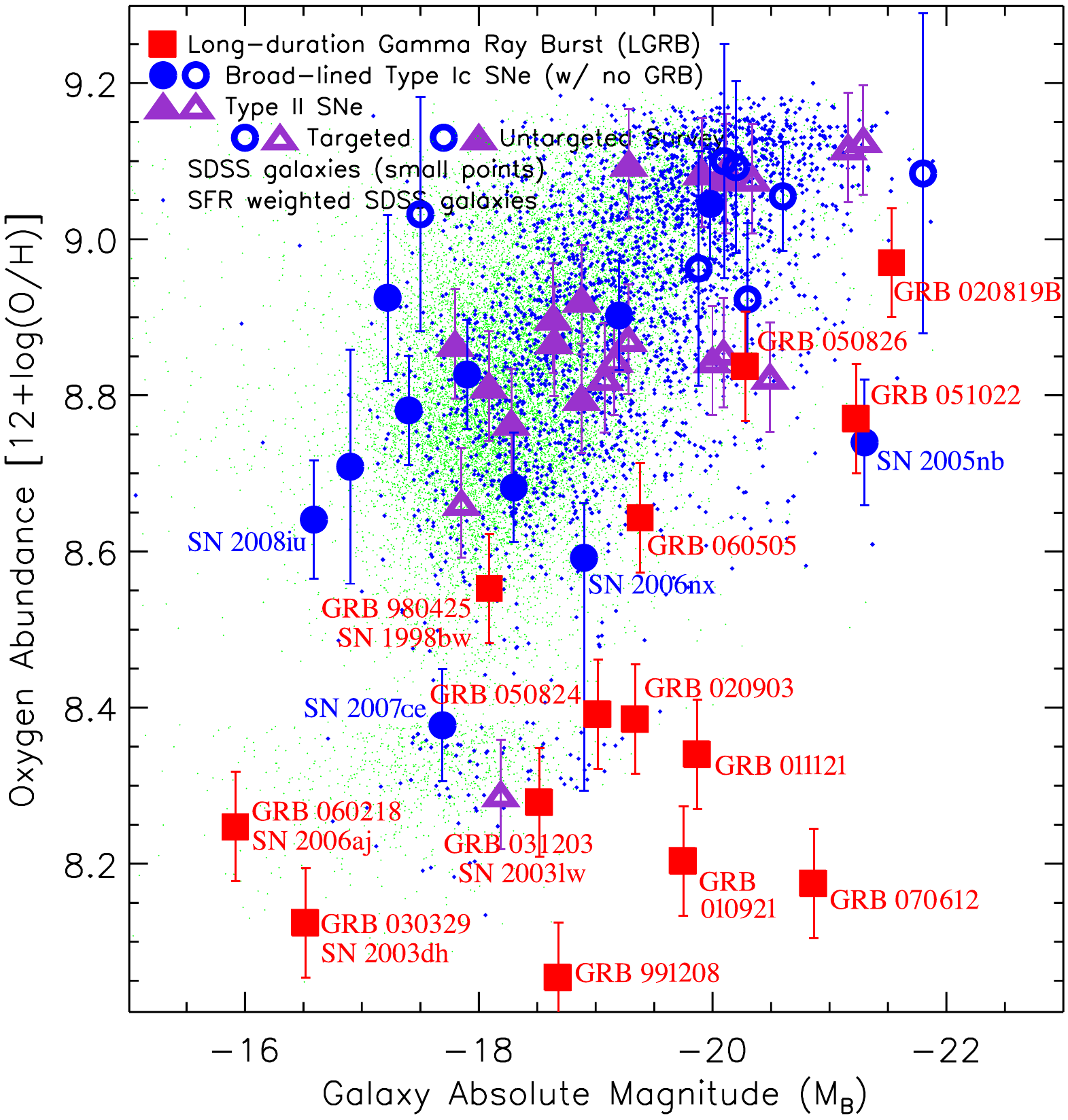}
\caption{\label{LZ_zcut} Metallicity vs.\thinspace \thinspace B band absolute galaxy luminosity.
The Type II SNe hosts (purple triangles), SDSS galaxies (small green points) and the star-formation weighted random SDSS galaxies synthetic population (small blue points) shown here are limited to the sample redshift range of $0.02 < z < 0.04$ discussed in Section \ref{zcut}.  In order to get reasonable comparison samples, no redshift cut has been applied to the Type Ic-bl host population (blue circles) or the LGRB population (red squares).  This may produce a small bias toward brighter hosts for the broad-line Type Ic SNe; however as the most distant broad-line Type Ic SNe has a redshift of 0.137, it will not produce a significant bias in metallicity.  For LGRBs, to some extent, both biases will occur and we discuss the small effect of an unrestricted redshift distribution on LGRBs in Section \ref{redshift}.  As in Figure \ref{LZ} filled and open symbols represent SNe selected in untargeted and targeted surveys respectively.  In order to explicitly trace star-formation we introduce the star-formation weighted synthetic population (small blue points) discussed in Section \ref{synthetic}.  This population tracks the SNe quite well in luminosity-metallicity space, as one would expect with SNe tracking star-formation.} 

\end{center}
\end{figure*}

While the LGRB and Type Ic-bl SNe populations may themselves be (biased) tracers of star-formation (presumably a galaxy with twice the star-formation would, all else remaining equal, have twice LGRB and SNe events) this is not the case for the SDSS population where the brighter galaxies have more star-formation but are still plotted singularly in Figure \ref{LZ}.  Thus, in Figure \ref{LZ_zcut}, we also plot a sub-selection of the redshift cut SDSS population selected randomly after weighting by star-formation (see Section \ref{synthetic} for details).  
 
Nature however, also provides a star-formation weighted galaxy sample through the hosts of Type II SNe.  These SNe are probably largely unbiased with respect to metallicity (though their discovery will be affected by dust), and they are sufficiently abundant that we can obtain a reasonably-sized population of hosts which are entirely contained in the redshift-cut SDSS sample.  This has the added benefit that the spectroscopic observations and analysis are identical for the object hosts and the SDSS population.

While in Section \ref{siteLZ} and \cite{Modjaz2008} the metallicity of the SNe and LGRB site within the host was used, here and subsequently we switch to using the central metallicities of the host galaxies to allow for a better comparison with the general star forming galaxy population.  Since galaxies are not uniform metallicity objects, there will frequently be an offset in metallicity between the center of the host and that of the explosion site.  In a few cases we have the metallicity of both the explosion site and the center of the host.  In most cases do not have a measurement of the metallicity offset, but can estimate it using the method described Appendix \ref{metal_grad_fit} .  Examining Table \ref{Ic_table}, one sees observed and expected offsets up to 0.3 dec in metallicity between the explosion site and the host center for the broad-line Type Ic SNe.  \label{perils}

As discussed in Section \ref{LGRBs}, LGRBs tend to occur in the brightest regions \citep{Fruchter} of their predominately irregular host galaxies.  Furthermore due to the distance to the LGRB hosts, and their generally compact sizes, most of the light of the host goes down a single ground-based slit (see Section \ref{LGRB_slc}).  Thus in most cases we would not expect a significant offset in metallicity between the host site and the measured host metallicity.  However, in the case of the three well-resolved LGRB hosts, LGRBs 980425, 020819B, \& 060505, we are able to directly address this issue, and we discuss their (small) metallicity offsets in Section \ref{LGRBs}.

In examining Figure \ref{LZ_zcut}, one notes that the agreement between the luminosity - metallicity relations of the SNe host and the red-shift cut star-forming SDSS sample is notably better than the agreement with the entire SDSS star-forming population (shown in Figure \ref{LZ}).  The agreement of the two-dimensional distribution is even better when one compares the SNe to the star-formation weighted sample (small blue dots in Figure \ref{LZ_zcut}).  The SNe and the synthetic star-forming 
galaxy population appear to follow the same luminosity-metallicity relation and have similar distributions of population density in the two-dimensional luminosity
metallicity plane.  Nonetheless, the LGRBs remain well-separated from all of the comparison galaxy samples.  In the following section, we will use a more powerful statistical methodology to compare the properties of the hosts of the SNe, LGRBs, and the star-forming SDSS galaxy population.

\subsection{A Statistical Comparison of Populations}\label{popstat}

In the previous section, we examined the host metallicity distribution with respect to luminosity through the use of the metallicity luminosity plot.  Here we take a more statistical view and ask specifically: do LGRBs and SNe track star-formation independent of host metallicity?  In the second section we have a first look at the question: does the host SFR itself bias the probability of LGRB or SNe star-formation, as one might expect if star-burst conditions enhanced the production LGRBs or SNe beyond the SFR itself, perhaps by altering the IMF towards higher mass objects?

\subsubsection{Star-formation Metallicity Distribution} \label{obj_sf_track}

If LGRBs and SNe events track star-formation, then their distribution in metallicity space should be identical to that of star-formation in general.  Thus if in the local universe twice as many stars are generated per unit time at metallicity $Z_1$ than $Z_0$, then we should find twice as many SNe and LGRBs at a metallicity of $Z_1$ than at $Z_0$.  To look at this issue we think of each host galaxy, or galaxy in the SDSS, as a particle of given metallicity and star-formation.  We then sort our objects, and galaxies by metallicity, and examine what fraction of star-formation, or object formation, occurs up to a given metallicity.  

\begin{figure*}[ht]
\begin{center}
\includegraphics[width=.49\textwidth]{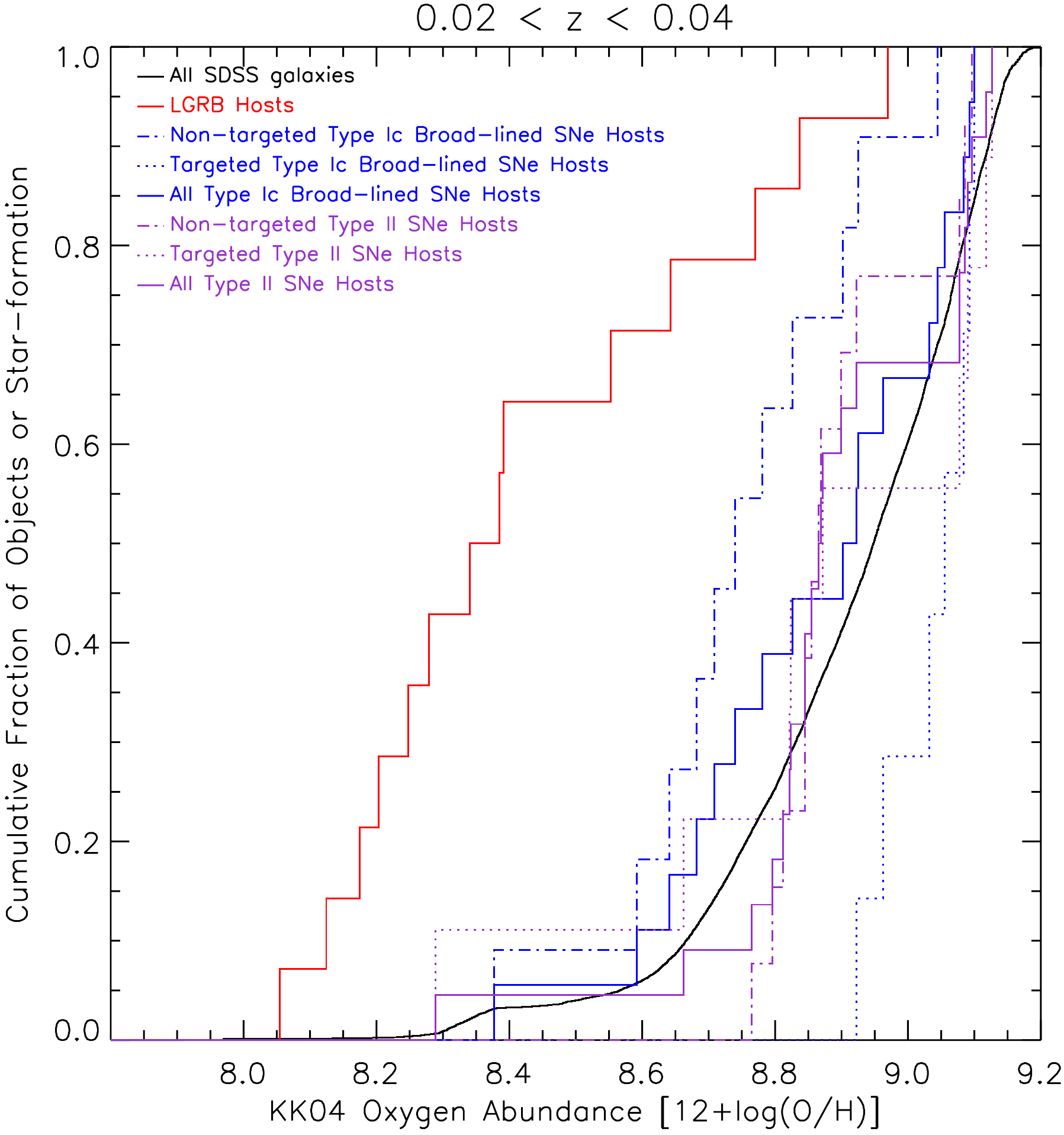}
\includegraphics[width=.49\textwidth]{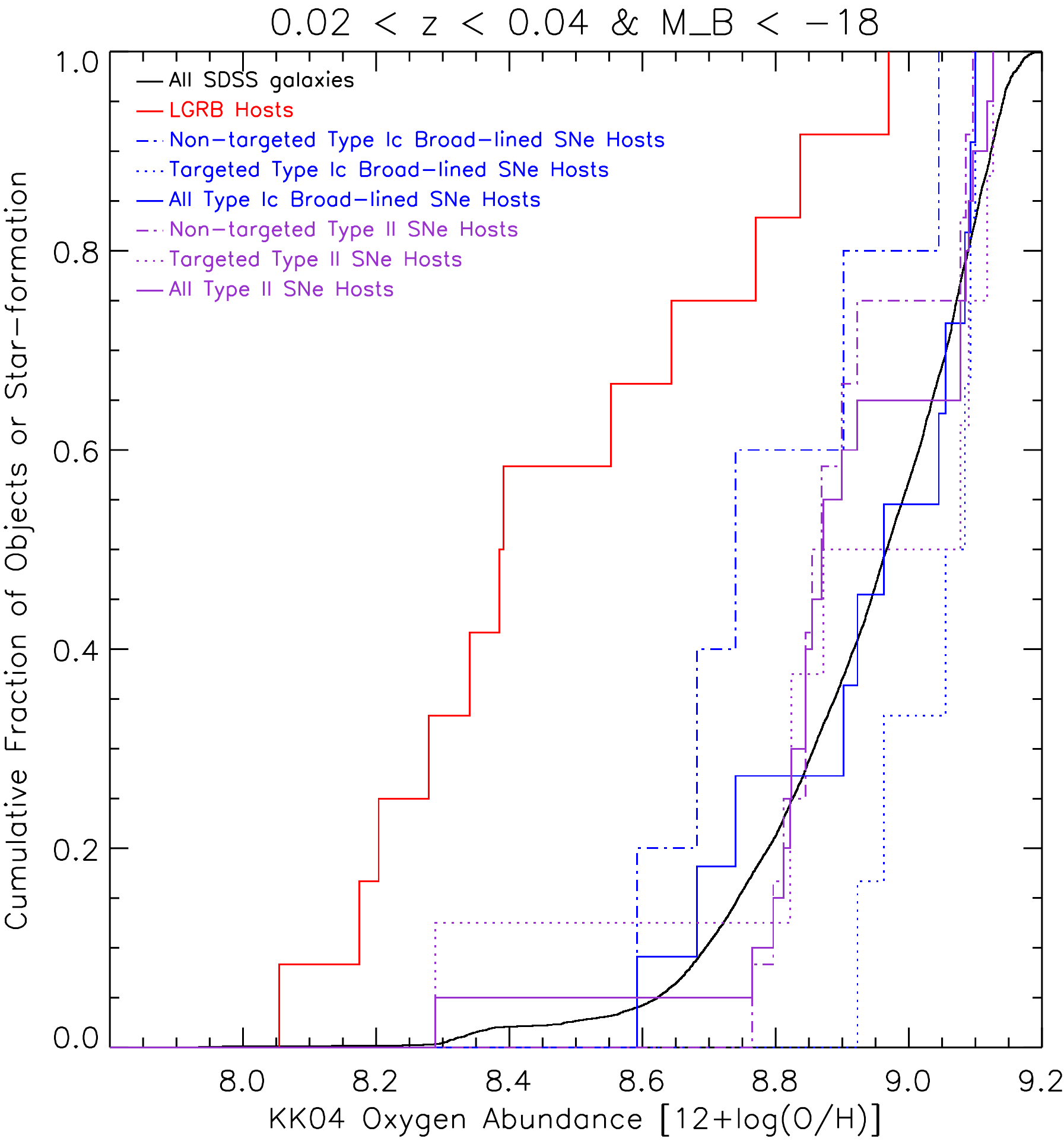}

\caption{\label{dSFR_oplot} Cumulative fraction of population (LGRBs and SNe) or total star-formation (redshift-cut SDSS star-forming sample) vs.\thinspace \thinspace galaxy central metallicity.  LGRB host galaxies are shown in red, Type Ic-bl SNe hosts in blue, and Type II SNe hosts in purple.  Non-targeted SNe are shown with a dashed line, targeted SNe with a dotted line and the union targeted and non-targeted populations with a solid line.  The redshift-cut SDSS star-forming galaxy population is shown in black.  Only the SDSS galaxies, and the Type II SNe hosts (which are all included in the SDSSS sample), are limited to the redshift cut range.  In the left-hand plot we use our full samples (subject to the redshift cuts).  In the right-hand plot we restrict both the host and SDSS samples to the magnitude the approximate completeness limit of the SDSS in our redshift range of $\rm M_B < -18$ to insure that host magnitude does not bias the metallicity selection.  A detailed description of the plotted populations is described in Section \ref{sample_pops}.  The LGRB hosts have metallicities considerably lower then would be obtained simply by following star-formation while both types of supernova are consistent with the star-formation distribution.  Thus the metallicity distribution of LGRBs cannot be explained only by association with star-forming galaxies.}

\end{center}
\end{figure*}

For the SDSS, we create a normalized cumulative distribution of star-formation versus metallicity, i.e. sort galaxies by their metallicity, and determine the fraction of total star-formation in the SDSS population contained in galaxies below a given metallicity.  This cumulative distribution can be seen in Figure \ref{dSFR_oplot}.  For the LGRBs and SNe we assume the null hypothesis that these objects track star-formation, and thus each explosion can be thought of as a ``quantum" of star-formation.  In Figure \ref{dSFR_oplot} we can therefore plot their cumulative distributions as a step function of objects.  A similar LGRB step function was first used in \cite{Fruchter} to compare the burst location with the distribution of host galaxy light, and has more recently been applied to the metallicity of SNe independently by \cite{Kelly}.  The primary innovation here is the direct comparison between the step functions of events with a normalized cumulative distribution function of the star-formation in the general galaxy population.  This analysis has the notable advantage of allowing comparison of events assumed to track star-formation without actually requiring information on the SFR of the hosts of these events themselves.  In our case this is primarily of benefit with the Type Ic-bl SNe and the LGRB's where such information is not readily available for a number of the objects.

The general agreement of both the Type Ic-bl and Type II SNe with the SDSS general star-forming galaxy population seen in the plots is what one would expect assuming the SNe are a product of the star-formation alone without additional selective pressures.  Indeed, if we restrict our SNe samples to only hosts with M$_B$ brighter than -18 magnitude, which is the completeness limit of our SDSS sample, all of the SNe populations agree reasonably well with the SDSS distribution of star-formation with KS values greater than $\sim$0.1 in all cases.  It is widely realized that targeted surveys will have selection effects.  Although our nearby sample may somewhat limit this effect, in general, targeted surveys overrepresent bright (and attractive) galaxies.  However, even untargeted surveys can have significant biases, and those used to create our non-targeted SNe samples are no exception.  For example, the Palomar Transient Factory (PTF), which contributes some of our Type Ic-bl SNe, has placed significant emphasis on typing SNe found in faint or missing hosts (Y. Arcavi private communication).  This will bias metallicities due to the mass-metallicity relationship of galaxies.  But perhaps the greatest issue for our untargeted samples is due to the difficulty of finding SNe on top of bright surface brightness hosts, or near the nuclei of massive hosts.  We find that this effect can be characterized fairly well by examining the fiber plug magnitudes from the centers of the SDSS galaxies, which act as an estimate of the galaxies' central surface brightnesses (see Appendix \ref{problem}).  Interestingly, the SNe sample that most closely tracks the SDSS star-formation metallicity function (KS = 0.88) is the combined sample of untargeted and targeted Type Ic-bl SN in hosts brighter than -18 (see Figure \ref{dSFR_oplot} right plot).

However, the LGRB population differs greatly from the distribution expected based on star-formation.  Indeed a K-S analysis suggests a probability of less than $10^{-5}$! From this comparison it is apparent that LGRB's occur much more preferentially than star-formation in low metallicity environments.  As can be seen in Figure \ref{dSFR_oplot}, while untargeted and targeted samples have differing biases (we will discuss possible biases in the untargeted population in greater detail in the next section), those biases are nowhere near sufficient to reproduce the deviation from star-formation seen in the LGRB population.  Indeed, 11 out of 14 LGRBs lie in the lowest metallicity tenth of the star-formation in the volume-limited SDSS sample.  This is in spite of the fact that the LGRB sample is not volume limited, and thus the requirement that we be able to measure the metallicity of the host, biases us towards including more luminous, and thus likely more metal rich, galaxies.  There is one bias on the LGRB population that tends toward lower-metallicites.  A fraction of our LGRB sample, however, extends up to a redshift of one.  This could bias us to lower metallicites.  However, as we show in Section \ref{redshift}, this bias is not at all sufficient to produce the observed effect.

\subsubsection{Fractional star-formation analysis} \label{frac}

Here we employ an alternate analysis directly comparing the star-formation of target host galaxies with those of the SDSS general star-forming galaxy population.  We ask for each SN or LGRB, what fraction of star-formation in the SDSS is contained in galaxies of equal or lesser star-formation than the host.  We do this to determine whether there is any apparent bias towards highly star-forming galaxies, beyond a linear correlation with SFR.  

We compare each of the hosts both to the SDSS population, and also to a subset of the SDSS galaxies with comparable metallicities.  However, we cannot just use the volume-limited redshift cut SDSS population for this procedure.  The LGRB sample is essentially a magnitude limited sample and is thus biased towards bright, and therefore highly star-forming galaxies.  The entire SDSS star-forming galaxy sample, however, is a magnitude limited sample and thus is the correct comparison with the LGRBs.  The Type II SNe are volume limited to the same range as the SDSS redshift cut and thus the volume limited SDSS sample is applied.  We also include in our comparison the Type Ic-bl SNe from our sample which have SDSS spectroscopy.  This allows us to obtain a SFR for their hosts.  However, we make no redshift cut on this sample, and allow redshifts both below and above the limits of the redshift cut SDSS sample.  Nonetheless, the Type Ic-bl SNe are far more magnitude limited by the brightness of their SNe than they are by their host, and thus the volume limited SDSS is a reasonable, but imperfect, comparison sample.  We therefore adopt it for determining the Type Ic-bl fractional star-formation values.

\begin{figure*}[ht]
\begin{center}

\includegraphics[width=.49\textwidth]{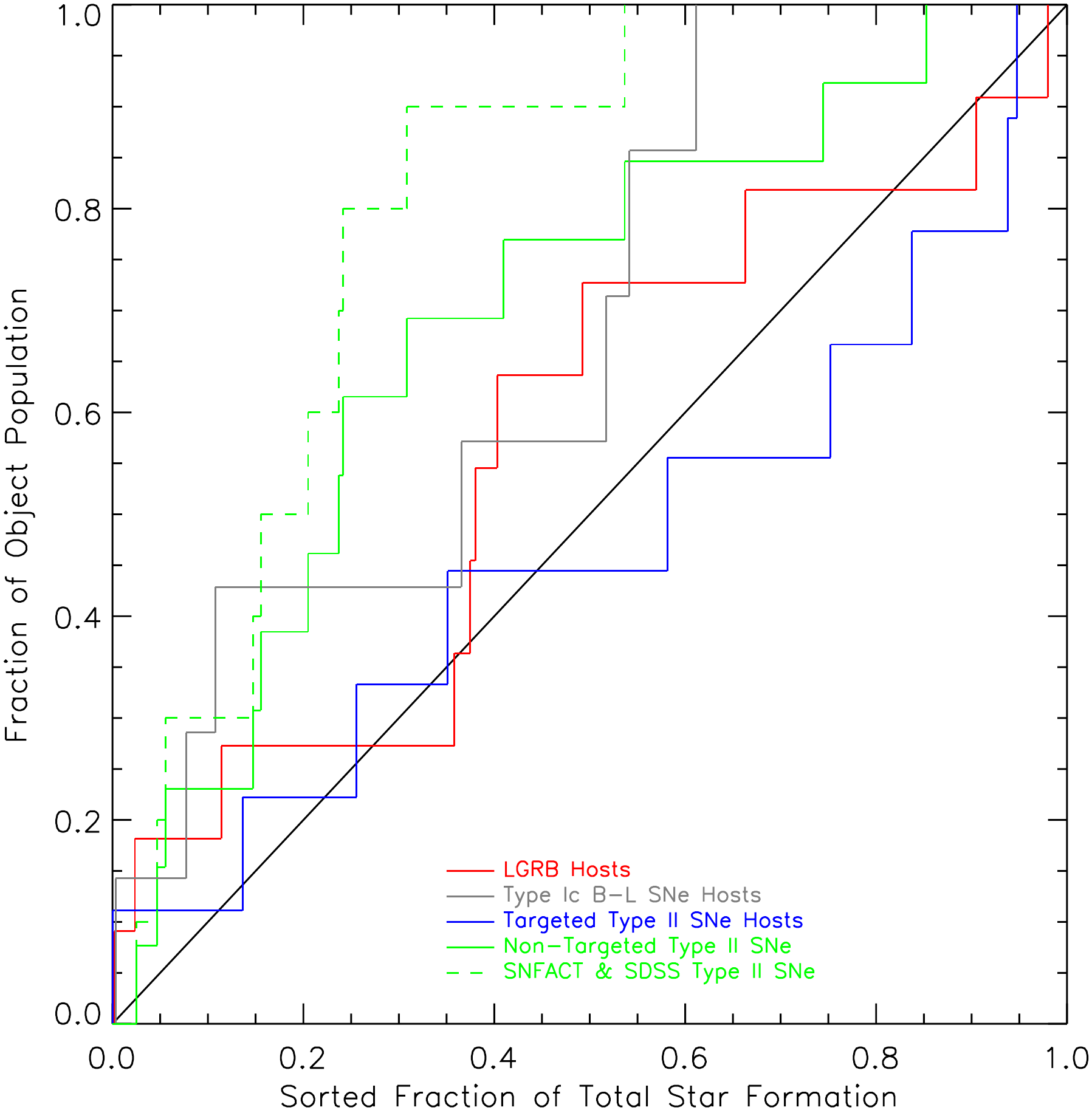}
\includegraphics[width=.49\textwidth]{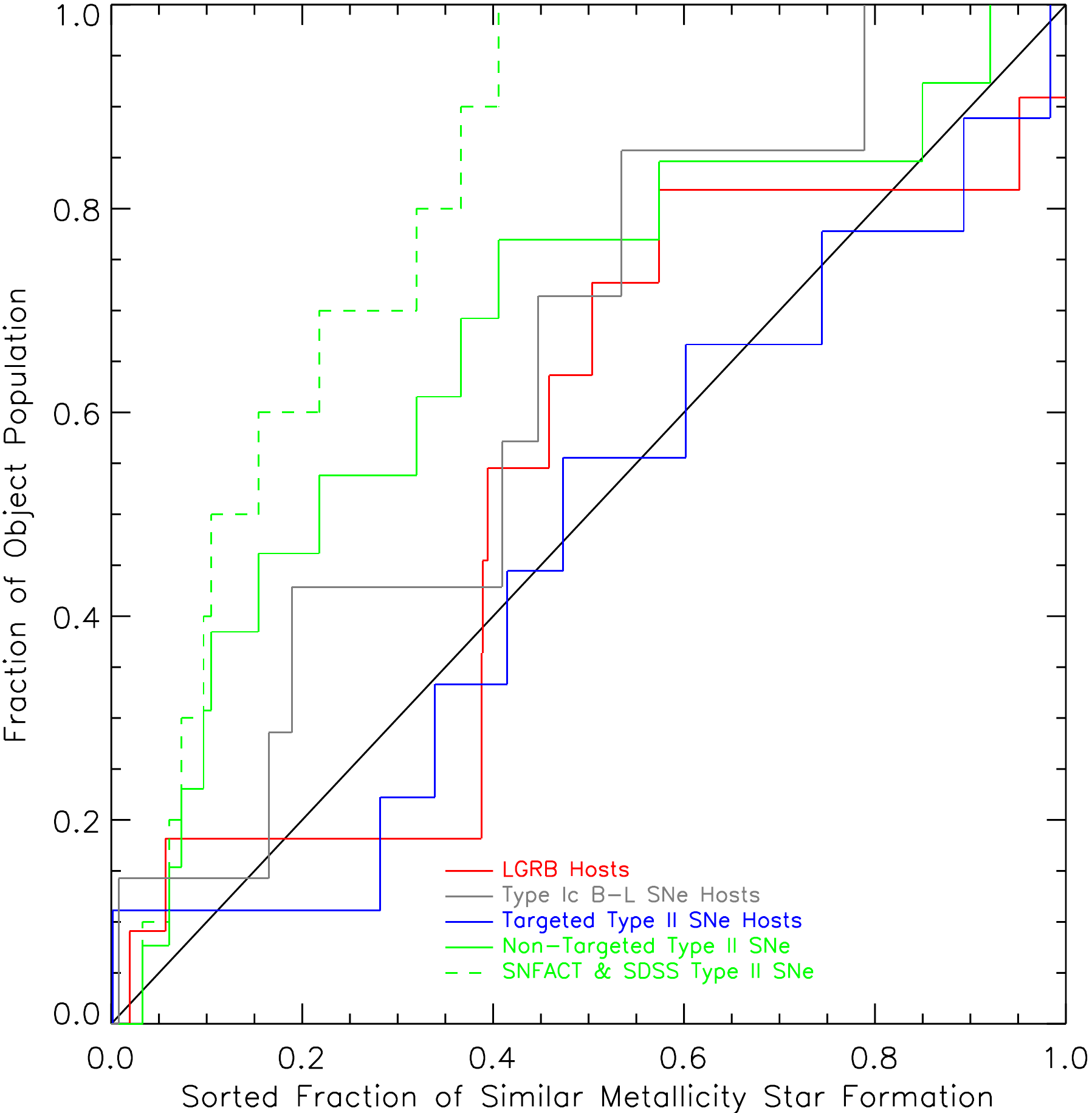}

\caption{\label{all_SFR} Normalized cumulative distribution of star-formation fractional values for the SNe and LGRBs.  (see Section \ref{frac}).  For each object we determine the fraction of star-formation contained in SDSS galaxies of equal or lesser star-formation than the object's host.  Were the objects distributed as star-formation the objects would be expected to follow a diagonal line on these plots.  The left plot uses star-formation across all metallicities to determine the fractional values whereas in the right plot uses only star-formation within $\pm$0.1 dex of the metallicity of the object.  For the SNe we use the volume limited SDSS sample as a comparison.  However, as the LGRB host sample is itself approximately a magnitude limited sample, the LGRBs are compared to the total SDSS star-forming sample.  These comparison choices are discussed in more detail in the main text.  The targeted Type II SNe (blue line) and LGRBs (red line) track the diagonal quite well indicating that both populations have fairly typical star-formation distributions.  However a similar comparison of the non-targeted Type II SNe sample (solid green line) and the Type Ic-bl SNe (grey line) show lower then expected host SFRs.  This is especially apparent when considering only Type II SNe found in the SDSS and SNFactory searches (dashed green line).  We suspect that this is due to limitations in the ability of non-targeted searches to identify SNe events on high surface brightness backgrounds, as is discussed in greater detail in Appendix \ref{problem}.  Thus while LGRB hosts do have higher than average SFRs, as noted by \cite{MannucciLGRBs}, their SFRs agree with what one would expect of a population that tracks star-formation.  Thus metallicity, rather than SFR, must be the primary source of the discordant results shown in Figure \ref{dSFR_oplot}.}

\end{center}
\end{figure*}

In Figure \ref{all_SFR} we show the results of this procedure.  In the left-hand plot, for each object (LGRB, SNe, etc.) we divide the total star-formation in SDSS galaxies with individual SFRs less then or equal to that of the target's host galaxy with the total star-formation in SDSS galaxies to create a fractional star-formation value.  
In the right-hand plot, the same procedure is used, but hosts are compared only to galaxies of similar metallicity, rather than the entire star-forming sample.  These sorted fractional values are shown in a standard normalized cumulative distribution plot.  A diagonal reference line corresponding to perfect tracking of the star-formation distribution is shown for comparison.  One would expect a distribution of objects that followed star-formation irrespective of metallicity to agree well with the diagonal of the left-hand plot.  However, a deviation from that line, might be expected of a population that followed star-formation, but was also biased by metallicity.  However, such a population should follow the diagonal line in the left-hand plot, as should a population, that has no metallicity bias.  The synthetic star-formation population was tested against the diagonal reference line as a check on methodology and implementation and was found to be is good agreement, but we do not show the synthetic population in this plot.  

\begin{figure*}[ht]
\begin{center}

\includegraphics[width=.49\textwidth]{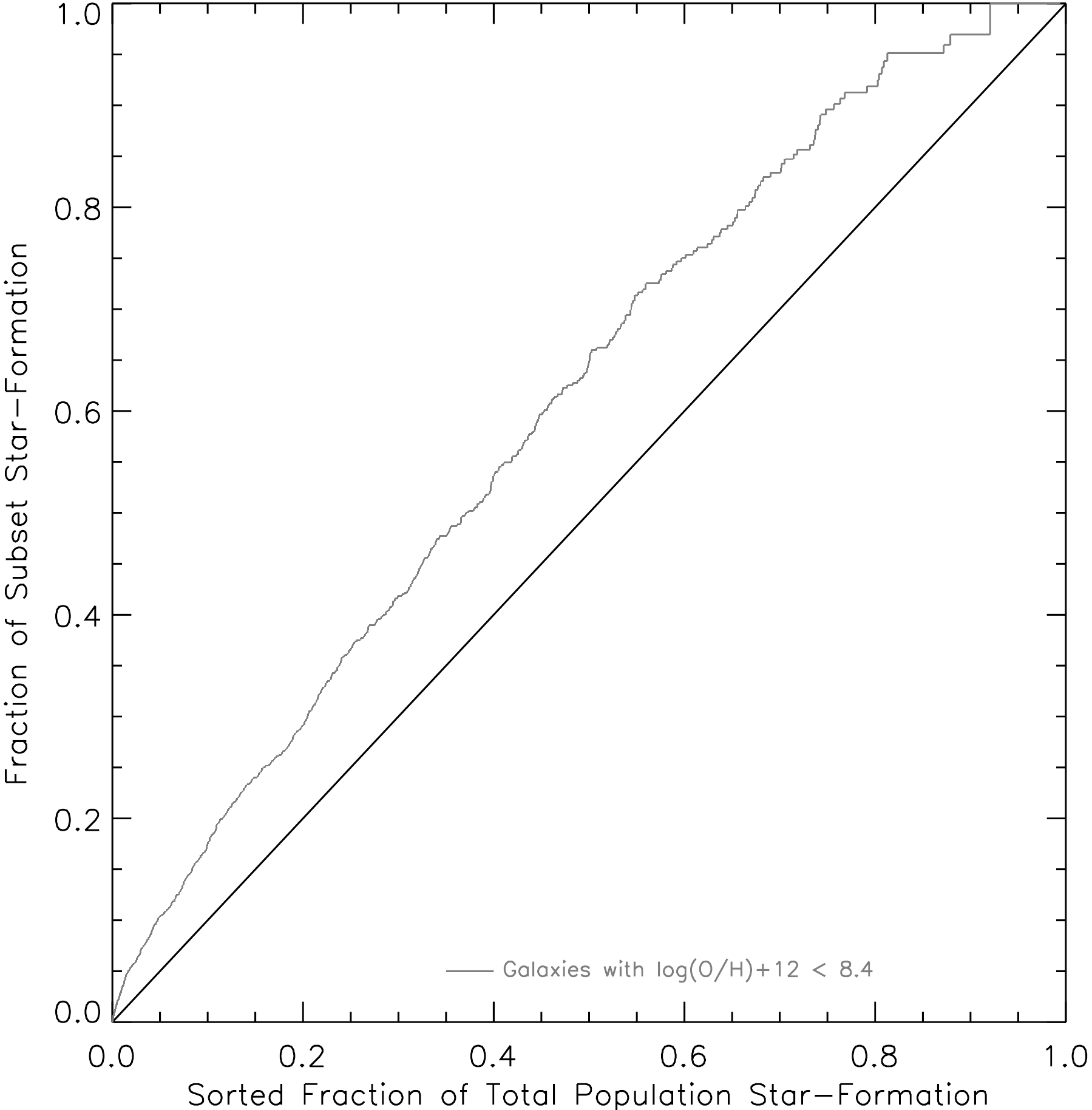}
\includegraphics[width=.49\textwidth]{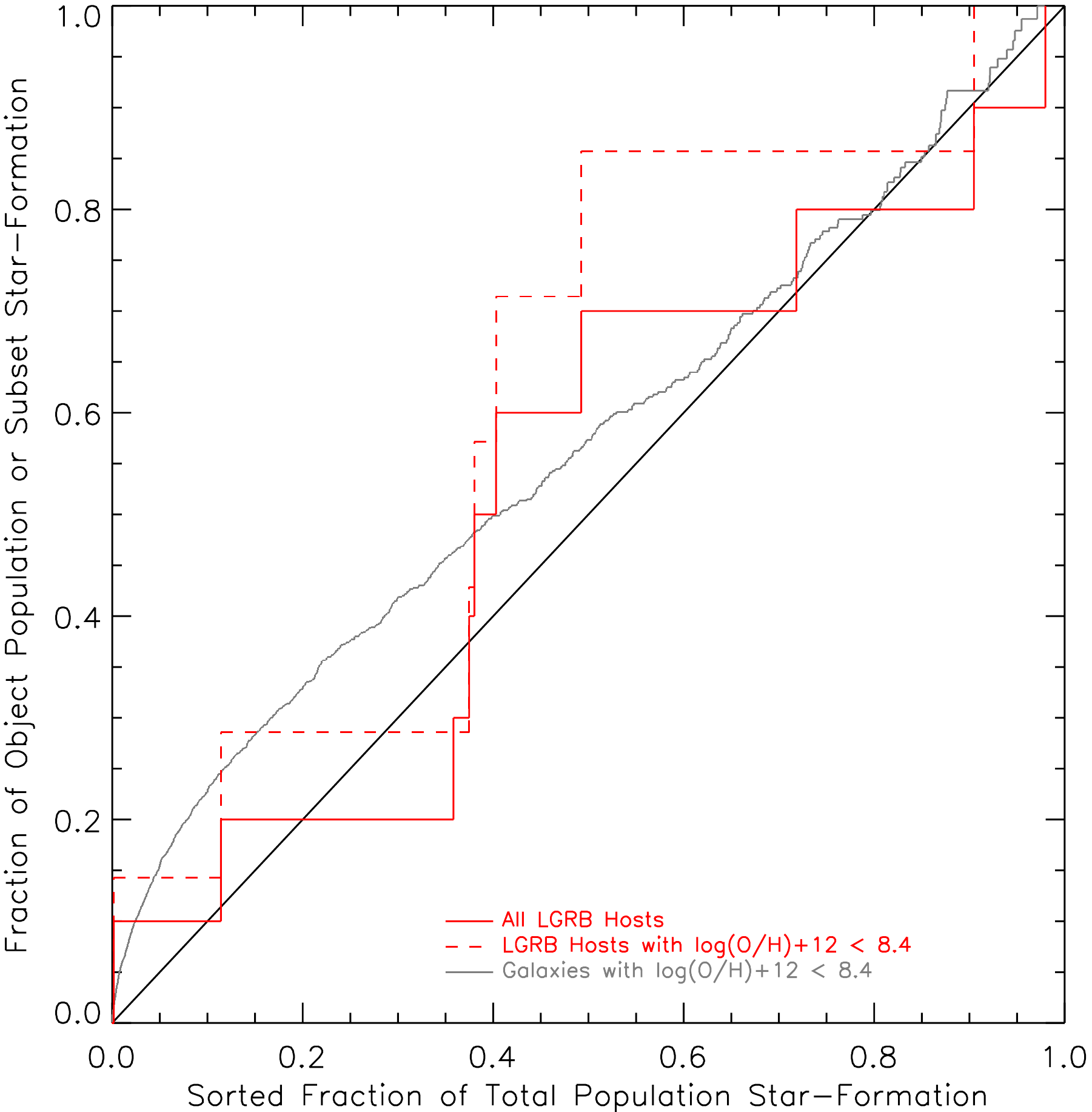}

\caption{\label{low_Z_sfr} Normalized cumulative distribution of star-formation of low metallicity galaxies (log(O/H)+12 $<$ 8.4) versus the entire SDSS population.  Left plot is for the volume limited SDSS population, right plot is for the entire SDSS population with the LGRB population (red lines) overplotted.  For each low metallicity SDSS galaxy we determine the fraction of star-formation contained in SDSS galaxies of equal or lesser star-formation in the low metallicity sample and in the entire SDSS.  We then plot the former as the abscissa and the latter as the ordinate.  The two samples are perhaps surprisingly similar, though as one might expect, the larger SDSS sample has a somewhat higher fraction of extreme star-forming galaxies.  Both the entire LGRB host sample and the subset of low-metallicity hosts are both plotted (solid vs. dashed red line respectively).  The star-formation distributions are sufficiently similar (and our LGRB sample sufficiently small) that neither is obviously a better match for the LGRB population.}

\end{center}
\end{figure*}

With the exception of the untargeted Type II SNe, all of the SN and LGRB population shown reasonable agreement with the general star-forming galaxy population (when the latter population is weighted by SFR) shows that the individual SFRs of these host galaxies tracks the general population.  Their distributions are representative of events occurring as would be expected where they simply a product of star-formation.  For the non-LGRB SNe this appears to be the entirety of the explanation necessary, however the trend of LGRBs to occur in low metallicity hosts means that while metallicity is obviously a critical factor in burst occurrence the star-formation distribution does not alter nor is altered by the low metallicity nature of these host galaxies.  Thus the observed metallicity bias can not be the incidental product of a deviation towards a host galaxy population of atypical SFR.  The untargeted Type II SNe appear to suffer from a strong bias against being found on high-surface brightness galaxies, and we believe this is the cause of their deviation from the expected star-formation track.  The evidence for this bias is discussed in detail in Appendix \ref{problem}.

As a further check to insure that metallicity has not biased our results, in the right hand plot of Figure \ref{all_SFR}, we isolate the underlying star-formation by removing the metallicity component from the analysis.  To effect this we reduce the SDSS sample used for the fraction value computation to only those galaxies whose metallicity matches ($\pm$ 0.1 dex) of the target object.  Thus for a low metallicity LGRB we would limit comparison to other low metallicity objects.  This isometallicity analysis allows a comparison of the star-formation tracking as influenced by all remaining factors.  The good agreement between the LGRB, targeted Type II SNe population, and the general star-forming galaxy population (when the latter population is weighted by SFR) shows that the individual SFRs of these host galaxies are typical of galaxies of similar metallicity, and indeed of the entire population, as shown previously.

The original similarity between the star-formation distribution of LGRB hosts, a population heavily biased towards low metallicity, and SDSS galaxies, whose star-formation we showed in the previous section to be primarily at high metallicity is perhaps surprising.  We further explore this by plotting the SFR distribution of the low metallicity SDSS galaxies weighted by their star-formation against the similarly weighted star-formation distribution of the entire SDSS sample (see Figure \ref{low_Z_sfr}).  This is equivalent to plotting fractional values (just as in Figure \ref{all_SFR} left) for a low-metallicity subset of the star-formation weighted galaxy sample.  From this we find that the SFR distribution of the low-metallicity SDSS galaxies is reasonably similar to that of the entire SDSS population.  This is actually done twice to show that the similarity is present in both our full SDSS sample and the volume limited subset.  We also overplot the LGRB population on Figure \ref{low_Z_sfr} showing the full SDSS sample.  This is done with both the full LGRB sample and after discarding the three high metallicity objects to better match the low metallicity SDSS population.  However the LGRB population is too small to differentiate between the star-formation distribution of the low-metallicity SDSS galaxies or the full SDSS galaxy sample.

\subsection{Host Metallicity Luminosity Redshift Relation} \label{redshift}

While our SNe samples are all restricted to local events, LGRBs tend to be cosmological, and thus local events are rare, precluding a similar redshift restriction.  Thus our LGRB sample extends out to nearly a redshift of one, while our volume limited SDSS sample only extends to $z< 0.04$ and the entire SDSS to about $z< 0.4$.  While the intrinsic metallicity preferences of LGRB and SNe should be independent of redshift, the metallicities of galaxies in general change with redshift, and thus any understanding of the effects of metallicity on LGRB formation must also include the possible biases introduced by the metallicity evolution of galaxies with redshift.  To extend our general star-forming galaxy comparison population out to a comparable redshift as the LGRB population we therefore add the TKRS galaxy survey population as described in Section \ref{TKRS}.  The TKRS population covers the redshift range $0.3 < z < 1.0$, and thus the the SDSS and the TKRS together span the entire redshift range of the LGRB sample.  

\begin{figure*}[p!]
\begin{center}
\includegraphics[width=1.0\textwidth]{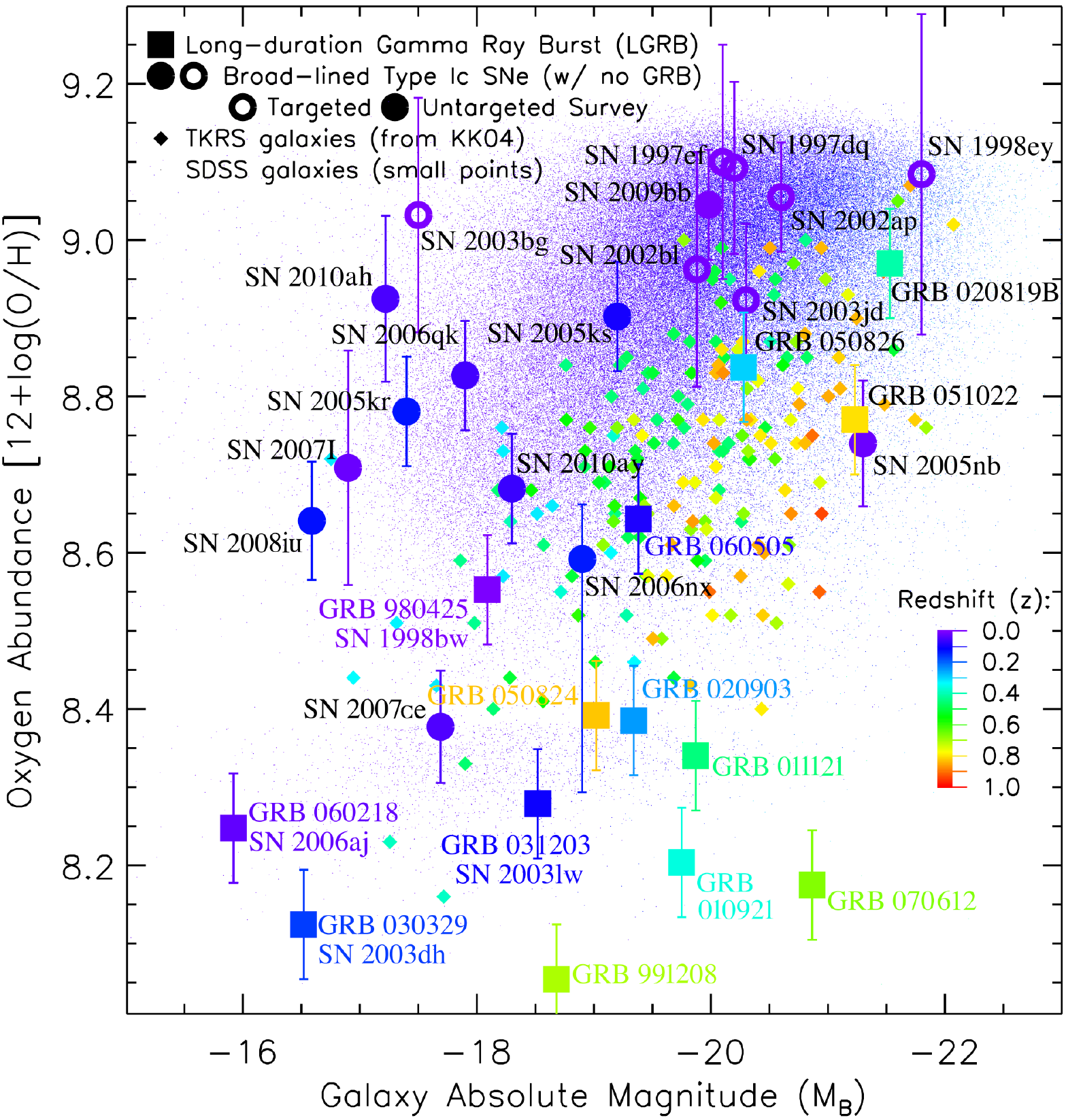}
\caption{\label{LZz} Central metallicity vs.\thinspace \thinspace B band absolute magnitude of LGRB (squares) and Type Ic-bl SNe hosts (circles).  As before (figures \ref{LZ} and \ref{LZ_zcut}) filled \& open circles represent SNe selected in an untargeted and targeted manner respectively.  Star-forming galaxies from the SDSS (small dots) and TKRS (diamonds) are plotted in the background to provide low and high redshift comparison samples respectively.  While in the previous plots, objects are color coded by type, here color is used to index redshift.  The different redshift distributions of the SNe and LGRBs populations mean that they are best compared not with each other but with the general star-forming populations at similar redshifts.  While as discussed earlier, the SNe we'll match their comparison sample; however, with the exception of a few relatively high metallicity LGRB hosts, which match the metallicities of similar magnitude galaxies in the TRKS, the LGRB population as a whole remains at far lower metallicities than even the TKRS.}

\end{center}
\end{figure*}

In Figure \ref{LZz}, we plot the host central metallicity luminosity redshift relation in a similar manner to previous plots, but instead of using color to separate object classes, we instead us it to index the redshifts of the objects.  In our previous plots, we implicitly ignored the effect of galaxy chemical evolution over time, and thus assumed this effect to be much smaller than the metallicity offset of LGRBs from the general star-forming galaxy population.  Here we explicitly show this to be the case.  

The different redshift distributions of the SNe and LGRBs populations mean that they are best compared not with each other but with the general star-forming populations at similar redshifts.  As shown in Section \ref{snapshot} the SNe are consistent with the general star-forming galaxy population at similar redshifts.  The LGRB population however is not consistent with the general star-forming galaxy population at any redshift.  Despite the existence of three relatively high metallicity LGRB hosts, the LGRB population is on average at far lower metallicities than the distribution of our comparison galaxy samples across any redshift range It should again be noted that the TRKS population above a redshift of z $\gtsim$ 0.5 (where the [N II] and H$\alpha$ lines move into the infrared) is assumed by \cite{KobulnickyKewley} to be on the upper branch of the $R_{23}$ metallicity diagnostic.  Thus the $ \sim 10\%$ of the sample that would be expected to be on the lower branch is missed (see Section \ref{TKRS} for a more detailed discussion of this issue).  However it is apparent in Figure \ref{LZz} that a ten percent correction in the TKRS galaxy numbers is insufficient the bring the galaxy population anywhere near agreement with the metallicity distribution of the LGRB hosts.

Interestingly, the three high metallicity LGRBs (051022, 020819B, \& 050826) are consistent with the general star-forming galaxy population for galaxies of their luminosity and redshift.  If the metal aversion effect was evident in the high metallicity LGRB population we might expect them to be close to, but still biased below, comparable galaxies.  However, for the three cases available they seem to be neither outliers with respect to the comparison population or even among the lowest galaxies available within them.  It therefore seems possible that above some metallicity, any further bias towards lower metallicities is muted.

\section{Summary}

In this paper, we compare the metallicity distribution of the hosts of Long duration Gamma Ray Bursts (LGRB) with that of several other related populations: Type II supernovae (SNe) (which can serve as markers of star-formation), broad-lined Type Ic (Ic-bl) SNe (which as discussed earlier are closely related to LGRBs), and the general star-forming galaxy population.  We perform this comparison across these different populations not primarily by number of galaxies but rather by weighting by their star-formation rates (SFRs).  Our results show that, not only are the hosts of LGRBs at lower metallicites than either SNe hosts or general star-forming galaxies, but also that while Type Ic-bl and Type II SNe track star-formation (within our statistical ability to measure), more then three-quarters of our LGRBs are clustered in the lowest metallicity tenth of the star-formation.

To the end, we have complied all spectra for LGRB and Type Ic-bl SNe hosts with host emission spectroscopy sufficient to allow metallicity measurement.  To relate these events to the general star-forming galaxy population we have extracted the approximately 137 thousand galaxies from the Sloan Digital Sky Survey (SDSS) with line measurements for suitable metallicity measurement.  This has also allowed us to compile a sample of the more common Type II SNe hosts simply by selecting the galaxies within our existing volume limited SDSS sample which are known to have hosted such Type II SNe events with all the expected advantages of inter-sample consistency.  Additionally, in order to better match the redshift range of our LGRB population, we extend our general star-forming galaxy population out to a redshift range of approximately unity via the addition of the higher redshift Team Keck Redshift Survey (TKRS) galaxy population.  

Our analyzes are based on four physical properties redshift, metallicity, rest frame B band absolute magnitude, and (for most of our galaxies) the SFR.  To maximize inter-sample consistency we have calculated independent metallicities using a common metallicity diagnostic, scale, and code via the R$_{23}$ method.  The method uses the ratio of oxygen to hydrogen line strengths (and the ratio of two oxygen lines to characterize the degree of ionization) to estimate the oxygen abundance in HII regions as a proxy of the total galaxy metallicity.  R$_{23}$ however has a metallicity degeneracy issue where metal line emission cools the election temperature causing the oxygen line strength, that was originally increasing with metallicity, to subsequently decrease yielding undifferentiable oxygen line strengths for both a high and low metallicity value.  To resolve this degeneracy we have used observations of the [N II]/H$\alpha$ line ratio, itself a crude metallicity indicator, to select between the upper and lower metallicity branch.  For a few Type Ic-bl SNe we are forced to rely on this [N II]/H$\alpha$ diagnostic exclusively with a corresponding increase in error.  When comparing galaxies we use the B band galaxy luminosity rather than galaxy mass as reliable host mass estimates were not available for a substantial fraction of the objects in our samples.  We determine the SFRs from the galaxy's H$\alpha$ emission via the \cite{K98SFR} metric.  The primary difficulty with this is adjusting the H$\alpha$ line flux for slit and fiber losses.  

As shown in \cite{Modjaz2008} the low metallicity bias of LGRBs is visually apparent in a simple scatterplot of host metallicity versus absolute magnitude, when compared to that of the hosts of Type Ic-bl SNe and the general star-forming galaxy population, represented by the SDSS.  This result becomes even more impressive with the approximate three fold increase the the number of LGRBs and a 50 \% increase in number of Type Ic-bl SNe presented in Figure \ref{LZ}.  In this figure, we also add a second general star forming galaxy population, the TKRS, to better reflect the higher redshift distribution of the LGRBs.  This point is discussed in much greater detail in Figure \ref{LZz} and Section \ref{redshift}.  In our first presentation of the scatter plot, we present the metallicity measured at the location of the LGRBs and SNe within their hosts to show the actual metallicity of the environment responsible for their creation.  In later plots, however, we switch to plotting the host galaxy central metallicities in order to better compare with the general galaxy population.  To obtain a fair sample of the galaxy population, we use a subset of the SDSS star-forming population restricted to a volume limited sample (see Section \ref{zcut} for details).  Even this population, however, does not represent the way we expect SNe or LGRBs to choose their hosts.  Their probability of going off in a particular galaxy should be proportional to the rate of star-formation in that galaxy (all other things being equal).  To better emulate the expected occurrence of LGRBs and SNe, we then further select among these galaxies via random selection weighted by the underlaying star-formation of each galaxy.  These volume limited and star-formation weighted samples are plotted along with the similarly volume limited Type II SNe in Figure \ref{LZ_zcut}.  

As can be seen in Figure \ref{LZ_zcut}, the distribution of the star-formation weighted sample is significantly different from the distribution of galaxies by number with a large bias towards brighter galaxies having a greater fraction of star-formation.  As the LGRBs in our sample extend to a redshift of about one, far beyond the reach of the SDSS, let alone our smaller volume limited survey, we must consider the evolution of metallicity with redshift as a complicating factor.  Thus, later in the paper, in Figure \ref{LZz}, we expand our plotting by indexing object redshift with color to provide a visual representation of the evolution of metallicity over time.  From this plot it is clearly apparent that the effect of abundance evolution over the redshift range of our sample is nowhere near as dramatic as the metal aversion of the LGRB population as a whole.  While there are three LGRB hosts in the high metallicity range of our sample (LGRBs 051022, 020819B, \& 050826), is it clearly apparent in all three metallicity vs. luminosity scatterplots (figures \ref{LZ}, \ref{LZ_zcut}, \& \ref{LZz}) that while the SNe are distributed with the general star-forming galaxy population the LGRBs are on whole at the bottom of the metallicity distribution.  It is notable that the three high metallicity (out of 14 total) LGRBs do appear consistent with the general star forming galaxy population of comparable brightness and redshift.  This is intrinsically surprising as were the metal aversion to remain in effect for these objects we would expect their occurrence, if still in the high metallicity range, to be far lower then the typical metallicity for the population at that luminosity and redshift (i.e. either a outlier of said population or among the lowest galaxies available within it).

While the metallicity vs. luminosity scatterplots show quite conclusively that the hosts of LGRBs are at lower metallicites than either SNe hosts or general star-forming galaxies, they do not directly address whether this metallicity bias could be caused by the anti-correlation between SFR and metallicity of \cite{Mannucci} as claimed in \cite{MannucciLGRBs}.  To confront this issue, we plot in Figure \ref{dSFR_oplot} the integrated star-formation of the SDSS sample as a function of metallicity in comparison with the cumulative distributions of LGRBs and SNe.  A similar plot, with a much smaller sample LGRB sample and no comparison of SNE, was first shown by \cite{Stanek2006}.  Both the non LGRB Type Ic-bl SNe and Type II SNe track the distribution of star-forming SDSS galaxies quite well.  The LGRBs however display a profound preference for lower metallicities.  Thus the CCSNe track star-formation independent of its metallicity, while LGRBs do not.

However, this result could still be consistent with the \cite{Mannucci} relation, if the SFRs of the LGRB host were wildly discrepant from the other populations.  To exclude this possibility, in Figure \ref{all_SFR} (left plot) we directly compare the SFRs of the LGRB and SNe hosts to the general star-forming SDSS galaxy population.  This comparison is performed by taking the SFR of each LGRB and SNe host and determining the fraction of the total star-formation in the general SDSS galaxy population that occurs in galaxies with less star-formation than the host.  These fractional values are then sorted and plotted as a normalized cumulative histogram.  Should the distribution of star-formation for a given object type follow the general star-forming SDSS galaxy population then this histogram would track a diagonal line on the plot.  For the SNe, the SDSS comparison sample is volume limited.  However, for the LGRBs, which are intrinsically a magnitude limited sample, we use the entire SDSS, as a comparison magnitude limited sample.

Both the LGRBs and targeted Type II SNe population track the diagonal well, indicating a good correspondence between the SFRs of the two populations and the general SDSS galaxy population, and suggesting that SFR is correlated with LGRB and SNe formation.  However, as further refinement we can also generate fractional values for the LGRB and SNe hosts using only a subset of the SDSS galaxy population with similar metallicity to each host whose fractional value is being calculated.  This similar metallicity comparison, shown in Figure \ref{all_SFR} (right plot), yields a slightly closer agreement between the SFR distribution of the LGRBs and the SDSS galaxy population.  

Given the low metallicity nature of the LGRB population the absence of any profound differences in these distributions is itself interesting.  We explore this in Figure \ref{low_Z_sfr} by plotting the SFR distribution of the low metallicity star-formation versus the star-formation of the entire population.  One way to show this, would be to plot a low-metallicity subset of the randomly generated star-formation weighted galaxy population just as we have done for the LGRB and SNe hosts.  However, this then relies on our generating a very large number random population.  We can achieve the same effect, without relying on random selection, by using the whole subset of low-metallicity star-forming galaxies and comparing their fractional star-formation values in that subset against their values in the whole population, which is what we actually do in the figure.  From this we find that the SFR distribution of the low-metallicity SDSS galaxies is reasonably similar to that of the entire SDSS population and because of this similarity , the LGRB population is too small to determine whether it preferentially tracks the SFR distribution of the low-metallicity SDSS galaxies or the full SDSS galaxy sample.  Curiously the untargeted Type II SNe population fails to track the expected diagonal, even more so when the untargeted population is restricted to objects found trough untargeted SNe surveys (as apposed to incidental detections from other untargeted observations).  We believe this bias to be an issue with at least one of the untargeted surveys failing to detect SNe in high surface brightness backgrounds (see Appendix \ref{problem} and Figure \ref{plug_mag_plot}).

Thus the population distribution supports the scatter plots and rules out \cite{Mannucci} as an explanation.

\section{Conclusions}

In this paper, we have shown quite conclusively in various scatterplots that the hosts of LGRBs are at lower metallicites than either SNe hosts or general star-forming galaxies.  However, in order to directly address the question posed by \cite{MannucciLGRBs}, of whether SFR can explain metallicity, we plot as function of metallicity the integrated star-formation of the SDSS sample in comparison with the cumulative distributions of LGRBs and SNe.  There we find that three quarters of our LGRBs are found in the in the bottom low-metallicity tenth of the star-formation, with the remaining quarter (like our SNe populations) appearing to track star-formation independent of metallicity.  Assuming, all else being equal, that LGRBs (and SNe) occur proportionally to the allowable star-formation this bias indicates that LGRBs clearly prefer much lower metallicity host environments (as first suggested in \cite{Stanek2006} and shown with a much larger sample here).  We also consider the SFR distribution of the LGRB and SNe populations and find that they are consistent with the SFR distribution of the SDSS sample and especially consistent with this comparison is limited only to SDSS objects of similar metallicity.  Interestingly we find that the SFR distribution of the low-metallicity SDSS galaxies is reasonably similar to that of the entire SDSS population and cannot determine which the LGRB population best tracks.

In contrast with LGRBs, both our Type II and Type Ic-bl SNe populations appear to track the metallicity of the integrated SDSS star-formation.  This is what one would expect if star-formation alone is sufficient to explain the metallicity distribution of both SNe types.  Nonetheless, any metallicity bias, even one much more minor than that seen with LGRBs, has significant implications for a unified standard formation model for Type Ic SNe, Type Ic-bl SNe, and LGRBs.  \cite{Kelly, Sanders} have looked at the metallicity distribution of different SNe types vs. each other (without relation to the SDSS or LGRBs) and highlight a bias towards the Type Ic-bl SNe preferentially occurring in slightly lower metallicities than the other SNe populations (including the non Type Ic-bl SNe).  Due to our strict volume limits, imposed to allow comparison with the SDSS, our Type II SNe sample is far smaller than that used in these other works.  And we have made a decision to only look at two SNe types --- Type II SNe as hopefully a truly unbiased indicator of star-formation, and Type Ic-bl SNe, due to their close association with LGRBs.  Thus, while our work is well suited for its primary purpose of determining whether the metallicity distribution of LGRB hosts differs from other star-forming galaxies, it is much less powerful for distinguishing relative metallicities preference of SNe.  However, the metallicity differences between Type Ic-bl and other SNe, hinted at in our sample and perhaps seen more clearly in these other works, is dwarfed by the strong metallicity bias seen in LGRBs.  

The presence of a metallicity bias between LGRBs and Type Ic-bl SNe poses a problem for explanations of the LGRB metallicity bias being the incidental result of an IMF difference in their host galaxies.  In nearly all cases where one would have expected to detect a Type Ic underlying a LGRB to have been detected, one has been found, and where good spectroscopy is available, the Type Ic is broad-lined (c.f. \cite{Cano_thesis} for a good discussion of this point).  Thus, many and perhaps all Type Ic-bl SNe share a common progenitor type with LGRBs.  One would therefore expect the masses of Type Ic-bl SNe progenitors to be similar to those of LGRBs.  Thus while IMF differences between galaxies could play some role in determining where one finds LGRBs, the IMF is almost certainly far less important than galaxy metallicity.  

These observations do not agree with the suggestion of \cite{MannucciLGRBs} ``that the difference with the mass-metallicity relation is due to higher than average SFRs [of LGRB hosts] and that LGRBs with optical afterglows do not preferentially select low-metallicity hosts among the star-forming galaxies."  While the average SFR of LGRB hosts is indeed higher than that of typical SDSS galaxies, this is because LGRBs do not choose galaxies based on number but rather based on SFR (as well as metallicity).  The star-formation distribution of the LGRB hosts population tracks that of similar metallicity SDSS galaxies, Indeed, due to the fact that the star-formation distribution of galaxies in the SDSS is largely independent of metallicity, they track the star-formation distribution of the entire SDSS as best as can be determined with only the 11 LGRBs for which we have good SFRs.  The LGRB hosts population is explicitly concentrated in the low metallicity end of the available star-formation.  While the LGRB hosts themselves may remain consistent with the mass, metallicity, and SFR relation of \cite{Mannucci}, this relationship is not sufficient to explain the observed concentration of LGRBs in low metallicity star-formation.

Nonetheless, the preference of LGRBs for low-metallicity hosts is not absolute.  Three cases of LGRBs in roughly solar metallicity hosts (LGRBs 051022, 020819B, \& 050826) demonstrates that LGRBs can occur at high metallicity.  However, such events are quite rare.  These high-metallicity objects are all at substantial redshifts (99\% of the galaxies in the SDSS sample lie closer than the closest of these LGRBs), and thus the search volume needed to find them was large.  As a result, our sample overemphasizes high-metallicity LGRBs compared to the distribution likely to be found in a volume limited survey.  The presence of these objects does not substantially detract from our main conclusion, that on the whole, LGRBs prefer low metallicity.  However, the absence among the three high metallicity LGRBs for any apparent preference for a lower metallicity environment, compared with the general star forming galaxy population of comparable brightness \& redshift, may be a hint that the differential preference of LGRBs for lower metallicity environments is reduced at high metallicities.  We will discuss the implications of this in an upcoming paper on high metallicity LGRBs (\citealt{pregraham} in prep).

It is generally accepted that the emission from LGRBs is relativistically beamed from collimated outflows and that rotation is required to produce the collimation \citep{1997ApJ...482L..29M, 1999ApJ...524L..43S, 2002A&A...385..377Q, 2002A&A...385L..19M, 2006ApJ...647.1255N, 2007ApJ...659.1420T, 2007MNRAS.382.1029K, 2008Natur.452..966M, 2009MNRAS.394.1182K, 2009MNRAS.396.2038B, 2010A&A...516A..16L, 2011ApJ...731...80N, 2012ApJ...757...56Y}.  As increased metallicity produces mass loss through stellar winds (c.f. \citealt{Nugis, Crowther, Vink}), and these winds carry off angular momentum, it has been proposed \citep{Yoon, Langer} that the observed preference for low metallicity LGRB environments is a result of the need to ensure rapid progenitor rotation.  However the predicted metallicity range of 1/10$^{th}$ or 1/100$^{th}$ solar required to limit mass loss does not match where observations find a large drop in LGRB occurrence, at about a half to one-third solar metallicity.  

Any complete theory of LGRB formation must also account for the occasional LGRB at high metallicity.  \cite{Podsi} attempts to span this dichotomy through a binary common envelope ejection model in which low-metallicity helps drive mass transfer, but nuclear burning during a merger phase provides the energy to eject the stellar atmosphere.  However, this model also requires a much greater difference in metallicity (than present observationally) to yield an order of magnitude difference in the LGRB formation rate.  This model's requirement of hydrogen and helium layer ejection does explain the specific association of LGRBs with only Type Ic SNe but not the additional constraint that the SNe have the ultra fast ejecta of broad-line SNe events.  Again, we defer a more detailed discussion of high metallicity LGRBs and their implications upon theory to our upcoming paper which specifically looks at high-metallicity LGRB hosts (\citealt{pregraham} in prep).

We have shown that LGRBs have a strong intrinsic preference for low metallicity environments.  Although nearly all LGRBs are thought to be associated with Type Ic-bl SNe, we have also shown that any preference for low-metallicities in the general Type Ic-bl population is minor in comparison.  While LGRBs are clearly associated with star-formation, they are highly biased by metallicity.  Any use of LGRBs as tracers of star-formation must take this into account, though the fact that the production of LGRBs appears to turns on in force below one-half solar may give hope to those who wish to use LGRBs as tracers at high-redshift, where metallicities this low are the norm rather than the exception.

\acknowledgments

A number of people deserve great thanks for the help they have provided us in doing this work.  Lisa Kewley provided the \cite{KobulnickyKewley} scale metallicity code.  Jarle Brinchmann provided \cite{T04} scale metallicities for LGRB and SNe objects to match the SDSS object values in the MPA-JHU data products (though results in this metallicity diagnostic are not presented in this paper it provided a useful crosscheck throughout our analysis).  Greg Aldering \& Saul Perlmutter provided unreleased proprietary SNFactory data for 5 Type II SNe.  Kuntal Misra provided slit loss estimates for the LGRB population based on HST imaging data.  Kuang-Han Huang provided multi-band SExtractor photometry on Type Ic-bl SNe from SDSS data.  In addition to these individuals, Rebekah Hounsell, Norbert Langer, Mario Livio, Selma de Mink, Maryam Modjaz, Colin Norman, and Rosie Wyse also provided many useful discussions.  We thank them all.

\appendix

\section{A. \thinspace\thinspace Metallicity Gradient Fitting and Application} \label{metal_grad_fit} \label{mshifting}

LGRBs generally occur in the brightest regions of their host galaxies \citep{Fruchter} and have small enough hosts that the majority of the total host light is observed via slit spectroscopy.  SNe however are often found in larger galaxies where there is a measurable metallicity difference between the location within the galaxy where the event occurred and the galaxies center.  Since in most cases we lack observations of both metallicities, we follow the general methodology of \cite{Modjaz2008} and employ an expected metallicity gradient to estimate the metallicity at SNe location when it was not specifically measured (see Figure \ref{LZ}).  In this paper, to better compare the physical properties of the host galaxies, we also reverse this methodology to estimate the central galaxy metallicity based on the SNe location metallicity and gradient when an independent R$_{23}$ core galaxy metallicity measurement is not available (see Figure \ref{LZz}).  Here, we describe how we derive the galaxy metallicity gradients used for these estimates.

\begin{figure}[h]
\begin{center}
\includegraphics[width=.5\textwidth]{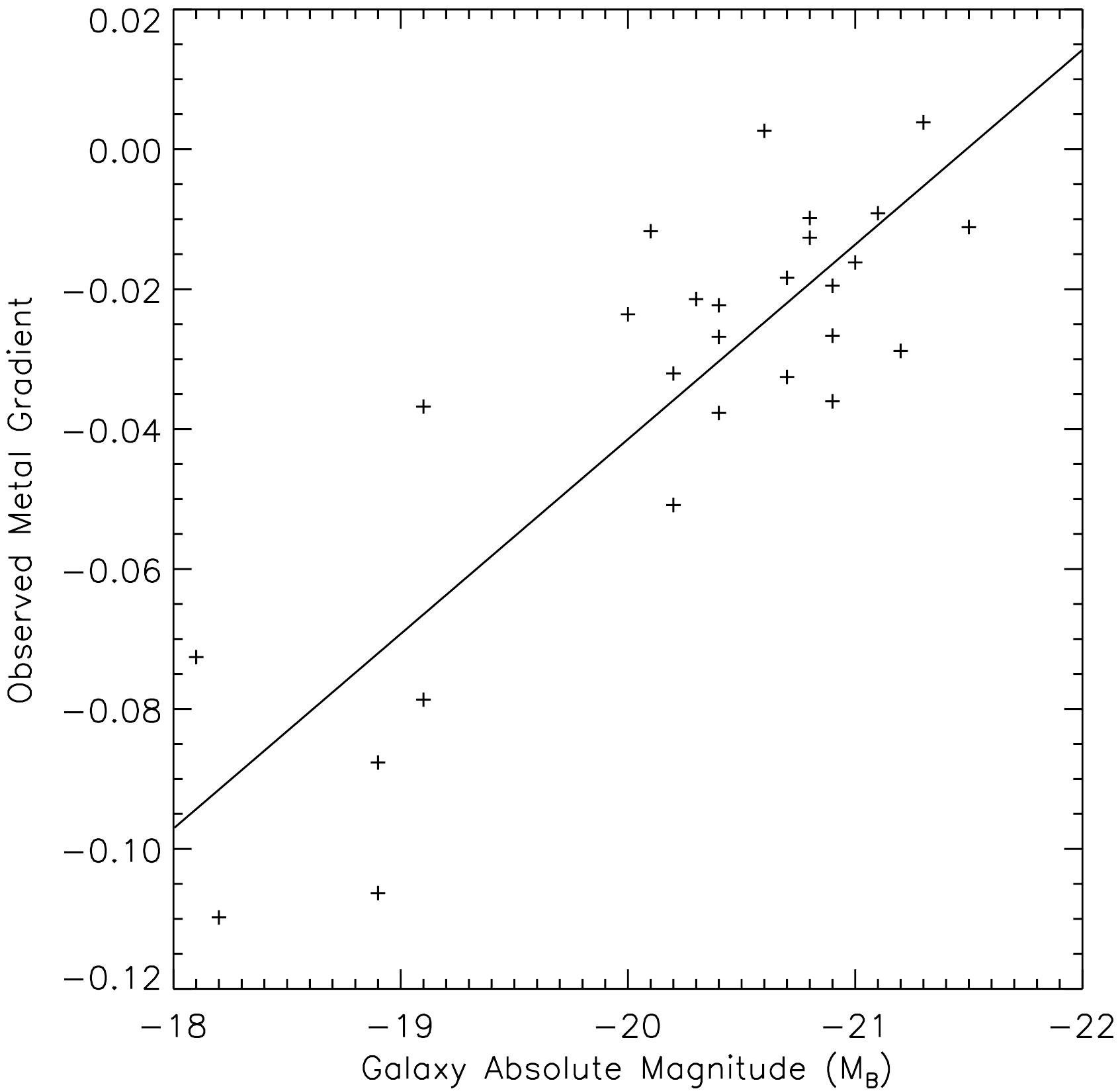}
\caption{\label{garnett_fit} Metallicity gradient values vs.\thinspace \thinspace B band absolute galaxy luminosity.  Gradient values calculated from the given metallicity values in \cite{Garnett} shifted to the KK04 \citep{KobulnickyKewley} scale via the appropriate \cite{KewleyEllison} transform and then estimated using the additional properties given in and methodology of \cite{Garnett} in order to update the given metallicity gradient values to the modern KK04 scale used here.  Fit is performed with the IDL linfit function and yields $\nabla$metallicity (dex/kpc) $= -0.028 \times M_{B} - 0.60$.}

\end{center}
\end{figure}

\cite{Garnett} Table 7 provides a compilation of metallicity gradients from a sample of thirty-two nearby spirals (see their Table 7).  However, the provided metallicities are calculated via the Z94 \citep{Z94} methodology.  While metallicity values can be converted from the Z94 scale to the KK04 scale used here via the transform provided in \cite{KewleyEllison}, this transform is for metallicity values not gradients.  Fortunately \cite{Garnett} also provides the central metallicites and the separation over which the gradient was observed.  Thus we back compute the Z94 metallicites at the given distance, convert both these and the central metallicity values to the KK04 scale (via the \citealt{KewleyEllison} transform) and then recalculate the metallicity gradients in the KK04 scale.  These gradient values are plotted as a function of M$_{B}$ values in Figure \ref{garnett_fit}.

We then fit the metallicity gradient values ($\nabla Z$) with the provided M$_{B}$ values generating the function given below:

\begin{equation}
\label{gradZfunc}
\nabla Z = -0.028 \times M_{B} - 0.60 \pm 0.016 \, \rm{ (dex/kpc)}
\end{equation}

We note that beyond M$_{B}$ values brighter then approximately -21.5, this function extrapolates to an inverted gradient with galaxies increasing in metallicity with radius.  To avoid this, we adopt a zero gradient for galaxies brighter than -21.5 (in practice this affects SN 1998ey only).  We use this equation to estimate central metallicites from site metallicities and vice versa in the obvious manner.  We estimate the metallicity gradient from the host M$_{B}$ magnitudes, multiply by the separation distance to determine the metallicity shift and use this differential to convert between progenitor site and galaxy core metallicities as needed.  R$_{23}$ metallicites are used where available and in a cases where only an [N II]/H$\alpha$ central metallicity value is available for the host and an R$_{23}$ values for the SNe site the shifted R$_{23}$ value is used for the central metallicity due to the respectively higher accuracy of the R$_{23}$ diagnostic.  

We note that this approach does not compensate for metallicity gradient differences as a function of metallicity and attempted to correct for this by fitting the metallicity gradient values of \cite{Garnett} as a function of both M$_{B}$ and central metallicity (and using iteration when solving for the central metallicity).  However this proved problematic.  We suspect the primary causes may be the small number of data points in the Garnett sample, the general difficulty of accurately transforming between metallicity indicators, and potentially problems with the \cite{Z94} itself, most particularly its failure to include an estimate of ionization.  A more accurate approach and an interesting exercise in itself would be to repeat the gradient analysis of \cite{Garnett} with the more modern metallicity diagnostics directly and then attempt a fit in M$_{B}$ and metallicity however this exceeds the scope of this paper.

Due to our inability to adjust our metallicity gradient calculations as a function of galaxy metallicity, we suspect that in cases with high SNe site metallicity (i.e.  SN 1997ef) we are likely over estimating the central metallicity.  It is apparent in the scatterplots (figures \ref{LZ}, \ref{LZ_zcut} and \ref{LZz}) that while the luminosity metallicity relation is linear over most if its range for the brightest objects the SDSS population appears to flatten in metallicity at about $\sim$2 solar.  This is to be expected as the metallicity will eventually reach a maximum yield value.  in terms of the metallicity gradient within the galaxies, as the central galaxy metallicites truncate at the yield while the enrichment continues in the remainder of the galaxy the metallicity gradient will progressively flatten from the center of the galaxies outward.  This effect is progressively more noticeable for larger galaxies as the enrichment reaches saturation at progressively increasing radii.  We also note that the [N II]/H$\alpha$ diagnostic seems to allow metallicities beyond the maximum of those commonly returned by R$_{23}$ of about log(O/H)+12 $\sim $9.1, although the values below this remain comparable.  This may be indicative of different yield values for Nitrogen vs. Oxygen or just an artifact of the different metallicity diagnostics.

While an investigation of these effects is beyond the scope of this paper we do attempt a crude correction for these effects as follows.  We limit the the measured [N II]/H$\alpha$ metallicities to a maximum log(O/H)+12 value of 9.1, thus forcing it into agreement with the R$_{23}$ range; however when downshifting the core metallicities to estimate a site value we begin with the actual [N II]/H$\alpha$ value (in practice this is relevant to SN 1997dq only).  We also place a variable limit on the maximum estimated upshifted central metallicities based on upon galaxy absolute magnitude.  To do this we we determine the 90th percentile metallicity of the SDSS galaxies as a function of galaxy absolute magnitude and use this as a celling for the gradient upshifted values.  Additionally in the case of SN 2006nx the [N II]/H$\alpha$ ratio (used to break the degeneracy between the R$_{23}$ upper and lower branches) is of sufficiently high error to not firmly exclude the lower branch solution thus the error bar on this object is extended to include this possibility.

\begin{figure}[ht]
\begin{center}
\includegraphics[width=.485\textwidth]{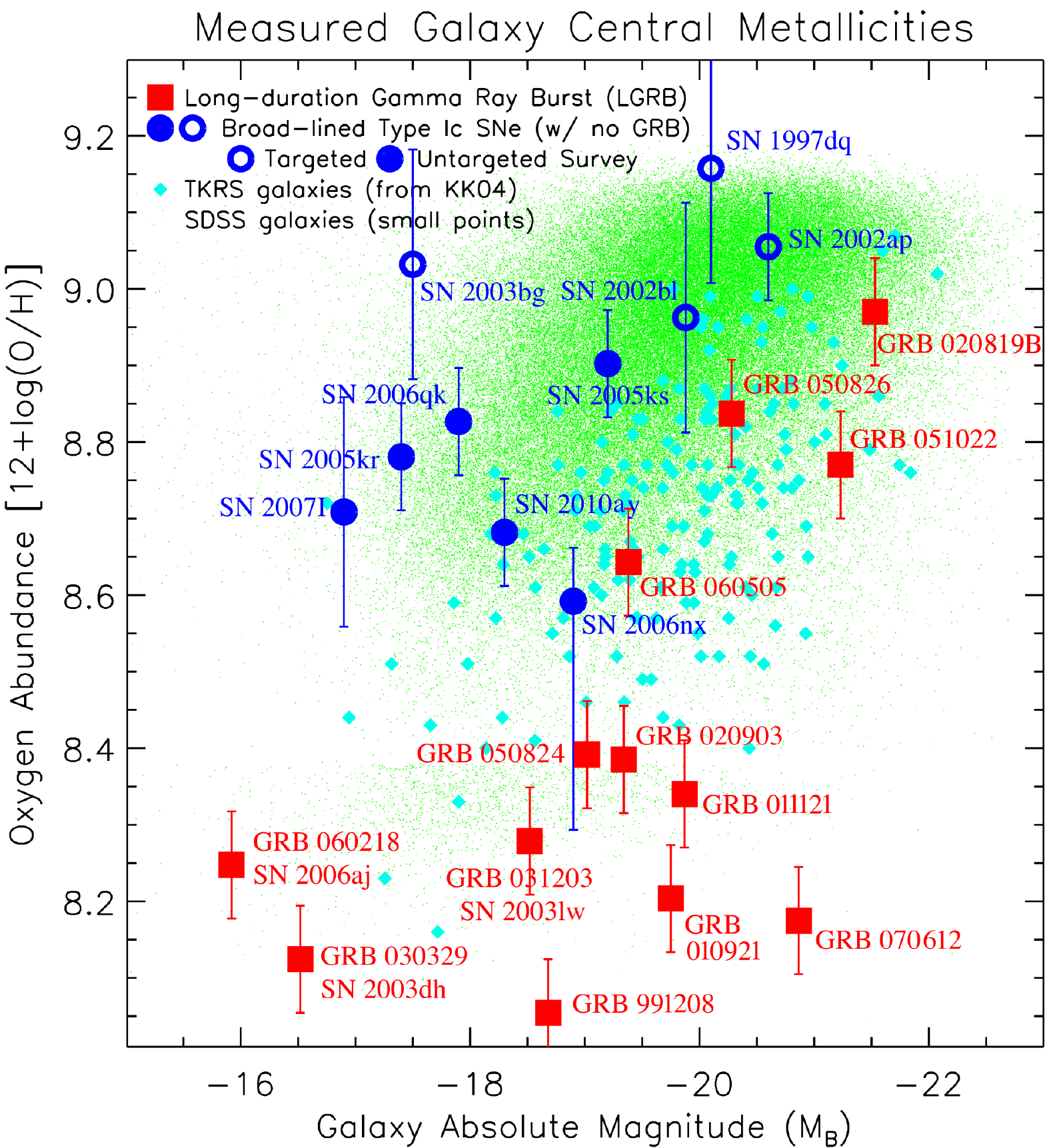}
\includegraphics[width=.49\textwidth]{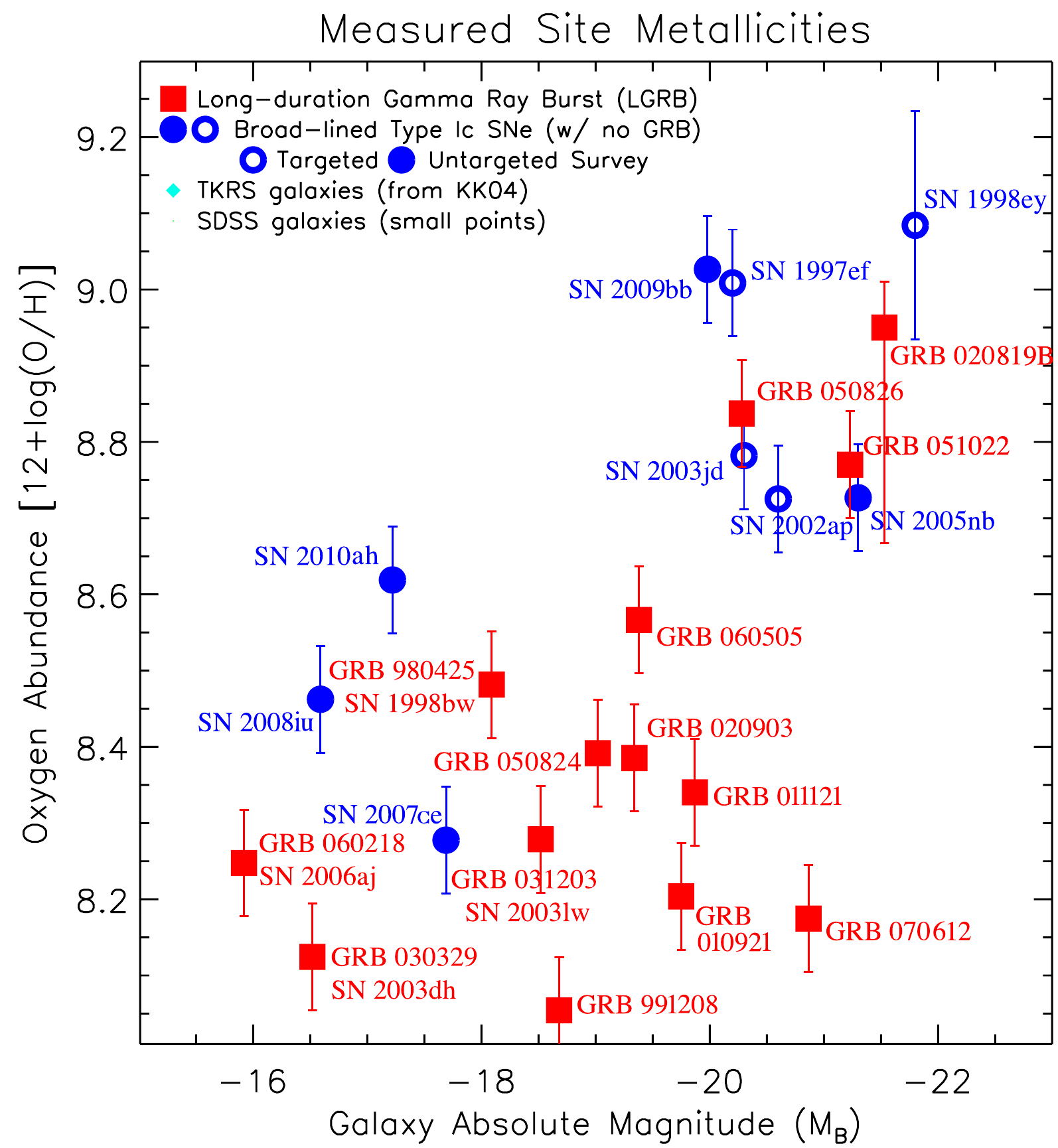}
\caption{\label{measured_plots} Measured central galaxy (left) and event site (right) metallicity vs.\thinspace \thinspace B band absolute galaxy luminosity of LGRB (red squares) and Type Ic-bl SNe hosts (blue circles) without associated LGRB events (filled circles represent SNe selected in an untaragted manner whereas open circlers were from galaxy targeted SNe surveys and thus may be biased in galaxy properties by target selection) as well as star-forming galaxies from the SDSS (green small dots) and TKRS (teal diamonds) to provide a low and high redshift comparison sample respectively.  Due to the absence of line flux error data on many objects a consistent error of $\pm$0.07 dex is assumed for R$_{23}$ metallicities and $\pm$0.15 for [N II]/H$\alpha$ metallicities with all metallicities consistently recalculated in the \cite{KobulnickyKewley} scale as described in Section \ref{metal_measurement}.  Subsequently these plots are combined such that when both measured site and central galaxy values are not available the missing value is estimated from the other using the metallicity gradient methodology outlined in Appendix \ref{mshifting} to generate a complete set of site and central galaxy values as plotted in figures \ref{LZ} \& \ref{LZz} respectively.  LGRBs are assumed to occur in the brightest regions of their host galaxies \citep{Fruchter} and used as both central and site values unless known otherwise (LGRBs 980425, 020819B, 060505).}

\end{center}
\end{figure}

\section{B. \thinspace\thinspace Apparent Biases of Non-targeted SNe Populations} \label{problem}

SNe found in surveys targeted at a population of galaxies are generally expected to be biased by the criteria used to select that galaxy population.  In general, target surveys preferentially search larger \& brighter galaxies.  However, one might expect that SNe surveys that are not targeted but instead observe an area of sky without emphasis on particular galaxies would produce truly unbiased SNe populations.  Here we find evidence that this is not the case.  

\begin{figure}[ht]
\begin{center}

\includegraphics[width=.49\textwidth]{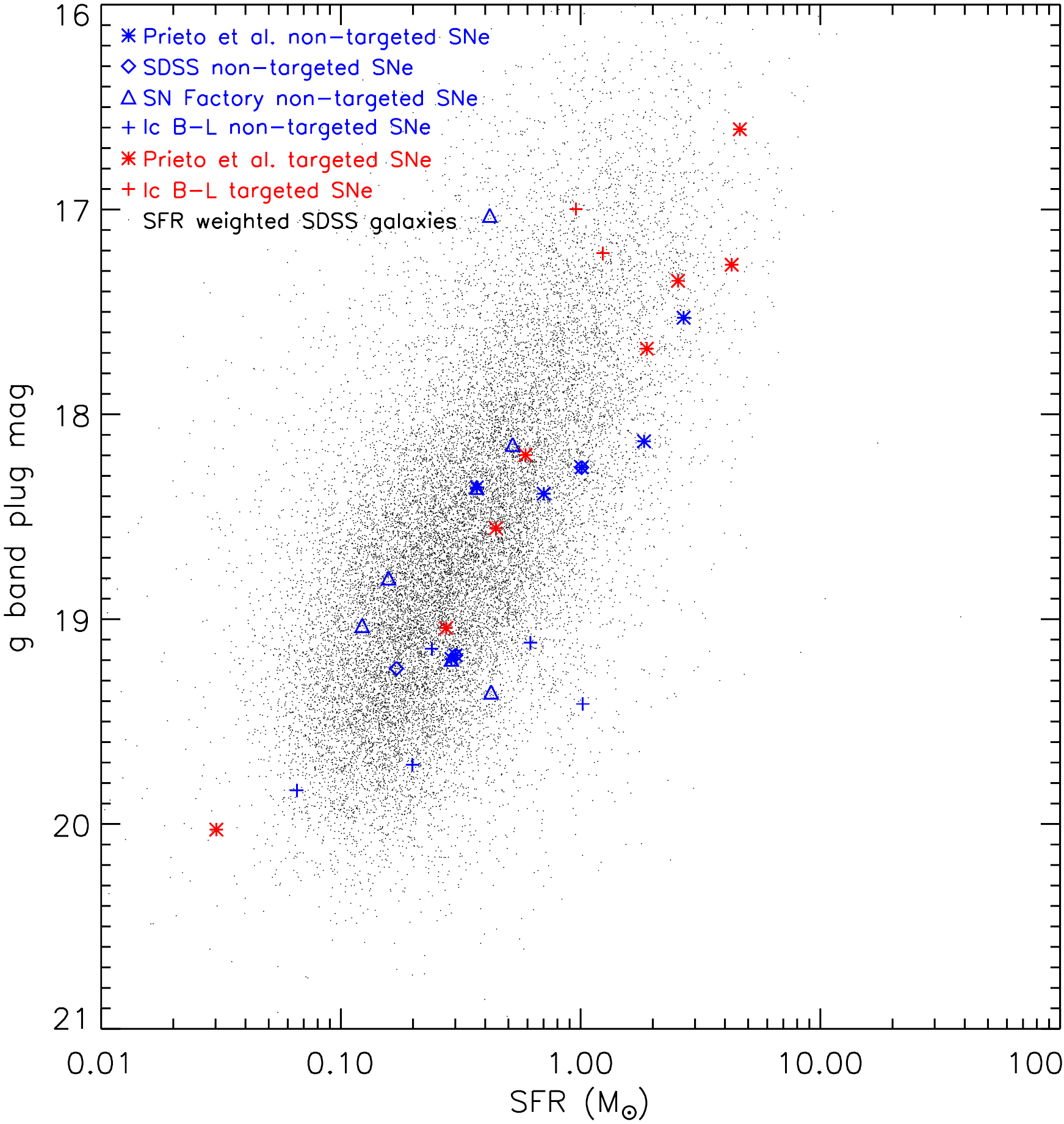}
\includegraphics[width=.49\textwidth]{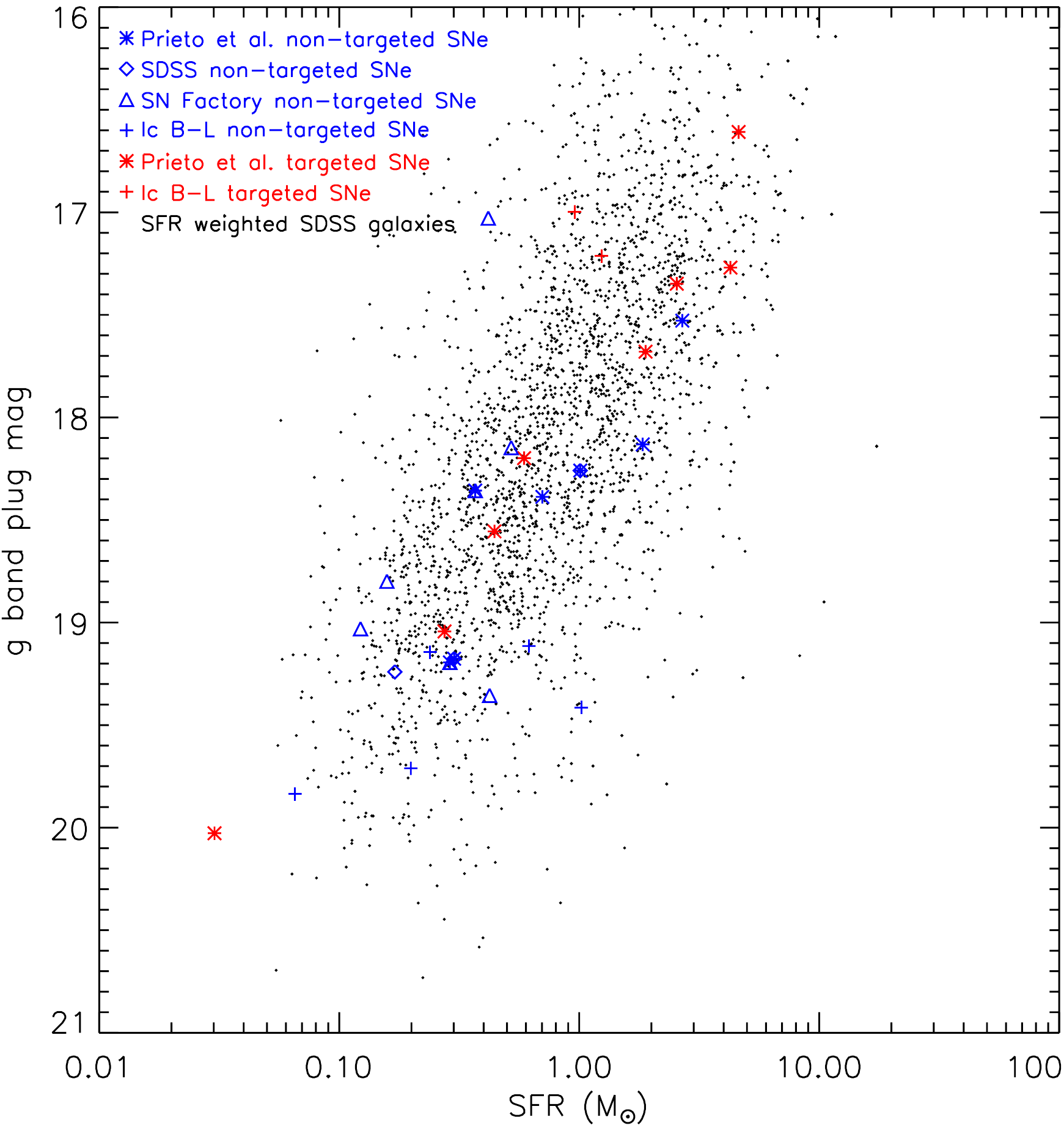}

\caption{\label{plug_mag_plot} SDSS g band fiber plug apparent magnitude vs.\thinspace \thinspace SFR for SNe.  Since the SDSS survey uses a consistent fiber size the apparent fiber plug magnitude functions as a proxy measure for galaxy surface brightness.  In the left plot the entire SDSS general star-forming galaxy sample meeting the criteria described within are plotted in the background.  In the right plot the background points are replaced with a one thousand object synthetic star-formation weighted random sample as described in Section \ref{synthetic}.  (The apparent duplication of two points in both the SDSS and Supernovae Factory with the \citealt{Prieto} objects is exactly that - the same SNe was detected in both surveys).  While the targeted sample (described in Section \ref{II_targ}) shows good tracking of the synthetic population the Supernovae Factory population (one of the two component samples in the non-targeted Type II SNe population - see Section \ref{II_non_targ}) only shows this tracking in dimmer galaxies with only one of seven SNe host galaxies with a g band plug mag brighter then 18.  A closer examination of this object shows a large galaxy where the SNe occurred on the outskirts in a region with much less surface brightness then the core where the SDSs fiber was located.  (The SDSS SNe search population may also show this effect but with only three data points we cannot be certain).}

\end{center}
\end{figure}

In Figure \ref{plug_mag_plot} we show that the untargeted surveys used by us appear to have a bias against finding SNe in high surface brightness hosts.  We can obtain a crude estimate of the central surface brightness of the host galaxies by using the plug magnitudes from the SDSS survey.  Given the SDSS fiber plugs are 3" in diameter, this corresponds to a physical scale of $\sim$1.2 to 2.3 kpc across our volume limited sample.  Thus for large galaxies which subtend the SDSS fiber, the fiber plug magnitudes can serve as a crude proxy for surface brightness.  None of the non-targeted Type Ic-bl and only 2 out of 13 non-targeted Type II SNe have a g band plug magnitude brighter than 18.  The latter ratio further reduces to 1 out of 10 non-targeted Type II SNe when only considering non-targeted SNe found in dedicated non-targeted SNe searches.  Compared with the 48\% of star-formation found in the volume-limited SDSS sample in galaxies with g band plug magnitudes brighter than 18 this suggests a strong, and we believe non physical, bias against SNe occurring in high surface brightness galaxies.  This claim is only further strengthened by a detailed examination of the two exceptions.  The higher surface brightness case (SN 2007es) is actually located on the outskirts of a large galaxy in a region with much less surface brightness than the core where the SDSS fiber observed.  The other exception (SN 2001fb) was found not in a dedicated non-targeted SNe survey but by incidental observation presumably involving an alert human rather than search software that often avoids the centers of galaxies \citep{IAUC7740}.

This surface brightness aversion could perhaps explain features seen in both our plots of cumulative population vs metallicity (Figure \ref{dSFR_oplot}) and SFR distribution (Figure \ref{all_SFR}).  In the latter figure, the untargeted Type II SNe deviate strongly from the expected SFR distribution.  (The small but predominately non-targeted set of Type Ic-bl SNe for which we have good SFRs also shows a similar but more muted bias).  The non-targeted Type II discrepancy becomes quite alarming when we consider only events detected via dedicated non-targeted SNe searches, namely the SN factory \& SDSS surveys.  (Some non-targeted SNe are the result of incidental discoveries where the original target was a region of sky or a galaxy unrelated to the SNe host).  That this bias in the non-targeted population is not even diminished when the analysis of Figure \ref{all_SFR} is limited to similar metallicities suggests that, at least for the Type II SNe, which is consistent with a surface brightness bias.  This bias could also be involved in the tendency of the non-targeted samples to lie at slightly lower metallicities than the SDSS population.

A bias against detecting non-targeted SNe in bight galaxies would also explain another curiosity; only a singe SNe in each of the non-targeted Type II and non-targeted Type Ic-bl SNe populations is in a host more luminous than -20 M$_B$ (whereas about a third of the star-formation in our volume limited SDSS sample is in galaxies more luminous then -20 M$_B$).  Furthermore introducing a maximum luminosity cut at an M$_B$ of -20 forces the non-targeted Type Ic-bl SNe and the general star-forming galaxy populations into good agreement.  While our minimum luminosity cut at -18 M$_B$ is justifiable given the completeness limit of the SDSS survey (at our volume limited redshift range), such a maximum luminosity would however be at best a crude compensation for our lack of plug magnitudes (or central surface brightness estimates) for the majority of our Type Ic sample.

\vspace{4pt}

\bibliographystyle{apj}
\bibliography{\jobname}

\begin{thebibliography}{83}
\expandafter\ifx\csname natexlab\endcsname\relax\def\natexlab#1{#1}\fi

\bibitem[{{Abazajian} {et~al.}(2009){Abazajian}, {Adelman-McCarthy},
  {Ag{\"u}eros}, {Allam}, {Allende Prieto}, {An}, {Anderson}, {Anderson},
  {Annis}, {Bahcall}, \& et~al.}]{SDSSdr7}
{Abazajian}, K.~N., {Adelman-McCarthy}, J.~K., {Ag{\"u}eros}, M.~A., {Allam},
  S.~S., {Allende Prieto}, C., {An}, D., {Anderson}, K.~S.~J., {Anderson},
  S.~F., {Annis}, J., {Bahcall}, N.~A., \& et~al. 2009, \apjs, 182, 543

\bibitem[{{Aldering} {et~al.}(2002){Aldering}, {Adam}, {Antilogus}, {Astier},
  {Bacon}, {Bongard}, {Bonnaud}, {Copin}, {Hardin}, {Henault}, {Howell},
  {Lemonnier}, {Levy}, {Loken}, {Nugent}, {Pain}, {Pecontal}, {Pecontal},
  {Perlmutter}, {Quimby}, {Schahmaneche}, {Smadja}, \&
  {Wood-Vasey}}]{SNfactory}
{Aldering}, G., {Adam}, G., {Antilogus}, P., {Astier}, P., {Bacon}, R.,
  {Bongard}, S., {Bonnaud}, C., {Copin}, Y., {Hardin}, D., {Henault}, F.,
  {Howell}, D.~A., {Lemonnier}, J.-P., {Levy}, J.-M., {Loken}, S.~C., {Nugent},
  P.~E., {Pain}, R., {Pecontal}, A., {Pecontal}, E., {Perlmutter}, S.,
  {Quimby}, R.~M., {Schahmaneche}, K., {Smadja}, G., \& {Wood-Vasey}, W.~M.
  2002, in Society of Photo-Optical Instrumentation Engineers (SPIE) Conference
  Series, Vol. 4836, Society of Photo-Optical Instrumentation Engineers (SPIE)
  Conference Series, ed. J.~A. {Tyson} \& S.~{Wolff}, 61--72

\bibitem[{{Allende Prieto} {et~al.}(2001){Allende Prieto}, {Lambert}, \&
  {Asplund}}]{solar}
{Allende Prieto}, C., {Lambert}, D.~L., \& {Asplund}, M. 2001, \apjl, 556, L63

\bibitem[{{Bastian} {et~al.}(2010){Bastian}, {Covey}, \& {Meyer}}]{Bastian}
{Bastian}, N., {Covey}, K.~R., \& {Meyer}, M.~R. 2010, \araa, 48, 339

\bibitem[{{Berger} {et~al.}(2007){Berger}, {Fox}, {Kulkarni}, {Frail}, \&
  {Djorgovski}}]{Berger2006}
{Berger}, E., {Fox}, D.~B., {Kulkarni}, S.~R., {Frail}, D.~A., \& {Djorgovski},
  S.~G. 2007, \apj, 660, 504

\bibitem[{{Blanton} {et~al.}(2005){Blanton}, {Lupton}, {Schlegel}, {Strauss},
  {Brinkmann}, {Fukugita}, \& {Loveday}}]{Blanton}
{Blanton}, M.~R., {Lupton}, R.~H., {Schlegel}, D.~J., {Strauss}, M.~A.,
  {Brinkmann}, J., {Fukugita}, M., \& {Loveday}, J. 2005, \apj, 631, 208

\bibitem[{{Blanton} \& {Roweis}(2007)}]{kcorrect}
{Blanton}, M.~R. \& {Roweis}, S. 2007, \aj, 133, 734

\bibitem[{{Bloom} {et~al.}(1998){Bloom}, {Djorgovski}, {Kulkarni}, \&
  {Frail}}]{Bloom}
{Bloom}, J.~S., {Djorgovski}, S.~G., {Kulkarni}, S.~R., \& {Frail}, D.~A. 1998,
  \apjl, 507, L25

\bibitem[{{Brinchmann} {et~al.}(2004){Brinchmann}, {Charlot}, {White},
  {Tremonti}, {Kauffmann}, {Heckman}, \& {Brinkmann}}]{SDSS-mpg}
{Brinchmann}, J., {Charlot}, S., {White}, S.~D.~M., {Tremonti}, C.,
  {Kauffmann}, G., {Heckman}, T., \& {Brinkmann}, J. 2004, \mnras, 351, 1151

\bibitem[{{Bucciantini} {et~al.}(2009){Bucciantini}, {Quataert}, {Metzger},
  {Thompson}, {Arons}, \& {Del Zanna}}]{2009MNRAS.396.2038B}
{Bucciantini}, N., {Quataert}, E., {Metzger}, B.~D., {Thompson}, T.~A.,
  {Arons}, J., \& {Del Zanna}, L. 2009, \mnras, 396, 2038

\bibitem[{{Cano}(2012)}]{Cano_thesis}
{Cano}, Z. 2012, ArXiv e-prints

\bibitem[{{Christensen} {et~al.}(2004){Christensen}, {Hjorth}, \&
  {Gorosabel}}]{Christensen}
{Christensen}, L., {Hjorth}, J., \& {Gorosabel}, J. 2004, \aap, 425, 913

\bibitem[{{Christensen} {et~al.}(2008){Christensen}, {Vreeswijk}, {Sollerman},
  {Th{\"o}ne}, {Le Floc'h}, \& {Wiersema}}]{Christensen980425}
{Christensen}, L., {Vreeswijk}, P.~M., {Sollerman}, J., {Th{\"o}ne}, C.~C., {Le
  Floc'h}, E., \& {Wiersema}, K. 2008, \aap, 490, 45

\bibitem[{{Crowther} {et~al.}(2002){Crowther}, {Dessart}, {Hillier}, {Abbott},
  \& {Fullerton}}]{Crowther}
{Crowther}, P.~A., {Dessart}, L., {Hillier}, D.~J., {Abbott}, J.~B., \&
  {Fullerton}, A.~W. 2002, \aap, 392, 653

\bibitem[{{Ferguson} {et~al.}(1998){Ferguson}, {Gallagher}, \& {Wyse}}]{M74}
{Ferguson}, A.~M.~N., {Gallagher}, J.~S., \& {Wyse}, R.~F.~G. 1998, \aj, 116,
  673

\bibitem[{{Frieman} {et~al.}(2008){Frieman}, {Bassett}, {Becker}, {Choi},
  {Cinabro}, {DeJongh}, {Depoy}, {Dilday}, {Doi}, {Garnavich}, {Hogan},
  {Holtzman}, {Im}, {Jha}, {Kessler}, {Konishi}, {Lampeitl}, {Marriner},
  {Marshall}, {McGinnis}, {Miknaitis}, {Nichol}, {Prieto}, {Riess}, {Richmond},
  {Romani}, {Sako}, {Schneider}, {Smith}, {Takanashi}, {Tokita}, {van der
  Heyden}, {Yasuda}, {Zheng}, {Adelman-McCarthy}, {Annis}, {Assef},
  {Barentine}, {Bender}, {Blandford}, {Boroski}, {Bremer}, {Brewington},
  {Collins}, {Crotts}, {Dembicky}, {Eastman}, {Edge}, {Edmondson}, {Elson},
  {Eyler}, {Filippenko}, {Foley}, {Frank}, {Goobar}, {Gueth}, {Gunn},
  {Harvanek}, {Hopp}, {Ihara}, {Ivezi{\'c}}, {Kahn}, {Kaplan}, {Kent},
  {Ketzeback}, {Kleinman}, {Kollatschny}, {Kron}, {Krzesi{\'n}ski}, {Lamenti},
  {Leloudas}, {Lin}, {Long}, {Lucey}, {Lupton}, {Malanushenko}, {Malanushenko},
  {McMillan}, {Mendez}, {Morgan}, {Morokuma}, {Nitta}, {Ostman}, {Pan},
  {Rockosi}, {Romer}, {Ruiz-Lapuente}, {Saurage}, {Schlesinger}, {Snedden},
  {Sollerman}, {Stoughton}, {Stritzinger}, {Subba Rao}, {Tucker}, {Vaisanen},
  {Watson}, {Watters}, {Wheeler}, {Yanny}, \& {York}}]{SDSS-SNe1}
{Frieman}, J.~A., {Bassett}, B., {Becker}, A., {Choi}, C., {Cinabro}, D.,
  {DeJongh}, F., {Depoy}, D.~L., {Dilday}, B., {Doi}, M., {Garnavich}, P.~M.,
  {Hogan}, C.~J., {Holtzman}, J., {Im}, M., {Jha}, S., {Kessler}, R.,
  {Konishi}, K., {Lampeitl}, H., {Marriner}, J., {Marshall}, J.~L., {McGinnis},
  D., {Miknaitis}, G., {Nichol}, R.~C., {Prieto}, J.~L., {Riess}, A.~G.,
  {Richmond}, M.~W., {Romani}, R., {Sako}, M., {Schneider}, D.~P., {Smith}, M.,
  {Takanashi}, N., {Tokita}, K., {van der Heyden}, K., {Yasuda}, N., {Zheng},
  C., {Adelman-McCarthy}, J., {Annis}, J., {Assef}, R.~J., {Barentine}, J.,
  {Bender}, R., {Blandford}, R.~D., {Boroski}, W.~N., {Bremer}, M.,
  {Brewington}, H., {Collins}, C.~A., {Crotts}, A., {Dembicky}, J., {Eastman},
  J., {Edge}, A., {Edmondson}, E., {Elson}, E., {Eyler}, M.~E., {Filippenko},
  A.~V., {Foley}, R.~J., {Frank}, S., {Goobar}, A., {Gueth}, T., {Gunn}, J.~E.,
  {Harvanek}, M., {Hopp}, U., {Ihara}, Y., {Ivezi{\'c}}, {\v Z}., {Kahn}, S.,
  {Kaplan}, J., {Kent}, S., {Ketzeback}, W., {Kleinman}, S.~J., {Kollatschny},
  W., {Kron}, R.~G., {Krzesi{\'n}ski}, J., {Lamenti}, D., {Leloudas}, G.,
  {Lin}, H., {Long}, D.~C., {Lucey}, J., {Lupton}, R.~H., {Malanushenko}, E.,
  {Malanushenko}, V., {McMillan}, R.~J., {Mendez}, J., {Morgan}, C.~W.,
  {Morokuma}, T., {Nitta}, A., {Ostman}, L., {Pan}, K., {Rockosi}, C.~M.,
  {Romer}, A.~K., {Ruiz-Lapuente}, P., {Saurage}, G., {Schlesinger}, K.,
  {Snedden}, S.~A., {Sollerman}, J., {Stoughton}, C., {Stritzinger}, M., {Subba
  Rao}, M., {Tucker}, D., {Vaisanen}, P., {Watson}, L.~C., {Watters}, S.,
  {Wheeler}, J.~C., {Yanny}, B., \& {York}, D. 2008, \aj, 135, 338

\bibitem[{{Fruchter} {et~al.}(2006){Fruchter}, {Levan}, {Strolger},
  {Vreeswijk}, {Thorsett}, {Bersier}, {Burud}, {Castro Cer{\'o}n},
  {Castro-Tirado}, {Conselice}, {Dahlen}, {Ferguson}, {Fynbo}, {Garnavich},
  {Gibbons}, {Gorosabel}, {Gull}, {Hjorth}, {Holland}, {Kouveliotou}, {Levay},
  {Livio}, {Metzger}, {Nugent}, {Petro}, {Pian}, {Rhoads}, {Riess}, {Sahu},
  {Smette}, {Tanvir}, {Wijers}, \& {Woosley}}]{Fruchter}
{Fruchter}, A.~S., {Levan}, A.~J., {Strolger}, L., {Vreeswijk}, P.~M.,
  {Thorsett}, S.~E., {Bersier}, D., {Burud}, I., {Castro Cer{\'o}n}, J.~M.,
  {Castro-Tirado}, A.~J., {Conselice}, C., {Dahlen}, T., {Ferguson}, H.~C.,
  {Fynbo}, J.~P.~U., {Garnavich}, P.~M., {Gibbons}, R.~A., {Gorosabel}, J.,
  {Gull}, T.~R., {Hjorth}, J., {Holland}, S.~T., {Kouveliotou}, C., {Levay},
  Z., {Livio}, M., {Metzger}, M.~R., {Nugent}, P.~E., {Petro}, L., {Pian}, E.,
  {Rhoads}, J.~E., {Riess}, A.~G., {Sahu}, K.~C., {Smette}, A., {Tanvir},
  N.~R., {Wijers}, R.~A.~M.~J., \& {Woosley}, S.~E. 2006, \nat, 441, 463

\bibitem[{{Fruchter} {et~al.}(1999){Fruchter}, {Thorsett}, {Metzger}, {Sahu},
  {Petro}, {Livio}, {Ferguson}, {Pian}, {Hogg}, {Galama}, {Gull},
  {Kouveliotou}, {Macchetto}, {van Paradijs}, {Pedersen}, \&
  {Smette}}]{Fruchter1999}
{Fruchter}, A.~S., {Thorsett}, S.~E., {Metzger}, M.~R., {Sahu}, K.~C., {Petro},
  L., {Livio}, M., {Ferguson}, H., {Pian}, E., {Hogg}, D.~W., {Galama}, T.,
  {Gull}, T.~R., {Kouveliotou}, C., {Macchetto}, D., {van Paradijs}, J.,
  {Pedersen}, H., \& {Smette}, A. 1999, \apjl, 519, L13

\bibitem[{{Galama} {et~al.}(1998){Galama}, {Vreeswijk}, {Pian}, {Frontera},
  {Doublier}, {Gonzalez}, {Lidman}, {Augusteijn}, {Hainaut}, {Boehnhardt},
  {Patat}, \& {Leibundgut}}]{IAUC6895}
{Galama}, T.~J., {Vreeswijk}, P.~M., {Pian}, E., {Frontera}, F., {Doublier},
  V., {Gonzalez}, J.-F., {Lidman}, C., {Augusteijn}, T., {Hainaut}, O.~R.,
  {Boehnhardt}, H., {Patat}, F., \& {Leibundgut}, B. 1998, \iaucirc, 6895, 1

\bibitem[{{Garnett} {et~al.}(1997){Garnett}, {Shields}, {Skillman}, {Sagan}, \&
  {Dufour}}]{Garnett}
{Garnett}, D.~R., {Shields}, G.~A., {Skillman}, E.~D., {Sagan}, S.~P., \&
  {Dufour}, R.~J. 1997, \apj, 489, 63

\bibitem[{{Graham}(2012)}]{pregraham}
{Graham}, J.~F. 2012, in prep

\bibitem[{{Graham} {et~al.}(2009){Graham}, {Fruchter}, {Levan}, {Melandri},
  {Kewley}, {Levesque}, {Nysewander}, {Tanvir}, {Dahlen}, {Bersier},
  {Wiersema}, {Bonfield}, \& {Martinez-Sansigre}}]{070714Bpaper}
{Graham}, J.~F., {Fruchter}, A.~S., {Levan}, A.~J., {Melandri}, A., {Kewley},
  L.~J., {Levesque}, E.~M., {Nysewander}, M., {Tanvir}, N.~R., {Dahlen}, T.,
  {Bersier}, D., {Wiersema}, K., {Bonfield}, D.~G., \& {Martinez-Sansigre}, A.
  2009, \apj, 698, 1620

\bibitem[{{Hammer} {et~al.}(2006){Hammer}, {Flores}, {Schaerer},
  {Dessauges-Zavadsky}, {Le Floc'h}, \& {Puech}}]{Hammer}
{Hammer}, F., {Flores}, H., {Schaerer}, D., {Dessauges-Zavadsky}, M., {Le
  Floc'h}, E., \& {Puech}, M. 2006, A\&A, 454, 103

\bibitem[{{Hjorth} {et~al.}(2003){Hjorth}, {Sollerman}, {M{\o}ller}, {Fynbo},
  {Woosley}, {Kouveliotou}, {Tanvir}, {Greiner}, {Andersen}, {Castro-Tirado},
  {Castro Cer{\'o}n}, {Fruchter}, {Gorosabel}, {Jakobsson}, {Kaper}, {Klose},
  {Masetti}, {Pedersen}, {Pedersen}, {Pian}, {Palazzi}, {Rhoads}, {Rol}, {van
  den Heuvel}, {Vreeswijk}, {Watson}, \& {Wijers}}]{Hjorth2003}
{Hjorth}, J., {Sollerman}, J., {M{\o}ller}, P., {Fynbo}, J.~P.~U., {Woosley},
  S.~E., {Kouveliotou}, C., {Tanvir}, N.~R., {Greiner}, J., {Andersen}, M.~I.,
  {Castro-Tirado}, A.~J., {Castro Cer{\'o}n}, J.~M., {Fruchter}, A.~S.,
  {Gorosabel}, J., {Jakobsson}, P., {Kaper}, L., {Klose}, S., {Masetti}, N.,
  {Pedersen}, H., {Pedersen}, K., {Pian}, E., {Palazzi}, E., {Rhoads}, J.~E.,
  {Rol}, E., {van den Heuvel}, E.~P.~J., {Vreeswijk}, P.~M., {Watson}, D., \&
  {Wijers}, R.~A.~M.~J. 2003, \nat, 423, 847

\bibitem[{{Jakobsson} {et~al.}(2005){Jakobsson}, {Frail}, {Fox}, {Moon},
  {Price}, {Kulkarni}, {Fynbo}, {Hjorth}, {Berger}, {McNaught}, \&
  {Dahle}}]{Jakobsson}
{Jakobsson}, P., {Frail}, D.~A., {Fox}, D.~B., {Moon}, D.-S., {Price}, P.~A.,
  {Kulkarni}, S.~R., {Fynbo}, J.~P.~U., {Hjorth}, J., {Berger}, E., {McNaught},
  R.~H., \& {Dahle}, H. 2005, \apj, 629, 45

\bibitem[{{Kelly} \& {Kirshner}(2012)}]{Kelly}
{Kelly}, P.~L. \& {Kirshner}, R.~P. 2012, \apj, 759, 107

\bibitem[{{Kennicutt}(1998)}]{K98SFR}
{Kennicutt}, Jr., R.~C. 1998, \araa, 36, 189

\bibitem[{{Kewley} {et~al.}(2007){Kewley}, {Brown}, {Geller}, {Kenyon}, \&
  {Kurtz}}]{Kewley2007}
{Kewley}, L.~J., {Brown}, W.~R., {Geller}, M.~J., {Kenyon}, S.~J., \& {Kurtz},
  M.~J. 2007, \aj, 133, 882

\bibitem[{{Kewley} \& {Dopita}(2002)}]{kd2002}
{Kewley}, L.~J. \& {Dopita}, M.~A. 2002, \apjs, 142, 35

\bibitem[{{Kewley} \& {Ellison}(2008)}]{KewleyEllison}
{Kewley}, L.~J. \& {Ellison}, S.~L. 2008, \apj, 681, 1183

\bibitem[{{Kobulnicky} \& {Kewley}(2004)}]{KobulnickyKewley}
{Kobulnicky}, H.~A. \& {Kewley}, L.~J. 2004, \apj, 617, 240

\bibitem[{{Kobulnicky} \& {Phillips}(2003)}]{KobulnickyPhillips}
{Kobulnicky}, H.~A. \& {Phillips}, A.~C. 2003, \apj, 599, 1031

\bibitem[{{Kocevski} \& {West}(2011)}]{Kocevski}
{Kocevski}, D. \& {West}, A.~A. 2011, \apjl, 735, L8

\bibitem[{{Komissarov} \& {Barkov}(2007)}]{2007MNRAS.382.1029K}
{Komissarov}, S.~S. \& {Barkov}, M.~V. 2007, \mnras, 382, 1029

\bibitem[{{Komissarov} {et~al.}(2009){Komissarov}, {Vlahakis}, {K{\"o}nigl}, \&
  {Barkov}}]{2009MNRAS.394.1182K}
{Komissarov}, S.~S., {Vlahakis}, N., {K{\"o}nigl}, A., \& {Barkov}, M.~V. 2009,
  \mnras, 394, 1182

\bibitem[{{Langer} \& {Norman}(2006)}]{Langer}
{Langer}, N. \& {Norman}, C.~A. 2006, \apjl, 638, L63

\bibitem[{{Le Floc'h} {et~al.}(2006){Le Floc'h}, {Charmandaris}, {Forrest},
  {Mirabel}, {Armus}, \& {Devost}}]{LeFloch2006}
{Le Floc'h}, E., {Charmandaris}, V., {Forrest}, W.~J., {Mirabel}, I.~F.,
  {Armus}, L., \& {Devost}, D. 2006, \apj, 642, 636

\bibitem[{{Le Floc'h} {et~al.}(2003){Le Floc'h}, {Duc}, {Mirabel}, {Sanders},
  {Bosch}, {Diaz}, {Donzelli}, {Rodrigues}, {Courvoisier}, {Greiner},
  {Mereghetti}, {Melnick}, {Maza}, \& {Minniti}}]{LeFlochblue}
{Le Floc'h}, E., {Duc}, P.-A., {Mirabel}, I.~F., {Sanders}, D.~B., {Bosch}, G.,
  {Diaz}, R.~J., {Donzelli}, C.~J., {Rodrigues}, I., {Courvoisier}, T.~J.-L.,
  {Greiner}, J., {Mereghetti}, S., {Melnick}, J., {Maza}, J., \& {Minniti}, D.
  2003, \aap, 400, 499

\bibitem[{{Le Floc'h} {et~al.}(2002){Le Floc'h}, {Mirabel}, \&
  {Duc}}]{LeFlochblue2002}
{Le Floc'h}, E., {Mirabel}, I.~F., \& {Duc}, P.-A. 2002, Journal of
  Astrophysics and Astronomy, 23, 119

\bibitem[{{Levesque} {et~al.}(2010{\natexlab{a}}){Levesque}, {Berger},
  {Kewley}, \& {Bagley}}]{Levesque051022}
{Levesque}, E.~M., {Berger}, E., {Kewley}, L.~J., \& {Bagley}, M.~M.
  2010{\natexlab{a}}, \aj, 139, 694

\bibitem[{{Levesque} {et~al.}(2010{\natexlab{b}}){Levesque}, {Kewley},
  {Graham}, \& {Fruchter}}]{Levesque020819B}
{Levesque}, E.~M., {Kewley}, L.~J., {Graham}, J.~F., \& {Fruchter}, A.~S.
  2010{\natexlab{b}}, \apjl, 712, L26

\bibitem[{{Liu} {et~al.}(2010){Liu}, {Liang}, {Gu}, {Zhao}, {Dai}, \&
  {Lu}}]{2010A&A...516A..16L}
{Liu}, T., {Liang}, E.-W., {Gu}, W.-M., {Zhao}, X.-H., {Dai}, Z.-G., \& {Lu},
  J.-F. 2010, \aap, 516, A16

\bibitem[{{Mannucci} {et~al.}(2010){Mannucci}, {Cresci}, {Maiolino}, {Marconi},
  \& {Gnerucci}}]{Mannucci}
{Mannucci}, F., {Cresci}, G., {Maiolino}, R., {Marconi}, A., \& {Gnerucci}, A.
  2010, \mnras, 408, 2115

\bibitem[{{Mannucci} {et~al.}(2011){Mannucci}, {Salvaterra}, \&
  {Campisi}}]{MannucciLGRBs}
{Mannucci}, F., {Salvaterra}, R., \& {Campisi}, M.~A. 2011, \mnras, 414, 1263

\bibitem[{{Marscher} {et~al.}(2008){Marscher}, {Jorstad}, {D'Arcangelo},
  {Smith}, {Williams}, {Larionov}, {Oh}, {Olmstead}, {Aller}, {Aller},
  {McHardy}, {L{\"a}hteenm{\"a}ki}, {Tornikoski}, {Valtaoja}, {Hagen-Thorn},
  {Kopatskaya}, {Gear}, {Tosti}, {Kurtanidze}, {Nikolashvili}, {Sigua},
  {Miller}, \& {Ryle}}]{2008Natur.452..966M}
{Marscher}, A.~P., {Jorstad}, S.~G., {D'Arcangelo}, F.~D., {Smith}, P.~S.,
  {Williams}, G.~G., {Larionov}, V.~M., {Oh}, H., {Olmstead}, A.~R., {Aller},
  M.~F., {Aller}, H.~D., {McHardy}, I.~M., {L{\"a}hteenm{\"a}ki}, A.,
  {Tornikoski}, M., {Valtaoja}, E., {Hagen-Thorn}, V.~A., {Kopatskaya}, E.~N.,
  {Gear}, W.~K., {Tosti}, G., {Kurtanidze}, O., {Nikolashvili}, M., {Sigua},
  L., {Miller}, H.~R., \& {Ryle}, W.~T. 2008, \nat, 452, 966

\bibitem[{{McBreen} {et~al.}(2002){McBreen}, {McBreen}, {Quilligan}, \&
  {Hanlon}}]{2002A&A...385L..19M}
{McBreen}, S., {McBreen}, B., {Quilligan}, F., \& {Hanlon}, L. 2002, \aap, 385,
  L19

\bibitem[{{Meszaros} \& {Rees}(1997)}]{1997ApJ...482L..29M}
{Meszaros}, P. \& {Rees}, M.~J. 1997, \apjl, 482, L29

\bibitem[{{Modjaz} {et~al.}(2008){Modjaz}, {Kewley}, {Kirshner}, {Stanek},
  {Challis}, {Garnavich}, {Greene}, {Kelly}, \& {Prieto}}]{Modjaz2008}
{Modjaz}, M., {Kewley}, L., {Kirshner}, R.~P., {Stanek}, K.~Z., {Challis}, P.,
  {Garnavich}, P.~M., {Greene}, J.~E., {Kelly}, P.~L., \& {Prieto}, J.~L. 2008,
  \aj, 135, 1136

\bibitem[{{Nagakura} {et~al.}(2011){Nagakura}, {Ito}, {Kiuchi}, \&
  {Yamada}}]{2011ApJ...731...80N}
{Nagakura}, H., {Ito}, H., {Kiuchi}, K., \& {Yamada}, S. 2011, \apj, 731, 80

\bibitem[{{Nagataki} {et~al.}(2006){Nagataki}, {Mizuta}, \&
  {Sato}}]{2006ApJ...647.1255N}
{Nagataki}, S., {Mizuta}, A., \& {Sato}, K. 2006, \apj, 647, 1255

\bibitem[{{Nugis} \& {Lamers}(2000)}]{Nugis}
{Nugis}, T. \& {Lamers}, H.~J.~G.~L.~M. 2000, \aap, 360, 227

\bibitem[{{Ofek} {et~al.}(2007){Ofek}, {Cenko}, {Gal-Yam}, {Fox}, {Nakar},
  {Rau}, {Frail}, {Kulkarni}, {Price}, {Schmidt}, {Soderberg}, {Peterson},
  {Berger}, {Sharon}, {Shemmer}, {Penprase}, {Chevalier}, {Brown}, {Burrows},
  {Gehrels}, {Harrison}, {Holland}, {Mangano}, {McCarthy}, {Moon}, {Nousek},
  {Persson}, {Piran}, \& {Sari}}]{Ofek}
{Ofek}, E.~O., {Cenko}, S.~B., {Gal-Yam}, A., {Fox}, D.~B., {Nakar}, E., {Rau},
  A., {Frail}, D.~A., {Kulkarni}, S.~R., {Price}, P.~A., {Schmidt}, B.~P.,
  {Soderberg}, A.~M., {Peterson}, B., {Berger}, E., {Sharon}, K., {Shemmer},
  O., {Penprase}, B.~E., {Chevalier}, R.~A., {Brown}, P.~J., {Burrows}, D.~N.,
  {Gehrels}, N., {Harrison}, F., {Holland}, S.~T., {Mangano}, V., {McCarthy},
  P.~J., {Moon}, D.-S., {Nousek}, J.~A., {Persson}, S.~E., {Piran}, T., \&
  {Sari}, R. 2007, \apj, 662, 1129

\bibitem[{{Pagel} {et~al.}(1979){Pagel}, {Edmunds}, {Blackwell}, {Chun}, \&
  {Smith}}]{Pagel1979}
{Pagel}, B.~E.~J., {Edmunds}, M.~G., {Blackwell}, D.~E., {Chun}, M.~S., \&
  {Smith}, G. 1979, \mnras, 189, 95

\bibitem[{{Pagel} {et~al.}(1980){Pagel}, {Edmunds}, \& {Smith}}]{Pagel1980}
{Pagel}, B.~E.~J., {Edmunds}, M.~G., \& {Smith}, G. 1980, \mnras, 193, 219

\bibitem[{{Peeples} {et~al.}(2009){Peeples}, {Pogge}, \& {Stanek}}]{Peeples}
{Peeples}, M.~S., {Pogge}, R.~W., \& {Stanek}, K.~Z. 2009, \apj, 695, 259

\bibitem[{{Podsiadlowski} {et~al.}(2010){Podsiadlowski}, {Ivanova}, {Justham},
  \& {Rappaport}}]{Podsi}
{Podsiadlowski}, P., {Ivanova}, N., {Justham}, S., \& {Rappaport}, S. 2010,
  \mnras, 406, 840

\bibitem[{{Prieto} {et~al.}(2008){Prieto}, {Stanek}, \& {Beacom}}]{Prieto}
{Prieto}, J.~L., {Stanek}, K.~Z., \& {Beacom}, J.~F. 2008, \apj, 673, 999

\bibitem[{{Quilligan} {et~al.}(2002){Quilligan}, {McBreen}, {Hanlon},
  {McBreen}, {Hurley}, \& {Watson}}]{2002A&A...385..377Q}
{Quilligan}, F., {McBreen}, B., {Hanlon}, L., {McBreen}, S., {Hurley}, K.~J.,
  \& {Watson}, D. 2002, \aap, 385, 377

\bibitem[{{Rest} {et~al.}(2001){Rest}, {Miceli}, \& {Covarrubias}}]{IAUC7740}
{Rest}, A., {Miceli}, A., \& {Covarrubias}, R. 2001, \iaucirc, 7740, 1

\bibitem[{{Sako} {et~al.}(2008){Sako}, {Bassett}, {Becker}, {Cinabro},
  {DeJongh}, {Depoy}, {Dilday}, {Doi}, {Frieman}, {Garnavich}, {Hogan},
  {Holtzman}, {Jha}, {Kessler}, {Konishi}, {Lampeitl}, {Marriner}, {Miknaitis},
  {Nichol}, {Prieto}, {Riess}, {Richmond}, {Romani}, {Schneider}, {Smith},
  {Subba Rao}, {Takanashi}, {Tokita}, {van der Heyden}, {Yasuda}, {Zheng},
  {Barentine}, {Brewington}, {Choi}, {Dembicky}, {Harnavek}, {Ihara}, {Im},
  {Ketzeback}, {Kleinman}, {Krzesi{\'n}ski}, {Long}, {Malanushenko},
  {Malanushenko}, {McMillan}, {Morokuma}, {Nitta}, {Pan}, {Saurage}, \&
  {Snedden}}]{SDSS-SNe2}
{Sako}, M., {Bassett}, B., {Becker}, A., {Cinabro}, D., {DeJongh}, F., {Depoy},
  D.~L., {Dilday}, B., {Doi}, M., {Frieman}, J.~A., {Garnavich}, P.~M.,
  {Hogan}, C.~J., {Holtzman}, J., {Jha}, S., {Kessler}, R., {Konishi}, K.,
  {Lampeitl}, H., {Marriner}, J., {Miknaitis}, G., {Nichol}, R.~C., {Prieto},
  J.~L., {Riess}, A.~G., {Richmond}, M.~W., {Romani}, R., {Schneider}, D.~P.,
  {Smith}, M., {Subba Rao}, M., {Takanashi}, N., {Tokita}, K., {van der
  Heyden}, K., {Yasuda}, N., {Zheng}, C., {Barentine}, J., {Brewington}, H.,
  {Choi}, C., {Dembicky}, J., {Harnavek}, M., {Ihara}, Y., {Im}, M.,
  {Ketzeback}, W., {Kleinman}, S.~J., {Krzesi{\'n}ski}, J., {Long}, D.~C.,
  {Malanushenko}, E., {Malanushenko}, V., {McMillan}, R.~J., {Morokuma}, T.,
  {Nitta}, A., {Pan}, K., {Saurage}, G., \& {Snedden}, S.~A. 2008, \aj, 135,
  348

\bibitem[{{S{\'a}nchez Almeida} {et~al.}(2010){S{\'a}nchez Almeida}, {Aguerri},
  {Mu{\~n}oz-Tu{\~n}{\'o}n}, \& {de Vicente}}]{SDSS_spec_class}
{S{\'a}nchez Almeida}, J., {Aguerri}, J.~A.~L., {Mu{\~n}oz-Tu{\~n}{\'o}n}, C.,
  \& {de Vicente}, A. 2010, \apj, 714, 487

\bibitem[{{Sanders} {et~al.}(2012){Sanders}, {Soderberg}, {Levesque}, {Foley},
  {Chornock}, {Milisavljevic}, {Margutti}, {Berger}, {Drout}, {Czekala}, \&
  {Dittmann}}]{Sanders}
{Sanders}, N.~E., {Soderberg}, A.~M., {Levesque}, E.~M., {Foley}, R.~J.,
  {Chornock}, R., {Milisavljevic}, D., {Margutti}, R., {Berger}, E., {Drout},
  M.~R., {Czekala}, I., \& {Dittmann}, J.~A. 2012, \apj, 758, 132

\bibitem[{{Sari}(1999)}]{1999ApJ...524L..43S}
{Sari}, R. 1999, \apjl, 524, L43

\bibitem[{{Savaglio} {et~al.}(2009){Savaglio}, {Glazebrook}, \& {Le
  Borgne}}]{Savaglio}
{Savaglio}, S., {Glazebrook}, K., \& {Le Borgne}, D. 2009, \apj, 691, 182

\bibitem[{{Schlegel} {et~al.}(1998){Schlegel}, {Finkbeiner}, \&
  {Davis}}]{Schlegel}
{Schlegel}, D.~J., {Finkbeiner}, D.~P., \& {Davis}, M. 1998, \apj, 500, 525

\bibitem[{{Sollerman} {et~al.}(2005){Sollerman}, {{\"O}stlin}, {Fynbo},
  {Hjorth}, {Fruchter}, \& {Pedersen}}]{Sollerman}
{Sollerman}, J., {{\"O}stlin}, G., {Fynbo}, J.~P.~U., {Hjorth}, J., {Fruchter},
  A., \& {Pedersen}, K. 2005, New Astronomy, 11, 103

\bibitem[{{Stanek} {et~al.}(2007){Stanek}, {Dai}, {Prieto}, {An}, {Garnavich},
  {Calkins}, {Serven}, {Worthey}, {Hao}, {Dobrzycki}, {Howk}, \&
  {Matheson}}]{Stanek2007}
{Stanek}, K.~Z., {Dai}, X., {Prieto}, J.~L., {An}, D., {Garnavich}, P.~M.,
  {Calkins}, M.~L., {Serven}, J., {Worthey}, G., {Hao}, H., {Dobrzycki}, A.,
  {Howk}, C., \& {Matheson}, T. 2007, \apjl, 654, L21

\bibitem[{{Stanek} {et~al.}(2006){Stanek}, {Gnedin}, {Beacom}, {Gould},
  {Johnson}, {Kollmeier}, {Modjaz}, {Pinsonneault}, {Pogge}, \&
  {Weinberg}}]{Stanek2006}
{Stanek}, K.~Z., {Gnedin}, O.~Y., {Beacom}, J.~F., {Gould}, A.~P., {Johnson},
  J.~A., {Kollmeier}, J.~A., {Modjaz}, M., {Pinsonneault}, M.~H., {Pogge}, R.,
  \& {Weinberg}, D.~H. 2006, Acta Astronomica, 56, 333

\bibitem[{{Stanek} {et~al.}(2003){Stanek}, {Matheson}, {Garnavich}, {Martini},
  {Berlind}, {Caldwell}, {Challis}, {Brown}, {Schild}, {Krisciunas}, {Calkins},
  {Lee}, {Hathi}, {Jansen}, {Windhorst}, {Echevarria}, {Eisenstein}, {Pindor},
  {Olszewski}, {Harding}, {Holland}, \& {Bersier}}]{Stanek2003}
{Stanek}, K.~Z., {Matheson}, T., {Garnavich}, P.~M., {Martini}, P., {Berlind},
  P., {Caldwell}, N., {Challis}, P., {Brown}, W.~R., {Schild}, R.,
  {Krisciunas}, K., {Calkins}, M.~L., {Lee}, J.~C., {Hathi}, N., {Jansen},
  R.~A., {Windhorst}, R., {Echevarria}, L., {Eisenstein}, D.~J., {Pindor}, B.,
  {Olszewski}, E.~W., {Harding}, P., {Holland}, S.~T., \& {Bersier}, D. 2003,
  \apjl, 591, L17

\bibitem[{{Svensson} {et~al.}(2010){Svensson}, {Levan}, {Tanvir}, {Fruchter},
  \& {Strolger}}]{Svensson}
{Svensson}, K.~M., {Levan}, A.~J., {Tanvir}, N.~R., {Fruchter}, A.~S., \&
  {Strolger}, L. 2010, \mnras, 479

\bibitem[{{Th{\"o}ne} {et~al.}(2008){Th{\"o}ne}, {Fynbo}, {{\"O}stlin},
  {Milvang-Jensen}, {Wiersema}, {Malesani}, {Della Monica Ferreira},
  {Gorosabel}, {Kann}, {Watson}, {Micha{\l}owski}, {Fruchter}, {Levan},
  {Hjorth}, \& {Sollerman}}]{060505}
{Th{\"o}ne}, C.~C., {Fynbo}, J.~P.~U., {{\"O}stlin}, G., {Milvang-Jensen}, B.,
  {Wiersema}, K., {Malesani}, D., {Della Monica Ferreira}, D., {Gorosabel}, J.,
  {Kann}, D.~A., {Watson}, D., {Micha{\l}owski}, M.~J., {Fruchter}, A.~S.,
  {Levan}, A.~J., {Hjorth}, J., \& {Sollerman}, J. 2008, \apj, 676, 1151

\bibitem[{{Toma} {et~al.}(2007){Toma}, {Ioka}, {Sakamoto}, \&
  {Nakamura}}]{2007ApJ...659.1420T}
{Toma}, K., {Ioka}, K., {Sakamoto}, T., \& {Nakamura}, T. 2007, \apj, 659, 1420

\bibitem[{{Tremonti} {et~al.}(2004){Tremonti}, {Heckman}, {Kauffmann},
  {Brinchmann}, {Charlot}, {White}, {Seibert}, {Peng}, {Schlegel}, {Uomoto},
  {Fukugita}, \& {Brinkmann}}]{T04}
{Tremonti}, C.~A., {Heckman}, T.~M., {Kauffmann}, G., {Brinchmann}, J.,
  {Charlot}, S., {White}, S.~D.~M., {Seibert}, M., {Peng}, E.~W., {Schlegel},
  D.~J., {Uomoto}, A., {Fukugita}, M., \& {Brinkmann}, J. 2004, \apj, 613, 898

\bibitem[{{Vink} \& {de Koter}(2005)}]{Vink}
{Vink}, J.~S. \& {de Koter}, A. 2005, \aap, 442, 587

\bibitem[{{Vreeswijk} {et~al.}(2001){Vreeswijk}, {Fruchter}, {Kaper}, {Rol},
  {Galama}, {van Paradijs}, {Kouveliotou}, {Wijers}, {Pian}, {Palazzi},
  {Masetti}, {Frontera}, {Savaglio}, {Reinsch}, {Hessman}, {Beuermann},
  {Nicklas}, \& {van den Heuvel}}]{Vreeswijk}
{Vreeswijk}, P.~M., {Fruchter}, A., {Kaper}, L., {Rol}, E., {Galama}, T.~J.,
  {van Paradijs}, J., {Kouveliotou}, C., {Wijers}, R.~A.~M.~J., {Pian}, E.,
  {Palazzi}, E., {Masetti}, N., {Frontera}, F., {Savaglio}, S., {Reinsch}, K.,
  {Hessman}, F.~V., {Beuermann}, K., {Nicklas}, H., \& {van den Heuvel},
  E.~P.~J. 2001, \apj, 546, 672

\bibitem[{{Wirth} {et~al.}(2004){Wirth}, {Willmer}, {Amico}, {Chaffee},
  {Goodrich}, {Kwok}, {Lyke}, {Mader}, {Tran}, {Barger}, {Cowie}, {Capak},
  {Coil}, {Cooper}, {Conrad}, {Davis}, {Faber}, {Hu}, {Koo}, {Le Mignant},
  {Newman}, \& {Songaila}}]{TKRS}
{Wirth}, G.~D., {Willmer}, C.~N.~A., {Amico}, P., {Chaffee}, F.~H., {Goodrich},
  R.~W., {Kwok}, S., {Lyke}, J.~E., {Mader}, J.~A., {Tran}, H.~D., {Barger},
  A.~J., {Cowie}, L.~L., {Capak}, P., {Coil}, A.~L., {Cooper}, M.~C., {Conrad},
  A., {Davis}, M., {Faber}, S.~M., {Hu}, E.~M., {Koo}, D.~C., {Le Mignant}, D.,
  {Newman}, J.~A., \& {Songaila}, A. 2004, \aj, 127, 3121

\bibitem[{{Wolf} \& {Podsiadlowski}(2007)}]{Wolf}
{Wolf}, C. \& {Podsiadlowski}, P. 2007, \mnras, 375, 1049

\bibitem[{{Woosley} \& {Bloom}(2006)}]{Woosley}
{Woosley}, S.~E. \& {Bloom}, J.~S. 2006, \araa, 44, 507

\bibitem[{{Yoon} \& {Langer}(2005)}]{Yoon}
{Yoon}, S.-C. \& {Langer}, N. 2005, \aap, 443, 643

\bibitem[{{Yuan} \& {Zhang}(2012)}]{2012ApJ...757...56Y}
{Yuan}, F. \& {Zhang}, B. 2012, \apj, 757, 56

\bibitem[{{Zaritsky} {et~al.}(1994){Zaritsky}, {Kennicutt}, \& {Huchra}}]{Z94}
{Zaritsky}, D., {Kennicutt}, Jr., R.~C., \& {Huchra}, J.~P. 1994, \apj, 420, 87

\end{thebibliography}

\end{document}